\documentclass[preprintnumbers,amssymb,amsmath,superscriptaddress,letterpaper,nofootinbib,showpacs,twocolumn,floatfix]{revtex4}[10pt]
\usepackage{graphicx}
\usepackage{dcolumn}
\usepackage{bm}
\usepackage{natbib}
\usepackage{calligra}
\usepackage[T1]{fontenc}
\usepackage{egothic}
\usepackage[T1]{fontenc}
\newfont{\rsfsten}{rsfs10 scaled 1200}
\newfont{\rsfsseven}{rsfs10 scaled 1200}
\newfont{\rsfsfive}{rsfs10 scaled 1200}
\usepackage{epsfig}
\usepackage{units}

\newcommand{\be}{\begin{equation}}
\newcommand{\ee}{\end{equation}}
\newcommand{\bea}{\begin{eqnarray}}
\newcommand{\eea}{\end{eqnarray}}

\def\lsim{\mathrel{\raise.3ex\hbox{$<$\kern-.75em\lower1ex\hbox{$\sim$}}}}
\def\gsim{\mathrel{\raise.3ex\hbox{$>$\kern-.75em\lower1ex\hbox{$\sim$}}}}

\begin{document}

\hspace*{130mm}{\large \tt FERMILAB-PUB-14-236-A}
\vskip 0.2in

\title{A New Determination of the Spectra and Luminosity Function of Gamma-Ray Millisecond Pulsars}

%{On The Gamma-Ray Spectra and Luminosity Function of Millisecond Pulsars}

\author{Ilias Cholis}
\affiliation{Fermi National Accelerator Laboratory, Center for Particle Astrophysics, Batavia, IL}
\author{Dan Hooper}
\affiliation{Fermi National Accelerator Laboratory, Center for Particle Astrophysics, Batavia, IL}
\affiliation{University of Chicago, Department of Astronomy and Astrophysics, 5640 S. Ellis Ave., Chicago, IL}
\author{Tim Linden}
\affiliation{University of Chicago, Kavli Institute for Cosmological Physics, Chicago, IL}
\date{\today}

\begin{abstract}

In this article, we revisit the gamma-ray emission observed from millisecond pulsars and globular clusters.  Based on 5.6 years of data from the Fermi Gamma-Ray Space Telescope, we report gamma-ray spectra for 61 millisecond pulsars, finding most to be well fit by a power-law with an exponential cutoff, producing to a spectral peak near $\sim$1-2 GeV (in $E^2 dN/dE$ units). Additionally, while most globular clusters exhibit a similar spectral shape, we identify a few with significantly softer spectra. We also determine the gamma-ray luminosity function of millisecond pulsars using the population found in the nearby field of the Milky Way, and within the globular cluster 47 Tucanae. We find that the gamma-ray emission observed from globular clusters is dominated by a relatively small number of bright millisecond pulsars, and that low-luminosity pulsars account for only a small fraction of the total flux. Our results also suggest that the gamma-ray emission from millisecond pulsars is more isotropic and less strongly beamed than the emission at X-ray wavelengths. Furthermore, the observed distribution of apparent gamma-ray efficiencies provides support for the slot gap or the outer gap models over those in which the gamma-ray emission originates from regions close to the neutron star's magnetic poles (polar cap models).

\end{abstract}


\maketitle

\section{Introduction}
\label{sec:intro}

Pulsars make up many of the brightest sources in the gamma-ray sky. These objects are understood to be rapidly spinning neutron stars whose radio, X-ray, and gamma-ray emission is made possible by the steady conversion of rotational kinetic energy into the acceleration of particles. Most observed pulsars are relatively young, with rotational periods in the range of 0.1 to 10 seconds, and magnetic fields of $\sim$$10^{11}$-$10^{13}$ G. As a result of magnetic-dipole braking, such pulsars slow down and become less luminous relatively quickly. The luminosities of the Crab and Vela pulsars, for example, will decrease by a factor of $\sim$$10^2$ over the course of the next few hundred thousand years.
% by a factor of $\sim$$10^3$ over the course of the the next million years.

%In addition to young pulsars, a population of much older and longer-lived pulsars has been observed. Such objects represent a later stage of evolution, 
In addition to this main population of young pulsars, there exists a second population of much older pulsars (with characteristic ages $\tau \equiv P/2\dot{P} \gsim 10^{8}$ yr, where $P$ is the rotational period and $\dot{P}$ is the derivative of the period with respect to time). These objects are thought to represent a later stage of evolution, in which a neutron star has been ``recycled'' by the transfer of angular momentum from a companion star. This process of ``spinning-up'' a neutron star can result in a pulsar with a much greater rotational speed ($P$$\sim$1.5-100 ms) and a much weaker magnetic field ($B$$\sim$$10^8$-$10^{10}$ G) than is found among young pulsars.  While such millisecond pulsars (MSPs) are often comparably luminous to young pulsars, their evolution proceeds at a much slower rate, allowing them to remain bright and rapidly spinning for billions of years. 

The catalog of MSPs observed in gamma-rays has grown significantly in recent years. More than 60 MSPs have been detected by the Fermi Gamma-Ray Space Telescope~\cite{TheFermi-LAT:2013ssa,Guillemot:2011th}, as well as many detected at X-ray energies~\cite{He:2013nua}. The population of known radio MSPs has increased by $\sim$50\% as a result of follow up searches coincident with unassociated Fermi sources, and multi-wavelength data now support the conclusion that most MSPs are significant gamma-ray emitters. This observational program has revealed a great deal about the nature of MSPs. For example, although most MSPs ($\gsim$ 75\% of those observed) are in binary systems, a non-insignificant fraction have been found in systems in which the (low-mass) companion has been destroyed by the pulsar wind (so-called ``black widow'' systems).  Significant progress has also been made in understanding the geometry and mechanisms for particle acceleration and emission from such objects~\cite{Johnson:2014qwa}.

In this paper, we make use of 5.6 years of data from the Fermi Telescope to determine the gamma-ray spectra of 61 MSPs. Although some pulsar-to-pulsar variation is observed among this population, such sources generally exhibit spectra that is well described by an exponentially suppressed power-law, peaking at an energy of $\sim$1-2 GeV (in $E^2 dN/dE$ units). 
The fact that this spectral shape is similar to that predicted from the annihilations of WIMPs with masses in the range of approximately 20-40 GeV make MSPs a potentially important background for indirect dark matter searches~\cite{Baltz:2006sv}. In a companion letter~\cite{letter}, we use the information presented in this study to revisit the question of whether MSPs could account for the excess GeV emission observed from the region of the Galactic Center~\cite{Daylan:2014rsa,Goodenough:2009gk,Hooper:2010mq,Hooper:2011ti,Abazajian:2012pn,Hooper:2013rwa,Gordon:2013vta,Abazajian:2014fta}.

In addition to spectral measurements, we have also used the current catalog of observed MSPs to derive the gamma-ray luminosity function for this class of objects. We do this though two independent and complementary methods. First, we use the nearby population of MSPs detected by Fermi to derive the luminosity function between $L \simeq 3\times 10^{31}$ and $3\times 10^{35}$ erg/s (integrated above 0.1 GeV). Second, we use observations of MSPs in the globular cluster 47 Tucanae (also known as 47 Tuc, or NGC 104), combined with an empirical correlation relating their X-ray and gamma-ray luminosities, to derive the gamma-ray luminosity function for the MSPs contained within that system. Despite their dependance on different systematic uncertainties and assumptions, these two approaches produce very similar results.  The derived luminosity function  implies that the gamma-ray emission from globular clusters is dominated by a small number of luminous millisecond pulsars, with low-luminosity pulsars accounting for little of the total flux. Our results also suggest that the gamma-ray emission from millisecond pulsars is somewhat more isotropic (coming from a larger emission region, and thus less strongly beamed) than is the emission produced at X-ray wavelengths. 

In comparing the isotropic gamma-ray luminosity to the spin-down power of MSPs, we find a larger number of such sources with moderate gamma-ray efficiencies, somewhat favoring the slot gap or the outer gap models over polar cap models. We have also searched for correlations between the spectral properties of MSPs and their physical properties, including gamma-ray luminosity, flux, distance, period, and age. However, we have not identified any significant correlations, other than those that appear to be due to selection effects.

\section{Fermi Data Analysis}
\label{data}

In order to calculate the observed fluxes and spectra of gamma-ray pulsars and globular clusters, we examine data taken over approximately 5.6 years of Fermi-LAT observations,\footnote{MET range 239557417 - 415926248} utilizing the Pass 7 Reprocessed photons in the energy range of 100 MeV to 100 GeV. We exclude events arriving at a zenith angle greater than 100$^\circ$,\footnote{Due to the potential for low-energy photons from the Earth's albedo to contaminate our data set, we have also performed this analysis using a more conservative zenith angle cut of 90$^{\circ}$. We find that this variation leads to no qualitative changes in our results.} as well as those which do not pass the ``Clean'' photon data selection. We also exclude events that were observed while the instrument was not in science survey mode, when the instrumental rocking angle was $>$52$^\circ$, or when the instrument was passing through the South Atlantic Anomaly. For each source, we examine the photons observed within a 14$^\circ$x14$^\circ$ box centered around the location of the source, and divide the photons into 140x140 angular bins and 15 evenly spaced logarithmic energy bins. We analyze the instrumental exposure throughout the region of interest using the {\tt P7REP\_CLEAN\_V15} instrumental response functions.

In order to model the photons observed within our region of interest, we employ the latest model for diffuse galactic gamma-ray emission, {\tt gll\_iem\_v05.fit}, the latest isotropic emission template for the Clean photon data selection {\tt iso\_clean\_v05.txt}, and include all point-sources listed in the Fermi 2FGL Catalog~\cite{Fermi:2011bm}. We allow the normalization of each source to float independently in each energy bin, and do not impose any parameterization on their spectral shape. We note that our model also includes 2FGL sources which lie nearby (but outside) of the region of interest. These sources may provide an important emission component if they lie near the edges of the region of interest and are relatively bright. We include these sources in our template model, but do not allow the spectrum or intensity of these sources to deviate from their best fit 2FGL values.

In order to calculate the best fit flux from each source (in each energy bin), we use the Fermi-LAT {\tt gtlike} code, utilizing the {\tt MINUIT} algorithm. We note that calculating the source flux independently of a spectral model can produce considerable uncertainty, especially at high energies where the source likely contributes very few photons. While this should be taken into account accurately by the likelihood fitting algorithm, the {\tt gtlike} code forces the best fit flux to be non-negative in each individual energy bin. Although this is physically true for any real source, it creates mathematical inconsistencies in the calculation of the best fit flux and flux error in regions where the total intensity is over-subtracted by the best-fit model. Due to this inconsistency, we also calculate the upper limit of the source flux using the {\tt pyLikelihood UpperLimits} tool in each energy bin, and present only the upper limit whenever the flux is smaller than the calculated flux error. 
                   
%%%%%%%%%%%%%%%%%%%%%%%%%%%%%%%%%%%%%%%%                   
                   
\section{Spectral Measurements}
\label{sec:spectral}

In this section, we follow the procedure and data set described in the previous section to determine the gamma-ray spectra of 61 MSPs and 16 globular clusters. The spectra of many of these sources have not been reported previously in the literature. For those sources with spectra previously reported in the 2FGL Catalog~\cite{Fermi:2011bm} and/or Second Pulsar Catalog (2PC)~\cite{TheFermi-LAT:2013ssa}, we provide an update using the current data set (5.6 years of data as opposed to the 2 or 3 years utilized in the previous catalogs). 

\subsection{Spectra of Individual and Stacked Millisecond Pulsars}

The Fermi Collaboration maintains a public list of pulsars that they have detected. At present, this list contains 147 pulsars, 62 of which exhibit millisecond-scale periods.
\footnote{For the complete public list of Fermi detected pulsars, see \url{https://confluence.slac.stanford.edu/display/GLAMCOG/} \\ \url{Public+List+of+LAT-Detected+Gamma-Ray+Pulsars}} The most recent pulsar spectra presented by the Fermi Collaboration can be found in the Second Fermi Catalog of Gamma-Ray Pulsars, which describes the gamma-ray spectra of 40 MSPs, as derived from three years of observation~\cite{TheFermi-LAT:2013ssa}.%In this section, we make use of 5.6 years of Fermi data to determine the gamma-ray spectrum from all of the millisecond pulsars in the Fermi Collaboration's public list.

\begin{figure*}
\includegraphics[width=3.40in,angle=0]{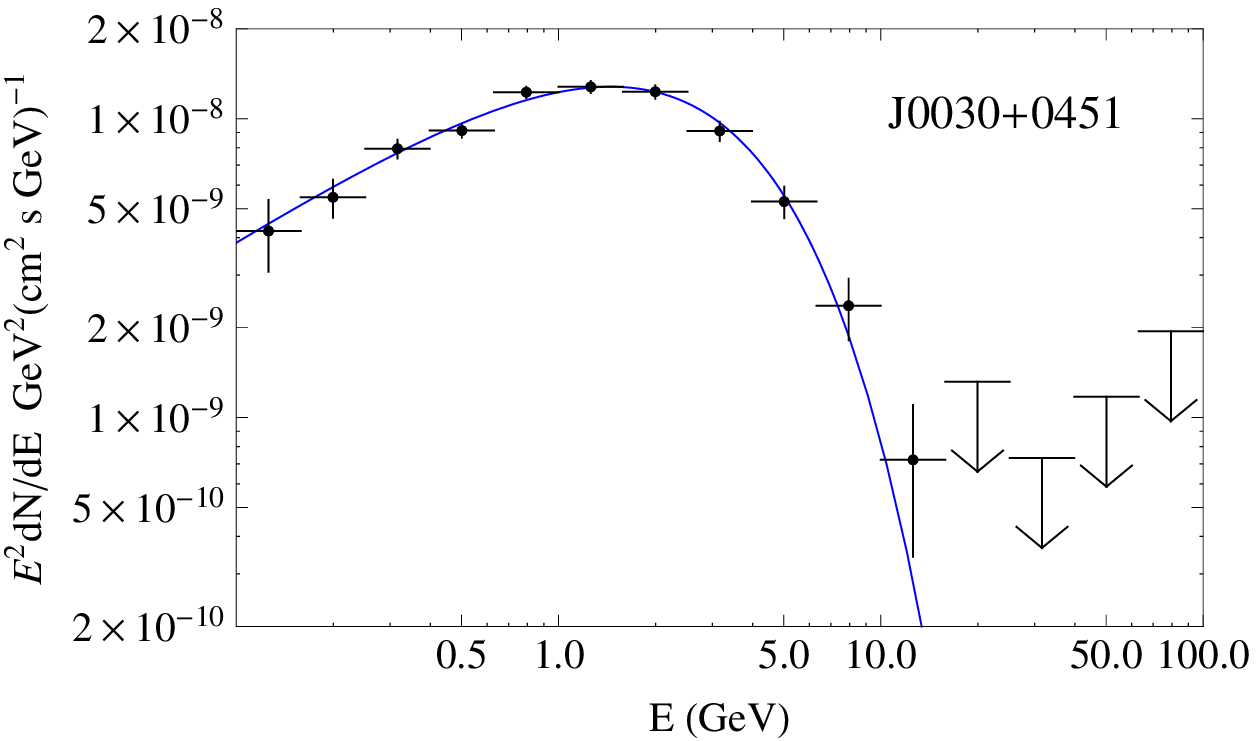} 
\includegraphics[width=3.40in,angle=0]{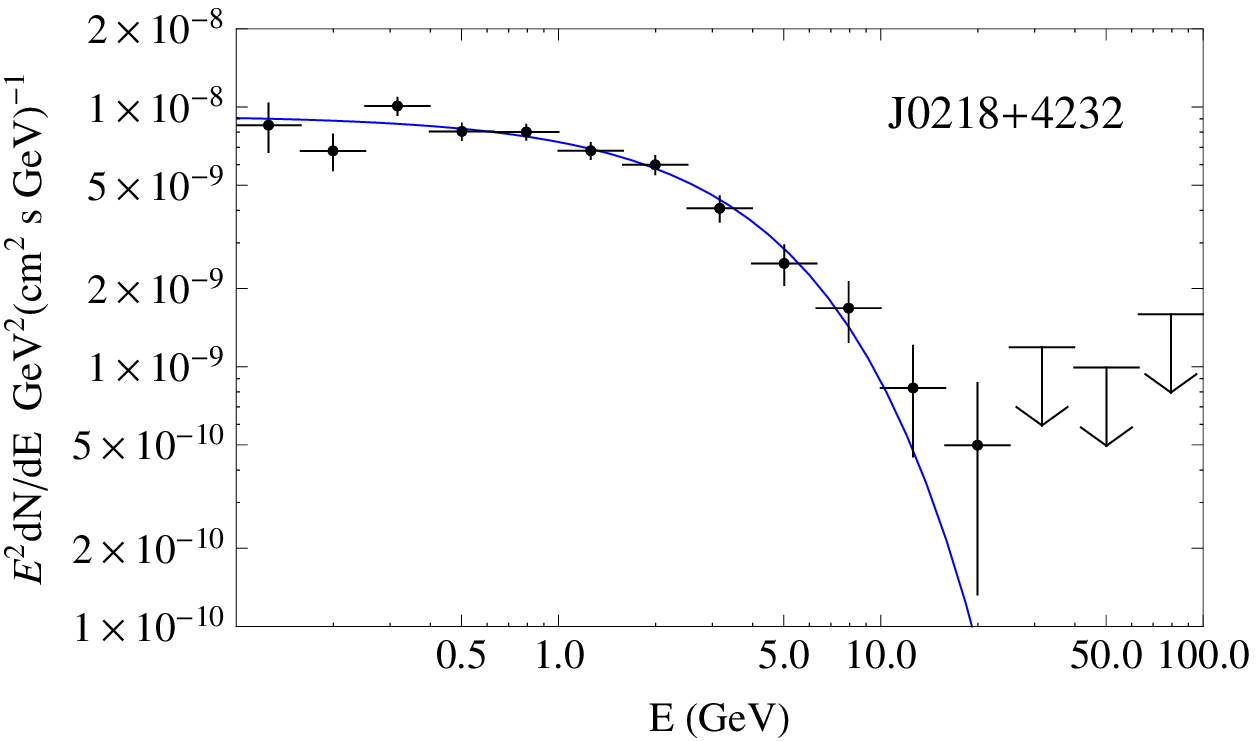} \\
\includegraphics[width=3.40in,angle=0]{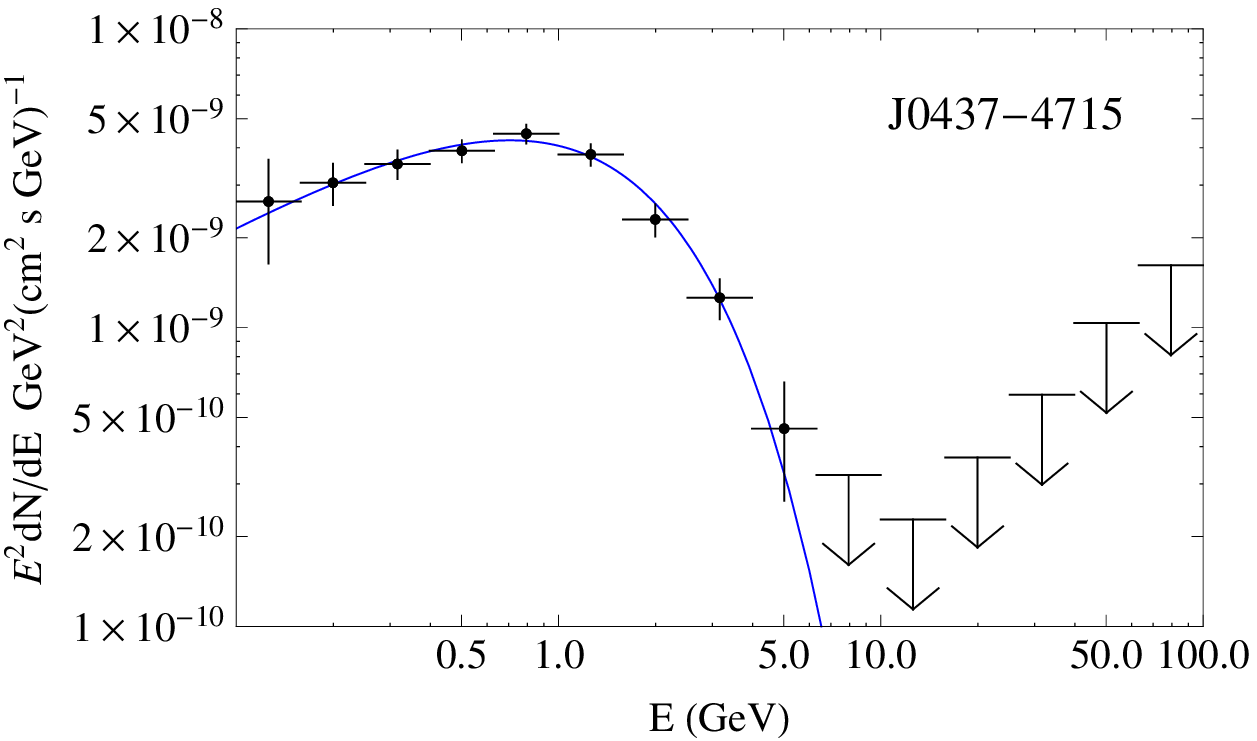} 
\includegraphics[width=3.40in,angle=0]{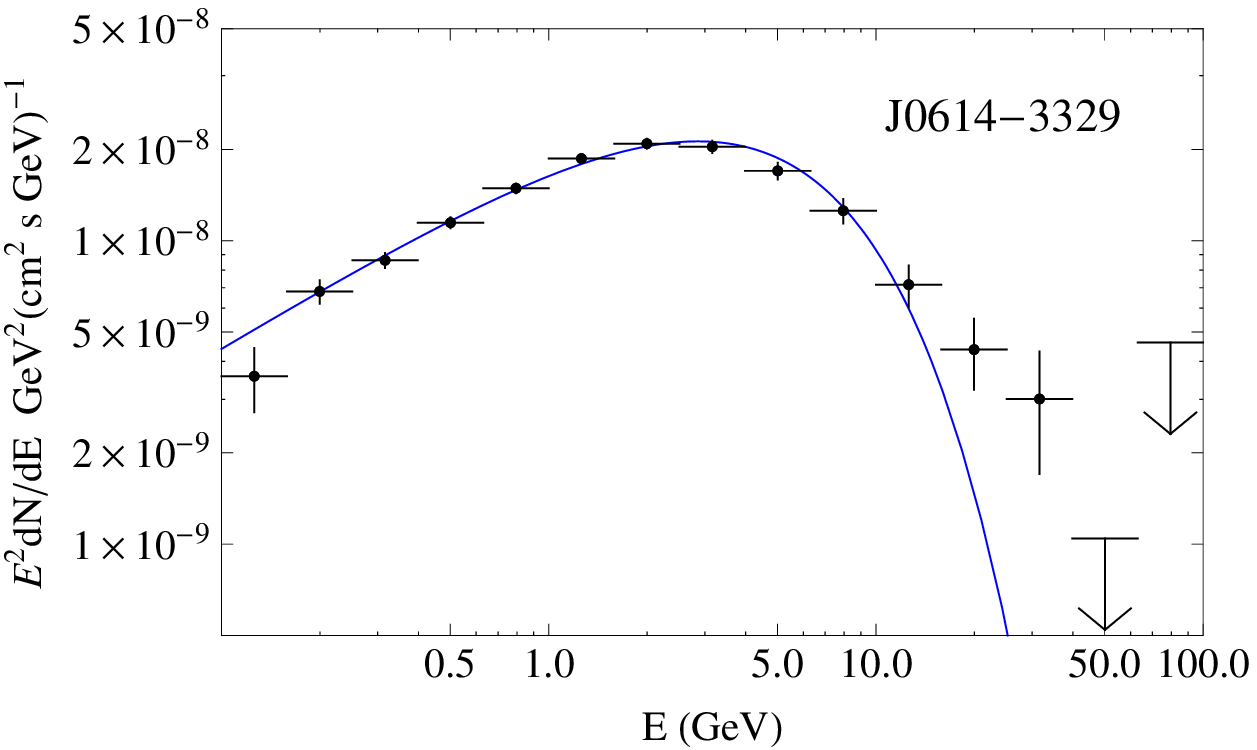} \\
\includegraphics[width=3.40in,angle=0]{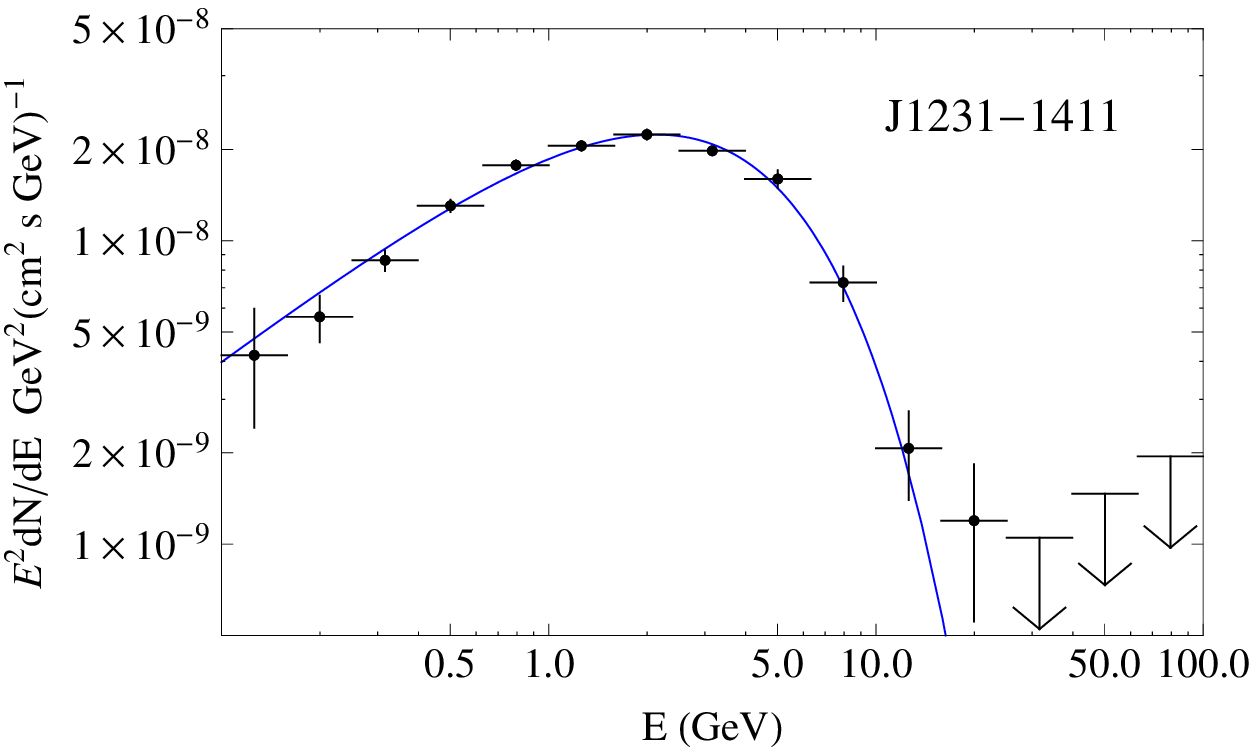} 
\includegraphics[width=3.40in,angle=0]{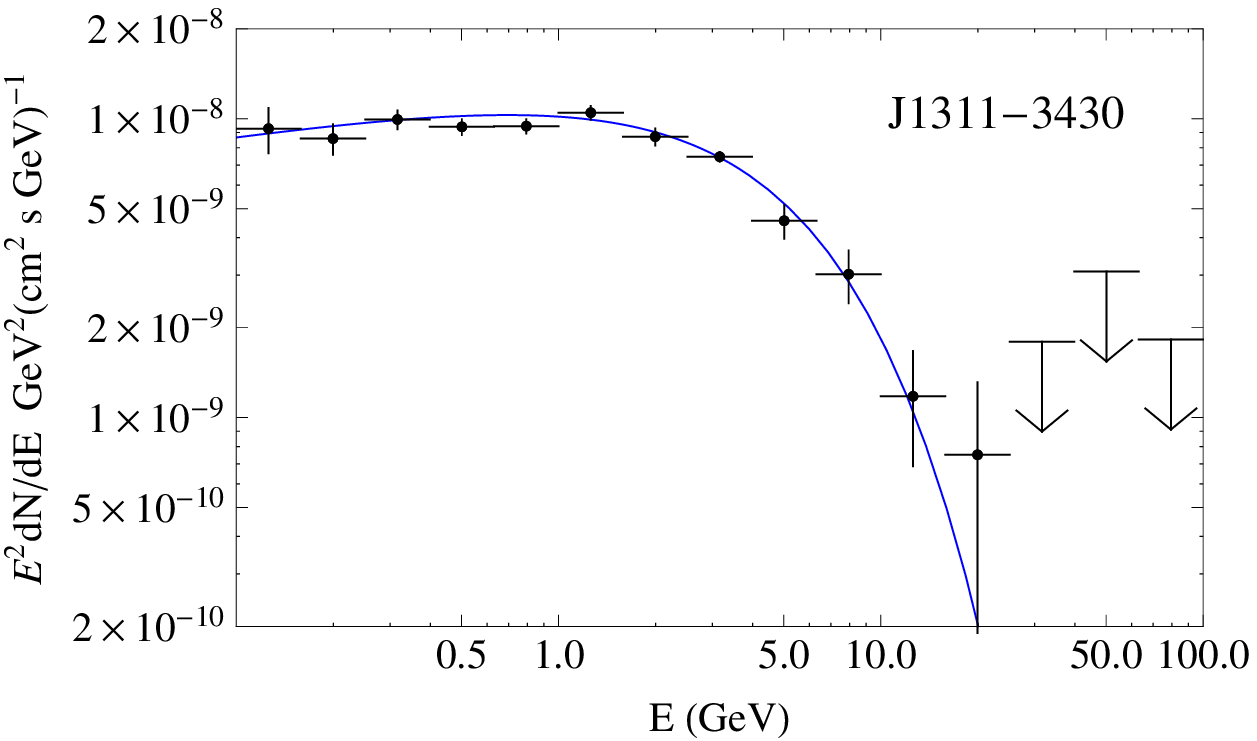} \\
\includegraphics[width=3.40in,angle=0]{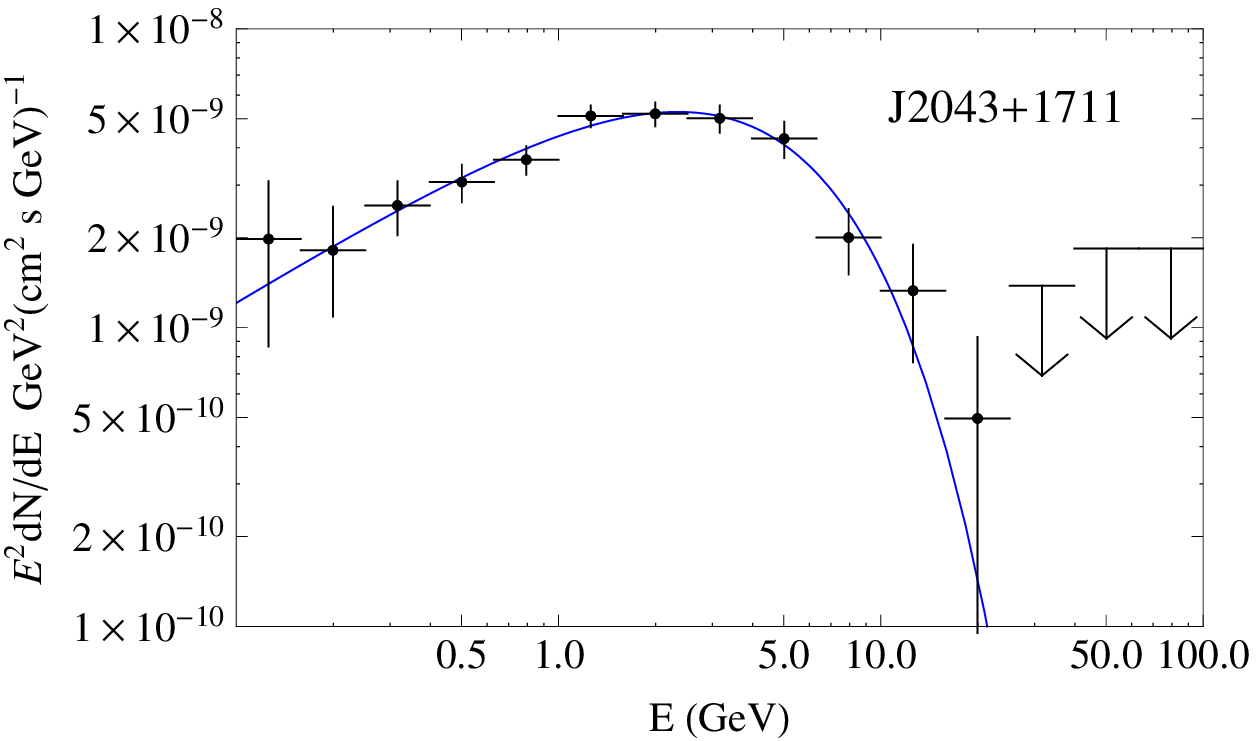} 
\includegraphics[width=3.40in,angle=0]{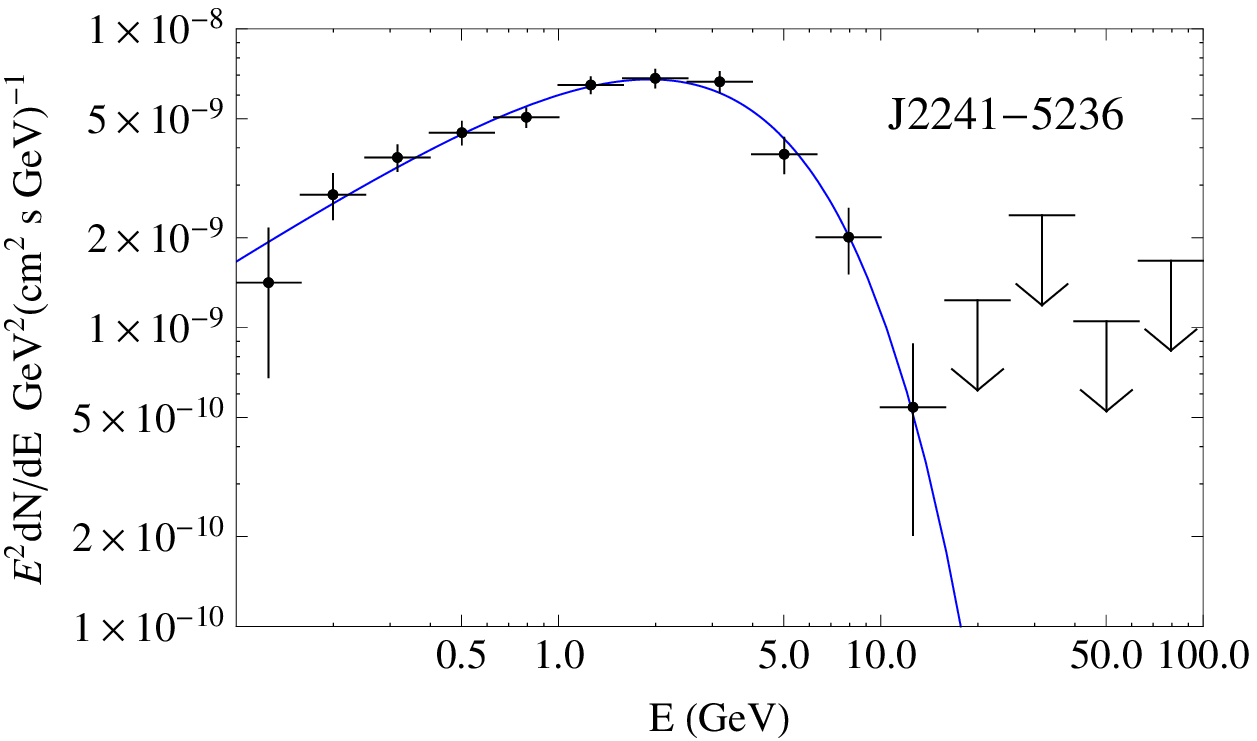} 
\caption{A sample of MSP spectra, along with their best fit model following the spectral form of Eq.~\ref{eq:MSP_Spect}. }
\label{fig:ExampleMSPs}
\end{figure*}

In Fig.~\ref{fig:ExampleMSPs}, we plot the gamma-ray spectrum from eight representative MSPs studied in our analysis, as found following the procedure described in Sec.~\ref{data}. This sample includes several spectra that strongly peak at energies of $\sim$0.5-3 GeV, as well as a couple with approximately flat low-energy spectra. For each of Fermi's MSPs, we fit the observed spectrum with the following analytic form:
\begin{equation}
\frac{dN}{dE} = A \cdot \frac{E^{\alpha}}{E_{\rm cut}^{1+\alpha}} \cdot e^{-E/E_{\rm cut}},
\label{eq:MSP_Spect}
\end{equation}
where $A$ is a normalization constant, and the spectral shape is described by the parameters $\alpha$ and $E_{\rm cut}$. In the fits we use the energy bins for which the calculated 1$\sigma$ errors are not larger than the calculated fluxes. For bins that do not meet this criterion, (typically the lowest and highest energy bins) we present only upper limits on the flux. 

For the majority of pulsars considered, we find that the parameterization of Eq.~\ref{eq:MSP_Spect} provides a good fit (with exceptions being J0101-6422, J0614-3329, J1614-2230, J1745+1017, J1858-2216 and J2215+5135).  In Appendix~\ref{app:MSPspectra} (Figs.~\ref{fig:MSPs1}-\ref{fig:MSPs8}), we present the measured gamma-ray spectra from the full set of the 61 MSPs considered, and in Table~\ref{MSP-table} we list the spectral parameters favored by our fit for these sources. 

\begin{figure*}
\includegraphics[width=5.40in,angle=0]{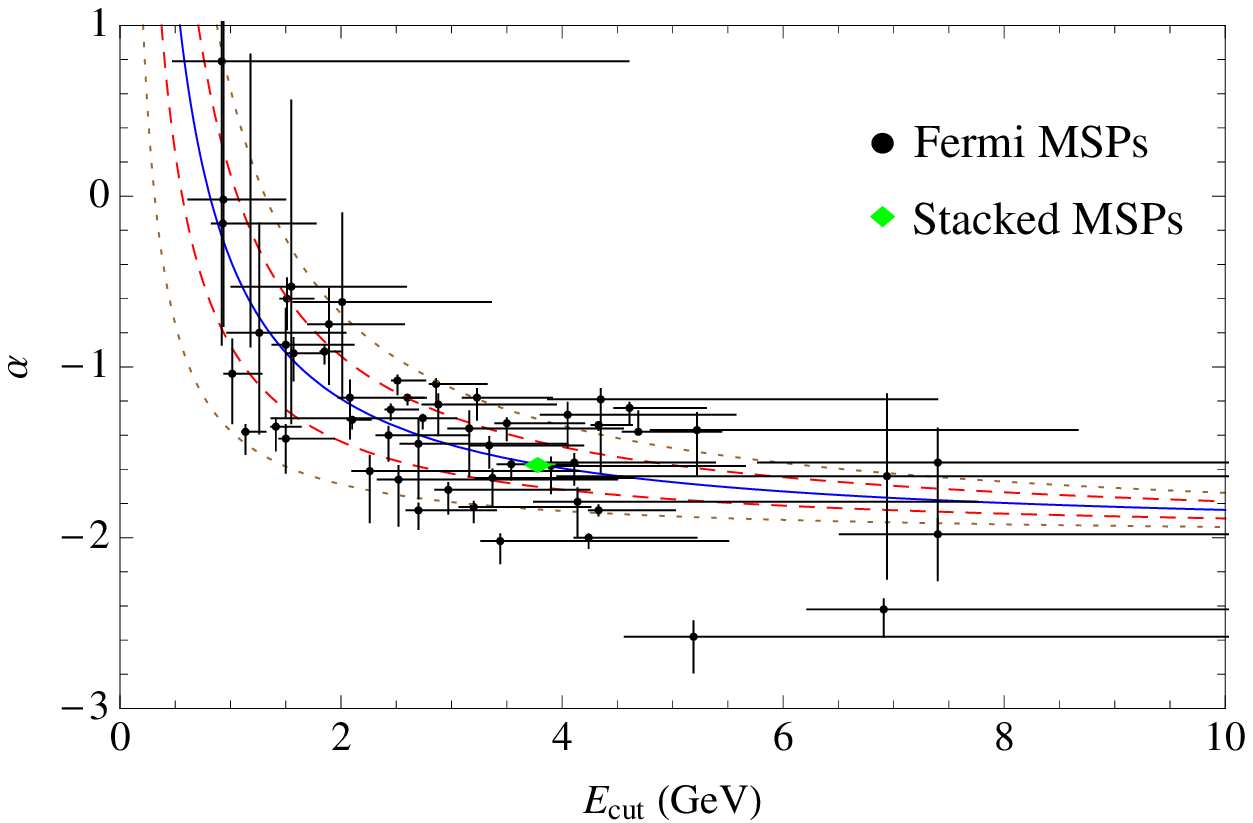}
\caption{The spectral properties of 61 millisecond pulsars observed by the Fermi Gamma-Ray Space Telescope, in terms of the parameterization of Eq.~\ref{eq:MSP_Spect}. The error bars denote the 68$\%$ confidence interval for each source. We also show the spectral shape of the stacked millisecond pulsar spectrum (green diamond). Although there is a significant degree of source-to-source variation in the spectra of this source population, these sources largely peak (in $E^2 dN/dE$ units) at an energy near $\sim$1-2 GeV.  The solid blue contour denotes spectra for which $E_{\rm peak}=1.625$ GeV, while the spectra between the red dashed (brown dotted) contours peak between 1.125-2.125 GeV (0.625-2.625 GeV).}
\label{fig:Alpha_VS_Ecut_MSPs}
\end{figure*}

\begin{figure*}[!t]
\includegraphics[width=3.40in,angle=0]{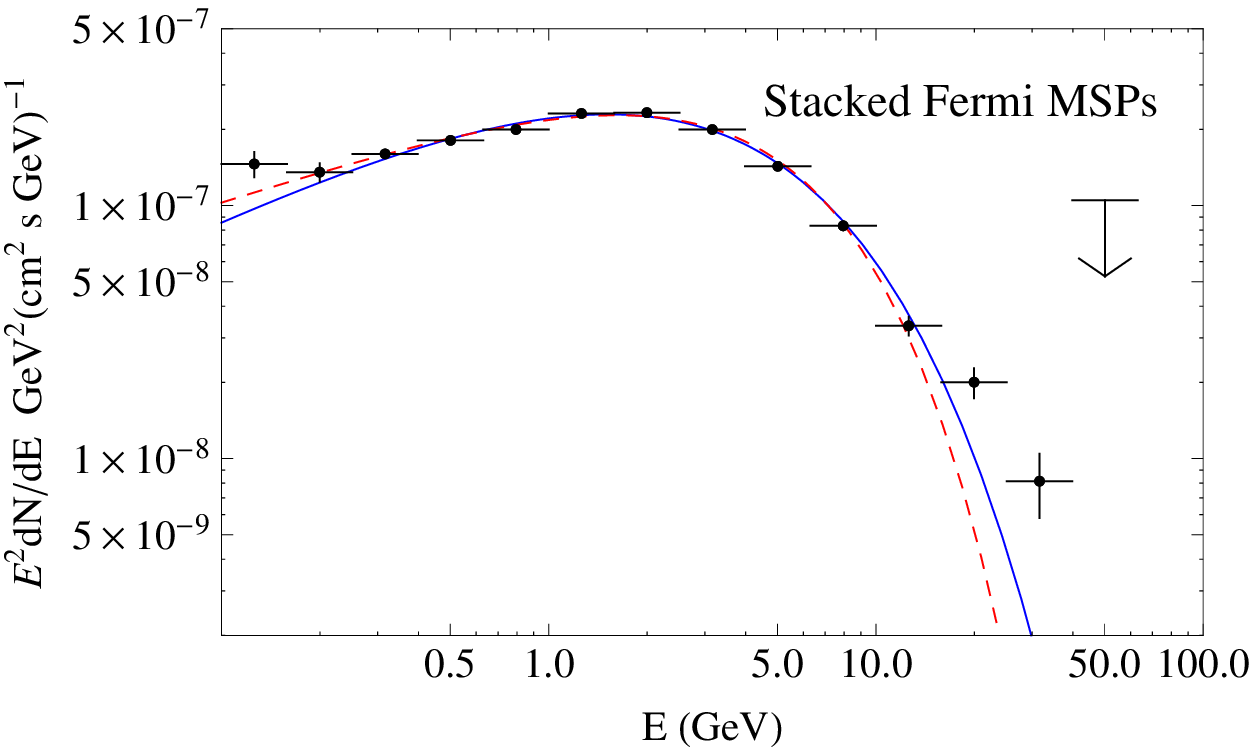}
\includegraphics[width=3.40in,angle=0]{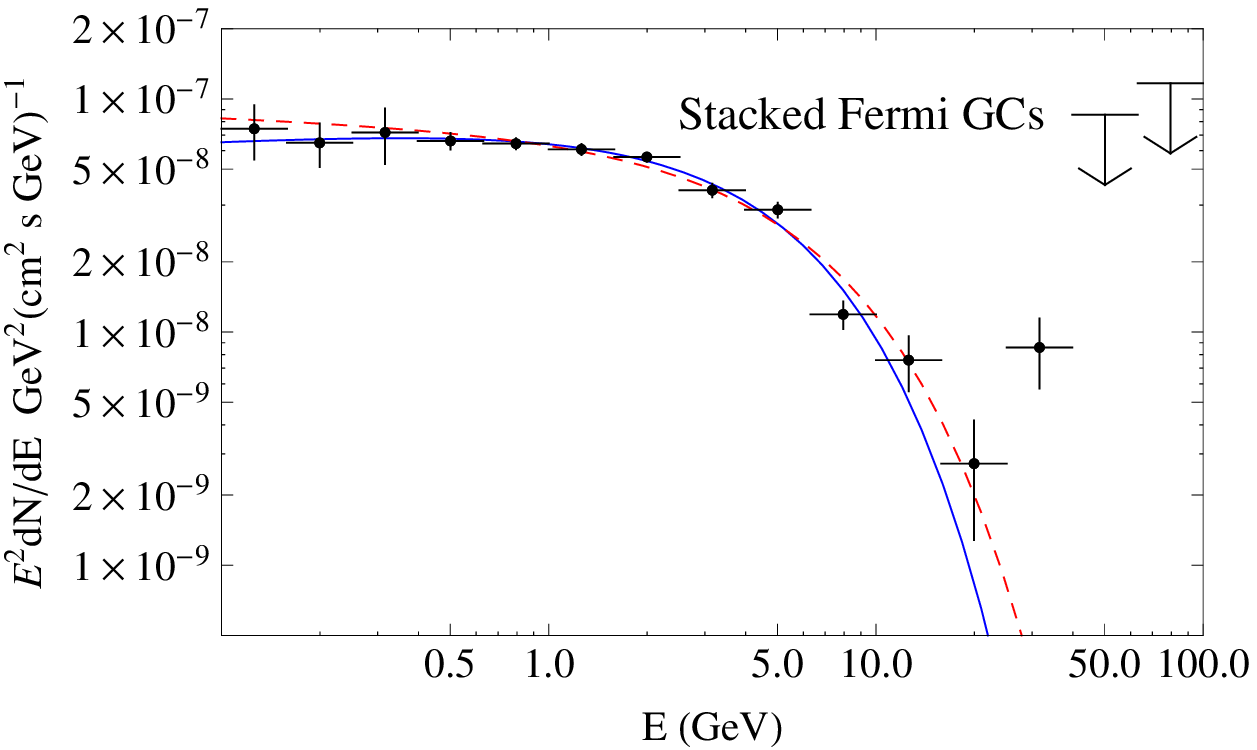}
%\caption{The spectral properties of the stacked Fermi Globular Clusters.  Dashed red line gives the exponential cut-off, while the solid line gives the sub-exponential cut-off fit.}
\caption{The spectrum of the stacked Fermi MSPs (left) and the stacked Fermi globular clusters (right).  The dashed red lines denote the best fit using an exponentially cut-off power-law form, while the solid lines denote the sub-exponential/super-exponential cut-off best fit.}
\label{fig:Stacked_MSPs}
\end{figure*}

\begin{table*}[t]
\begin{tabular}{|c|c|c|c|c|c|c|c|c|}
\hline
Name    &       $A$ (cm$^{-2}$ s$^{-1}$) & $\alpha$ & $\alpha_{\rm low}$(68$\%$) & $\alpha_{\rm high}$(68$\%$) & $E_{\rm cut}$ (GeV) & $E_{\rm cut_{low}}$(68$\%$) & $E_{\rm cut_{high}}$(68$\%$) & $\chi^{2}$/dof \\
\hline \hline
J0023+0923   &   4.30$\times 10^{-9}$       &         -1.04  &  -1.33     &      -0.84      &     1.02    &        0.94     &     1.28     &    1.43 \\
\hline
J0030+0451   &  1.57$\times 10^{-8}$        &        -1.31   &   -1.36     &      -1.30     &      2.10     &       2.06     &     2.27      &     0.94 \\
\hline
J0034-0534    &  2.06$\times 10^{-9}$      &        -1.72   &  -1.86       &    -1.68       &    2.97      &      2.85     &    4.25    &     0.76 \\
\hline
J0101-6422    &  2.49$\times 10^{-9}$        &         -1.40  &   -1.55      &     -1.35      &     2.43      &      2.32    &     3.16     &      2.23 \\
\hline
J0102+4839   &  1.78$\times 10^{-9}$      &          -1.65  &    -1.82     &      -1.60    &       3.37    &        3.22   &      4.65    &     1.59 \\
\hline
J0218+4232   &  2.19$\times 10^{-9}$        &        -2.00    &  -2.06       &    -1.98      &     4.24      &      4.11    &     5.22      &   1.49 \\
\hline
J0307+7443   &  7.20$\times 10^{-9}$        &          -0.92   &   -1.08      &     -0.83     &      1.57    &       1.51    &     1.82    &     0.15 \\
\hline
J0340+4130   &  3.70$\times 10^{-9}$        &         -1.18    &  -1.31       &    -1.13      &     3.23      &      3.10     &    3.91     &    1.15 \\
\hline
J0437-4715    &  9.29$\times 10^{-9}$        &         -1.38    &  -1.51     &      -1.34      &     1.14      &      1.106   &    1.32   &     0.52 \\
\hline
J0533+6759   &  1.11$\times 10^{-9}$      &          -1.28   &   -1.47     &      -1.21      &     4.05     &       3.81     &     5.57     &    0.80 \\
\hline
J0605+3757   &  3.13$\times 10^{-9}$        &         -0.16   &   -0.88     &      +0.74     &      1.18      &     0.83     &     1.77    &     0.49 \\
\hline
J0610-2100    &  1.25$\times 10^{-9}$      &          -1.66   &   -1.93     &      -1.58      &     2.52      &      2.33      &     4.50    &     0.93 \\
\hline
J0613-0200    &  4.04$\times 10^{-9}$        &         -1.57    &  -1.66      &     -1.54     &      3.54      &      3.42    &     4.14    &     0.74 \\
\hline
J0614-3329    &  1.25$\times 10^{-8}$        &        -1.34    &  -1.37      &     -1.33       &    4.33       &     4.27     &    4.63    &     2.32 \\
\hline
J0737-3039A  &   1.66$\times 10^{-10}$     &          -1.64   &   -2.24     &      -1.16      &     6.94      &      3.96    &     22.90     &     0.12 \\
\hline
J0751+1807    &  2.39$\times 10^{-9}$       &          -1.22    &  -1.40     &      -1.16     &      2.88      &     2.74   &        3.95   &      1.13 \\
\hline
J1024-0719    &  2.37$\times 10^{-9}$        &        -0.02   &   -0.76     &      +1.02     &      0.94       &      0.62      &    1.50     &      0.28 \\
\hline
J1124-3653    &  1.96$\times 10^{-9}$        &       -1.46   &   -1.59      &     -1.41      &     3.340     &       3.173      &     4.19   &      1.27 \\
\hline
J1125-5825    &  1.14$\times 10^{-11}$        &        -1.85    &  -1.99      &     -1.74     &      200      &    20   &      >1000    &     1.10 \\
\hline
J1137+7528    &  7.53$\times 10^{-13}$        &        -2.02   &   -2.30      &     -1.85      &     191    &      76.5    &      325     &   0.38 \\
\hline
J1142+0119   &   6.99$\times 10^{-10}$        &       -1.19   &   -1.62      &     -1.13      &     4.35      &      3.87    &     7.40     &      0.19 \\
\hline
J1231-1411   &   2.30$\times 10^{-8}$        &       -1.18   &   -1.22     &      -1.17     &      2.60      &      2.56     &      2.77     &      1.54 \\
\hline
J1301+0833   &   5.01$\times 10^{-10}$       &       -1.79   &   -1.99      &     -1.71     &      4.14      &      3.75    &     7.76   &     1.14 \\
\hline
J1302-3258   &   1.41$\times 10^{-9}$       &       -1.56    &  -1.69      &     -1.51     &      4.11      &      3.86     &    5.38    &     0.53 \\
\hline
J1311-3430   &   3.74$\times 10^{-9}$        &       -1.84   &  -1.87      &     -1.81     &      4.33      &      4.24     &    5.02    &     0.85 \\
\hline
J1312+0051   &   3.84$\times 10^{-9}$       &        -1.18   &   -1.42      &     -1.08     &      2.08      &      1.98     &      2.75    &     0.88 \\
\hline
J1446-4701   &   1.81$\times 10^{-9}$        &      -0.62    &  -1.18      &     -0.10     &      2.01       &     1.57     &    3.36     &    0.60 \\
\hline
J1514-4946   &   4.17$\times 10^{-9}$        &       -1.38   &   -1.26      &     -1.36     &      4.69      &      4.55    &     5.44    &     0.43 \\
\hline
J1543-5149   &   2.56$\times 10^{-10}$        &      -2.42   &   -2.58      &     -2.36      &     6.91      &      6.22      &     20.7     &     0.51 \\
\hline
J1544+4937  &    1.29$\times 10^{-12}$        &        -2.07   &   -2.29      &     -1.81     &      197.    &      20    &    >1000   &     0.55 \\
\hline
J1600-3053   &   7.35$\times 10^{-10}$         &     -1.37   &   -1.63       &    -1.27      &     5.22      &      4.80   &     8.67    &     0.50 \\
\hline
J1614-2230   &   1.13$\times 10^{-8}$         &      -0.60   &   -0.78      &     -0.48     &      1.51     &       1.45    &     1.75    &     2.21 \\
\hline
J1630+3734   &   1.23$\times 10^{-9}$        &       -1.45   &   -1.77       &    -1.31     &      2.70      &      2.54      &     4.05     &      1.47 \\
\hline
J1640+2224   &   7.96$\times 10^{-13}$      &        -2.08   &   -2.46      &     -1.60      &     199   &       20    &      >1000    &     0.81 \\
\hline
J1658-5324   &   1.46$\times 10^{-9}$        &     -2.02   &   -2.15     &      -1.98     &      3.44      &      3.27     &      5.50      &     0.85 \\
\hline
J1713+0747   &   1.36$\times 10^{-9}$       &         -1.36   &   -1.64     &      -1.26    &       3.16     &       2.97     &    4.55    &     1.95 \\
\hline
J1732-5049   &   2.58$\times 10^{-10}$      &         -1.56   &   -1.98     &      -0.56    &       7.40     &       1.48    &      >1000    &      1.80 \\
\hline
J1741+1351   &   2.49$\times 10^{-9}$         &        -0.80   &   -1.39     &      -0.16    &       1.26     &       0.97     &     2.04   &     1.01 \\
\hline
J1744-1134   &   1.52$\times 10^{-8}$          &       -1.35   &   -1.49     &      -1.30      &     1.41      &      1.37     &    1.64    &     1.52 \\
\hline
J1745+1017  &    1.46$\times 10^{-10}$         &        -2.05   &   -2.21     &      -1.99     &      12.3     &      10.7    &    36.8    &      2.87 \\
\hline
J1747-4036   &   3.25$\times 10^{-10}$        &        -1.98   &   -2.26     &      -1.67     &      7.40      &      3.70     &      58.5    &      0.30 \\
\hline
J1810+1744   &   2.44$\times 10^{-9}$       &         -1.84   &   -1.95    &       -1.80     &      2.70     &       2.59     &      3.40     &      0.78 \\
\hline
J1811-2405   &   2.06$\times 10^{-12}$          &       -2.63   &   -2.76      &     -2.47     &      93.5     &      18.7     &     860    &     1.23 \\
\hline
J1816+4510  &    1.08$\times 10^{-9}$        &         -1.58    &  -1.74      &     -1.53      &     3.90     &       3.71     &      5.66     &      0.78 \\
\hline
J1843-1113   &   2.13$\times 10^{-10}$          &       -2.58   &   -2.79     &      -2.49      &     5.19      &      4.57    &     17.9   &     0.61 \\
\hline
J1858-2216   &   4.43$\times 10^{-9}$         &        -0.87    &  -1.29      &     -0.66      &     1.50      &      1.38     &      2.12     &      2.06 \\
\hline
J1902-5105   &   1.89$\times 10^{-9}$        &        -1.82   &   -1.91     &      -1.79      &     3.20      &      3.07      &     4.25     &      1.69 \\
\hline
J1939+2134  &    1.43$\times 10^{-10}$       &         -1.41   &   -1.93      &     -0.41     &      30.0    &      3.60     &     >1000    &     1.22 \\
\hline
J1959+2048   &   5.98$\times 10^{-9}$         &        -1.42   &   -1.62      &     -1.34      &     1.50      &      1.44    &      1.94      &     1.68 \\
\hline
J2017+0603   &   4.30$\times 10^{-9}$        &         -1.24   &   -1.34     &      -1.21     &      4.61     &       4.47    &     5.30    &     1.25 \\
\hline
J2043+1711   &   3.85$\times 10^{-9}$         &        -1.33   &  -1.43     &      -1.30     &      3.50      &      3.40     &      4.20      &     0.49 \\
\hline
J2047+1053   &   1.70$\times 10^{-9}$       &         -0.53   &   -1.33     &      +0.56      &     1.55      &      1.01   &      2.59    &     0.96 \\
\hline
J2051-0827    &  1.08$\times 10^{-9}$        &        +0.79   &   -0.87     &      +2.79     &      0.92      &       0.48     &     4.60     &      1.64 \\
\hline
J2124-3358   &   1.46$\times 10^{-8}$        &         -0.91   &   -0.98     &      -0.88      &     1.85     &       1.81    &      2.00     &      1.27 \\
\hline
J2129-0429   &   1.33$\times 10^{-9}$        &        -1.61   &   -1.91     &      -1.52      &     2.26      &      2.10    &     4.16     &    1.00 \\
\hline
J2214+3000   &   7.83$\times 10^{-9}$       &          -1.25   &   -1.31     &      -1.22    &       2.45     &       2.40    &       2.70     &      0.86 \\
\hline
J2215+5135   &   6.97$\times 10^{-11}$       &         -2.31   &   -2.39     &      -2.28     &      17.5     &      16.1     &     39.2     &     5.32 \\
\hline
J2241-5236   &   6.39$\times 10^{-9}$        &         -1.30   &   -1.36     &      -1.28     &      2.74     &       1.37     &      3.04    &     0.69 \\
\hline
J2256-1024   &   2.19$\times 10^{-9}$        &         -0.75   &   -1.10     &      -0.54     &      1.89     &       1.70     &      2.57    &     1.05 \\
\hline
J2302+4442  &    9.07$\times 10^{-9}$       &          -1.08   &   -1.16    &       -1.05     &      2.51     &       2.46    &     2.76      &     1.66 \\
\hline
J2339-0533   &   5.44$\times 10^{-9}$        &         -1.10   &   -1.20      &     -1.07     &      2.86     &      2.80    &     3.32     &    0.89 \\
\hline \hline
\end{tabular}
\caption{The spectral parameters of 61 individual millisecond pulsars (see Eq.~\ref{eq:MSP_Spect}). The ``high'' and ``low'' values encompass the 68$\%$ confidence interval. In the last column, we give the quality of the fit ($\chi^{2}$ per degree-of-freedom) for the best fit values.}
\label{MSP-table}
\end{table*}

In Fig.~\ref{fig:Alpha_VS_Ecut_MSPs}, we plot the spectral parameters ($\alpha$, $E_{\rm cut}$) and their errors for this population of MSPs. Although considerable source-to-source variation is found among this population, most of these sources peak at an energy near $E_{\rm peak} \simeq 1-2$ GeV (in $E^2 dN/dE$ units).

In addition to considering individual MSPs, we have also derived a stacked spectrum of these 61 sources, which is shown in the left frame of Fig.~\ref{fig:Stacked_MSPs}. The stacking was done by summing the fluxes in each energy bin, and adding their errors in quadrature. Adopting the spectral parameterization of Eq.~\ref{eq:MSP_Spect}, the stacked analysis yields $\alpha = -1.57^{+0.01}_{-0.02}$ and $E_{\rm cut}=3.78^{+0.15}_{-0.08}$ GeV. At energies between approximately 0.2 and 10 GeV, the observed flux is well described by this parameterization. At $E>10$ GeV, however, more emission is observed than can be fit by the standard exponentially cut-off power-law. With this motivation in mind, we also fit the stacked MSP spectrum with a sub-exponential/super-exponential parametrization:
\begin{equation}
\frac{dN}{dE} = A \cdot \frac{E^{\alpha}}{E_{\rm cut}^{1+\alpha}} \cdot e^{-(E/E_{\rm cut})^{\beta}}.
\label{eq:MSP_Spect2}
\end{equation}
We find that the inclusion of this additional parameter ($\beta$) allows for a better fit to the combined spectrum of MSPs (with a $\Delta \chi^{2}$ of 6.6). This by itself could be explained by the existence of MSPs with high values of $E_{\rm cut}$ within the observed population. Alternatively, this may suggest that the simplified parametrization of Eq.~\ref{eq:MSP_Spect} is inadequate for describing the gamma-ray emission from these objects. 
%
%If the sub-exponential suppression indicates just a different power-law at high energies for MSPs is still to be resolved. 
We also note that VERITAS and MAGIC have each observed pulsed gamma-ray emission from the Crab pulsar at energies above 25 GeV, and also favor a sub-exponential suppression of its gamma-ray spectrum \cite{Aliu:2011zi, Aleksic:2011np, Aleksic:2011yx}. 
In Figs.~\ref{youngspec1} and \ref{youngspec2} of Appendix~\ref{app:MSPspectra}, we show the spectra of 16 regular (non-millisecond) pulsars for which we have a good measurement of the gamma-ray spectrum over a wide range of energies. These plots clearly demonstrate that, in some cases, the sub-exponential spectral parameterization allows for a better fit to the observed spectra. 

%\textbf{TIM: that gamma-ray spectra are not the on pulse spectra right? If they are not the on pulse emission there is not a direct connection with the Crab observation (even if it still may just be the same thing).}  

\subsection{Spectra of Globular Clusters}

\begin{figure*}

\includegraphics[width=3.40in,angle=0]{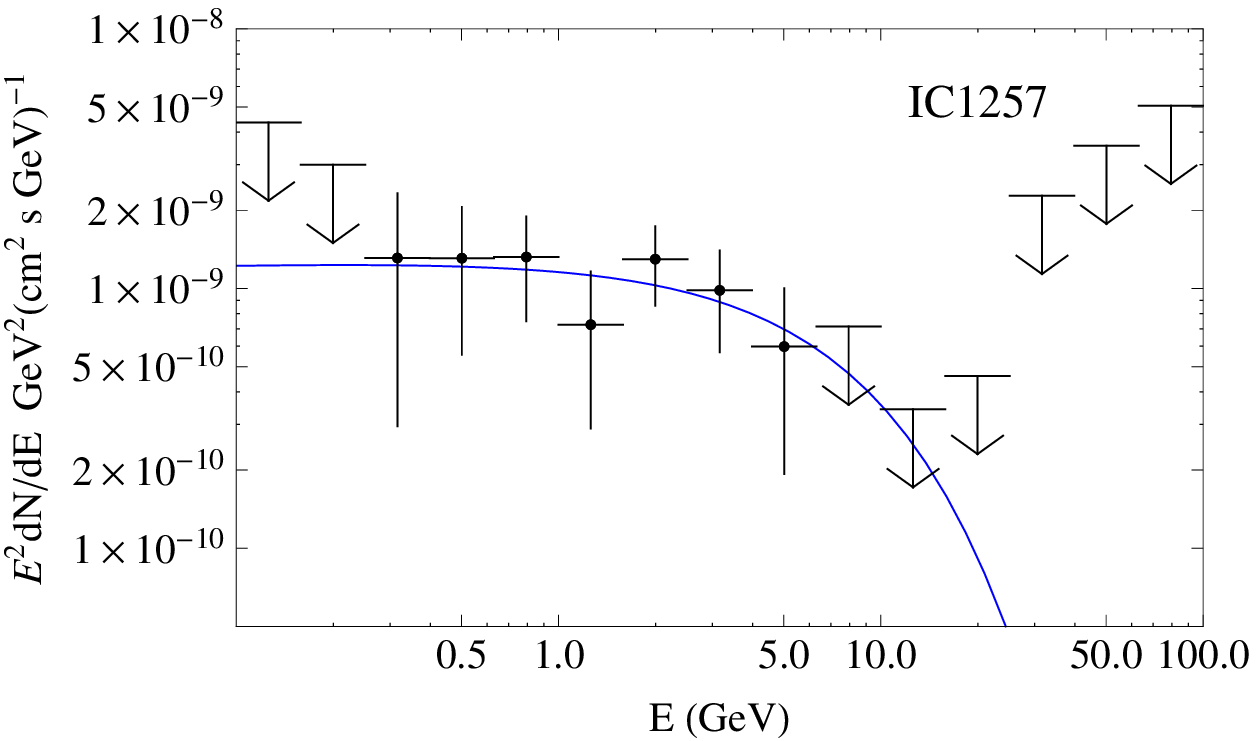}
\includegraphics[width=3.40in,angle=0]{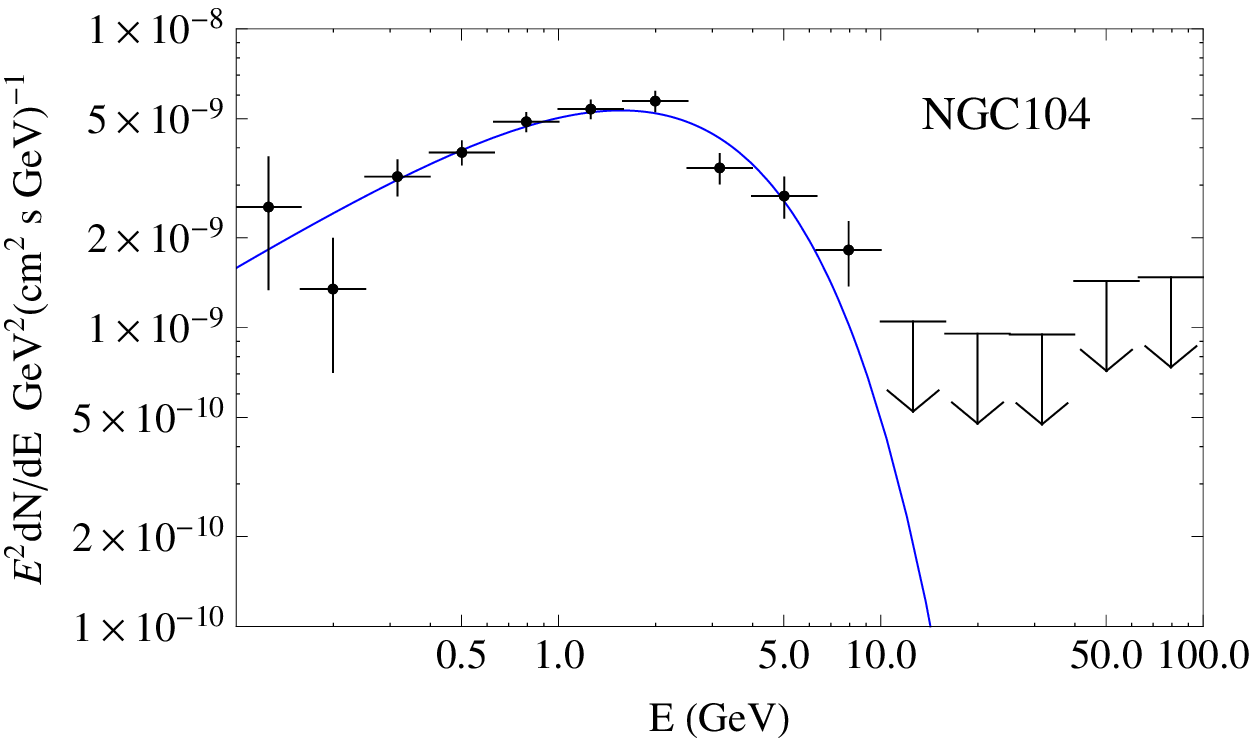}\\
\includegraphics[width=3.40in,angle=0]{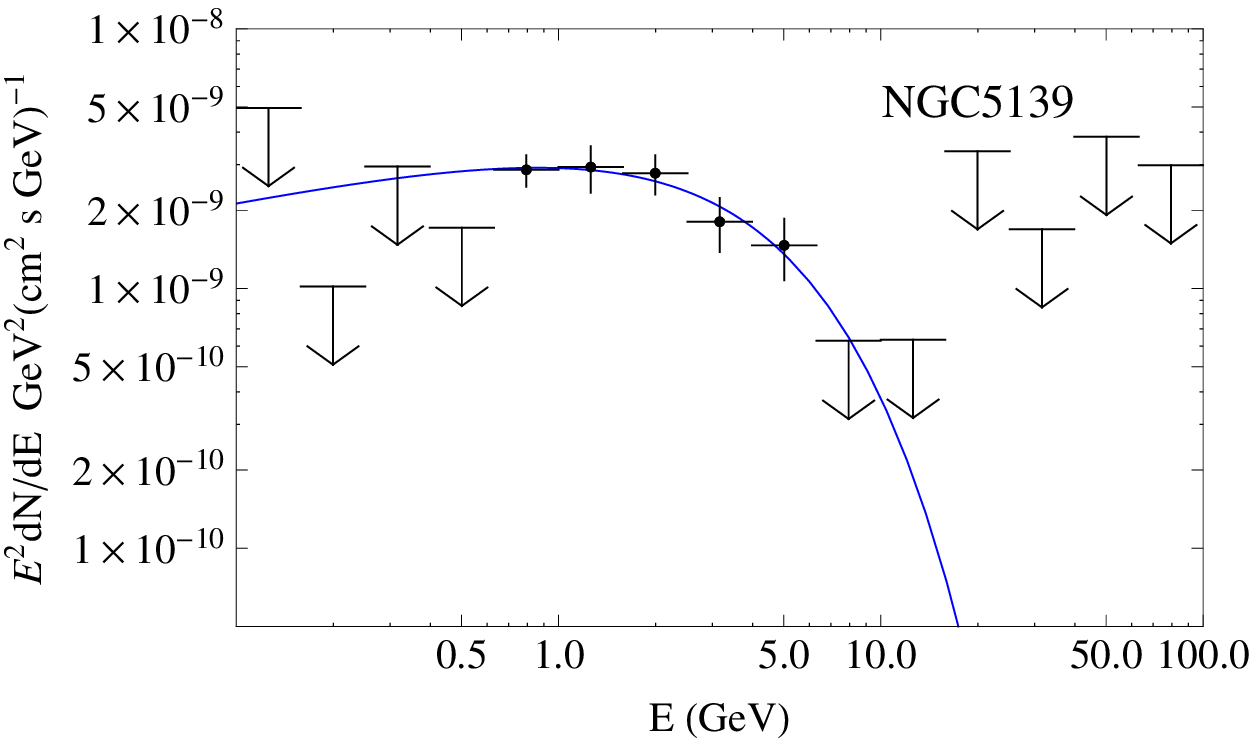} 
\includegraphics[width=3.40in,angle=0]{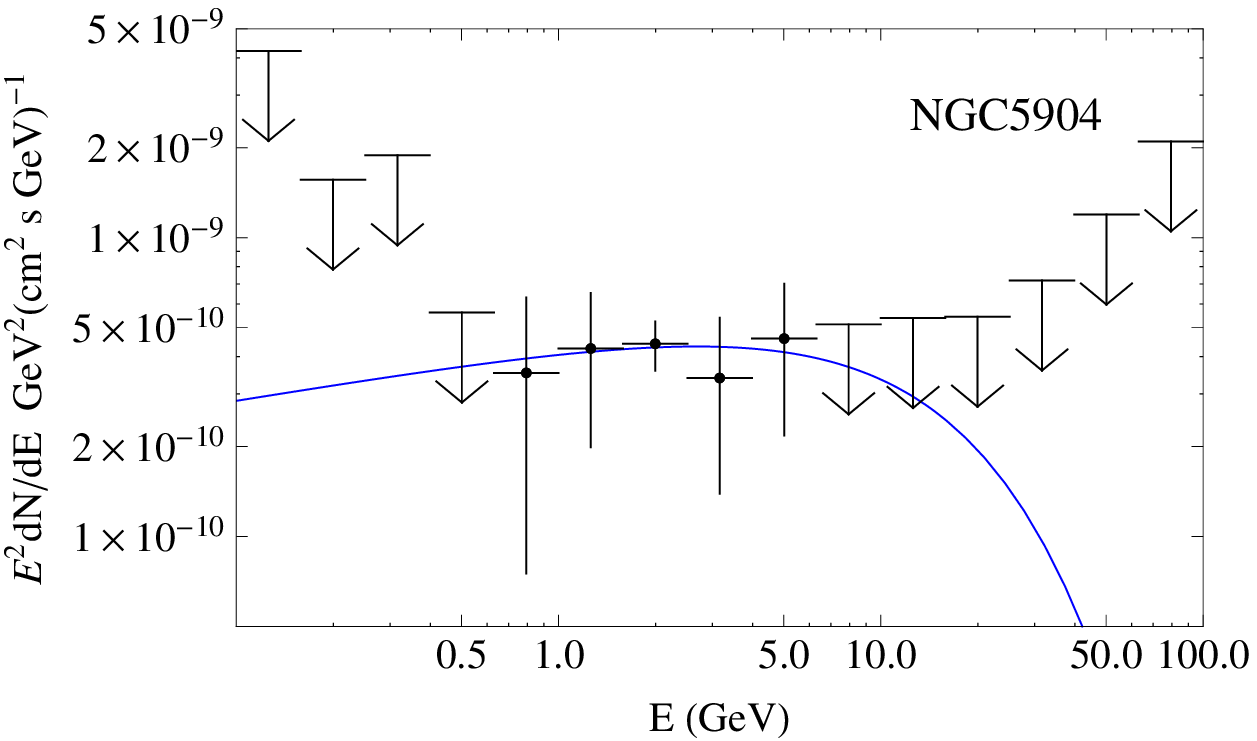} \\
\includegraphics[width=3.40in,angle=0]{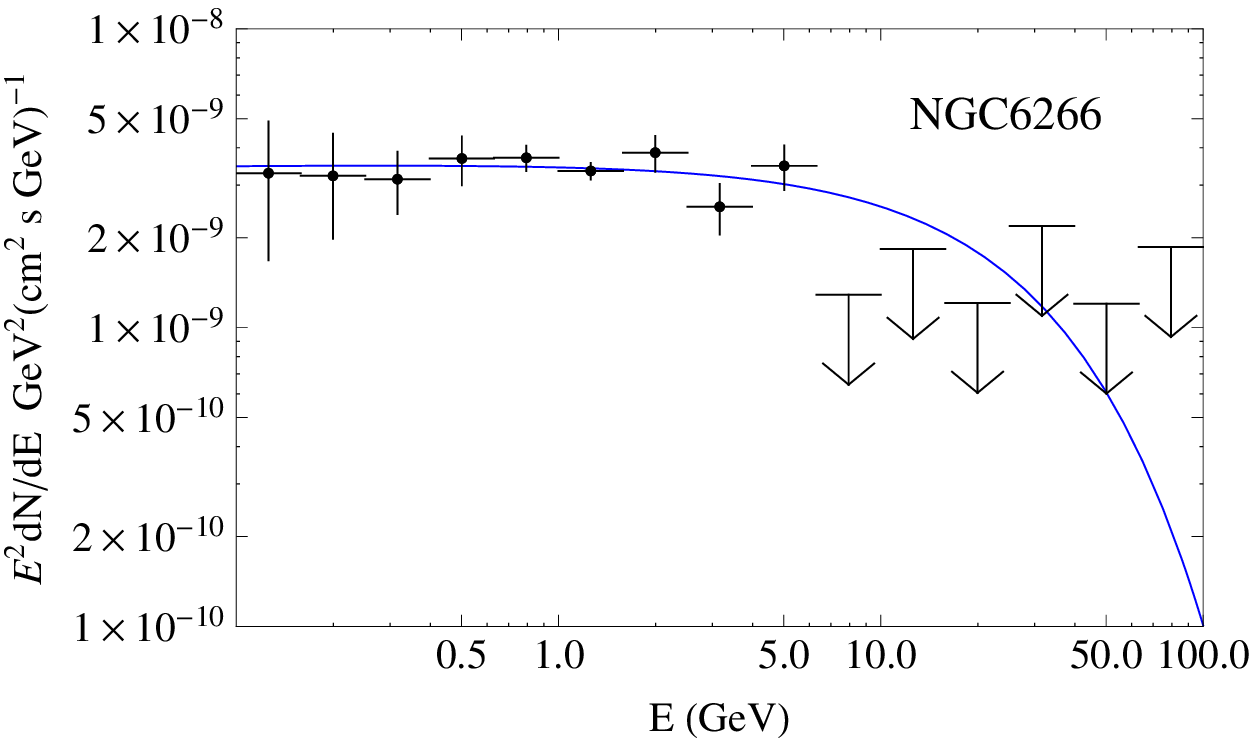} 
\includegraphics[width=3.40in,angle=0]{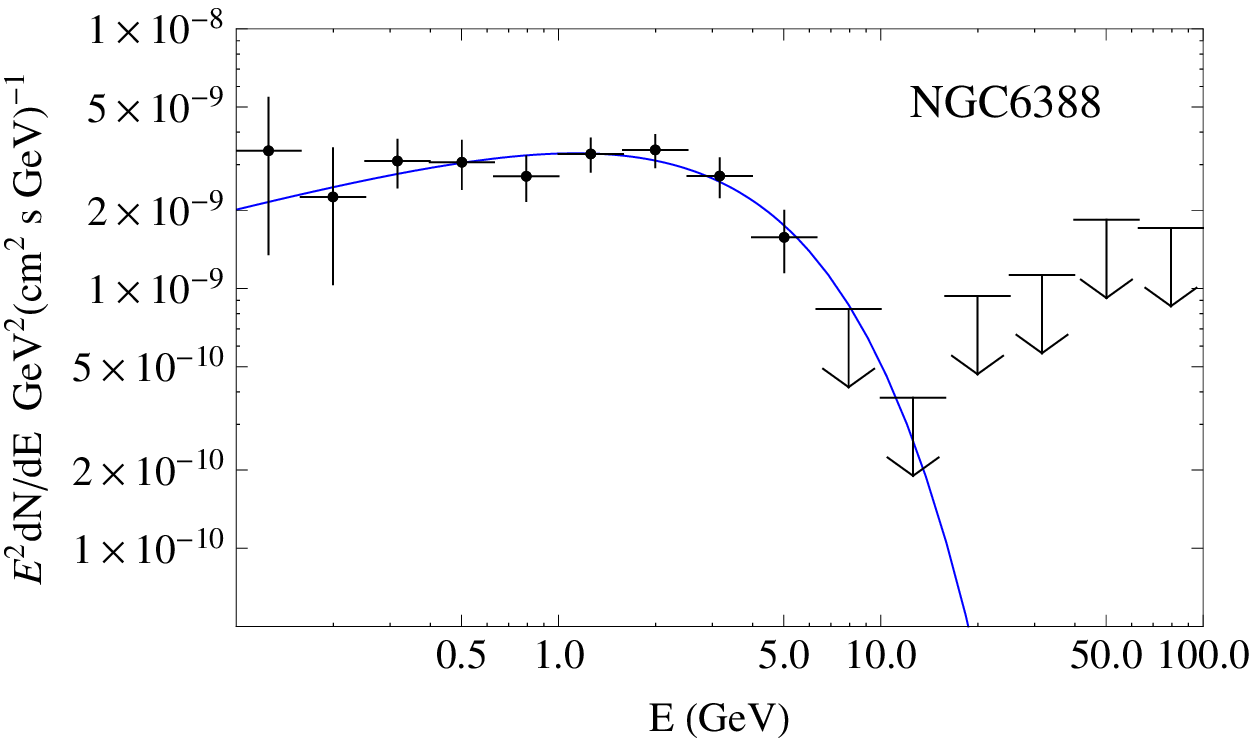} \\
\includegraphics[width=3.40in,angle=0]{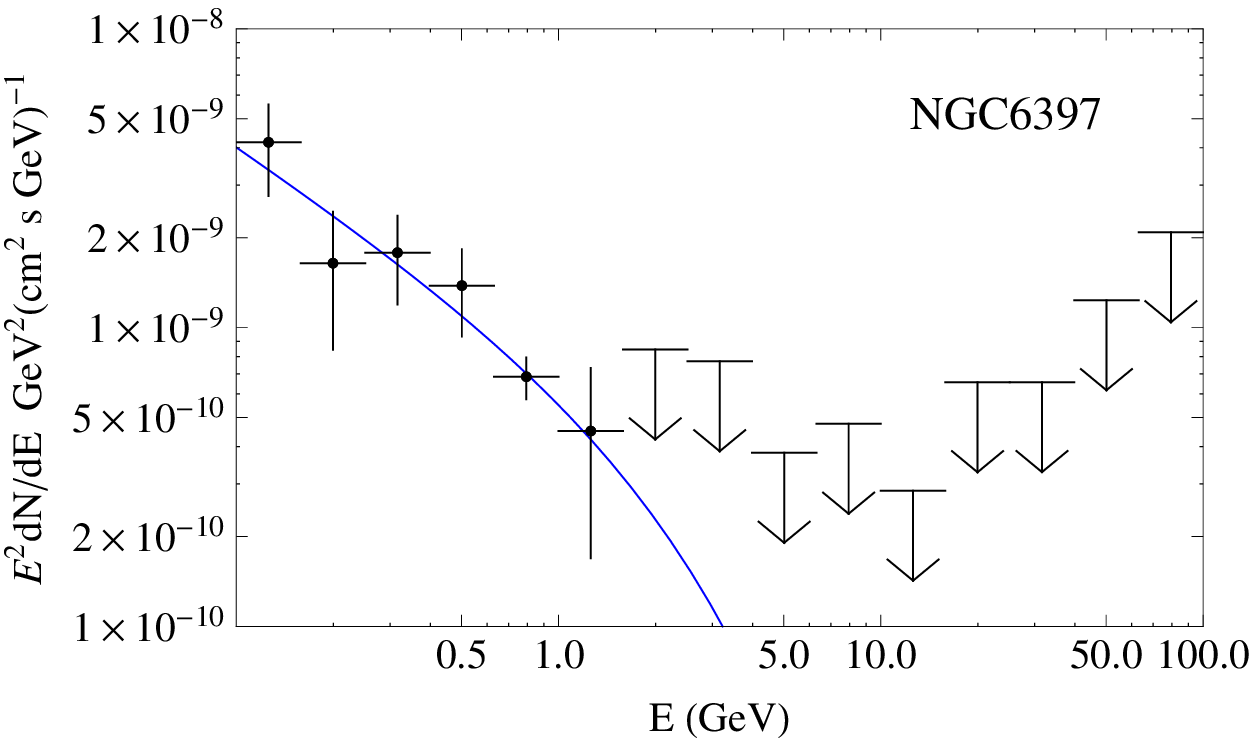} 
\includegraphics[width=3.40in,angle=0]{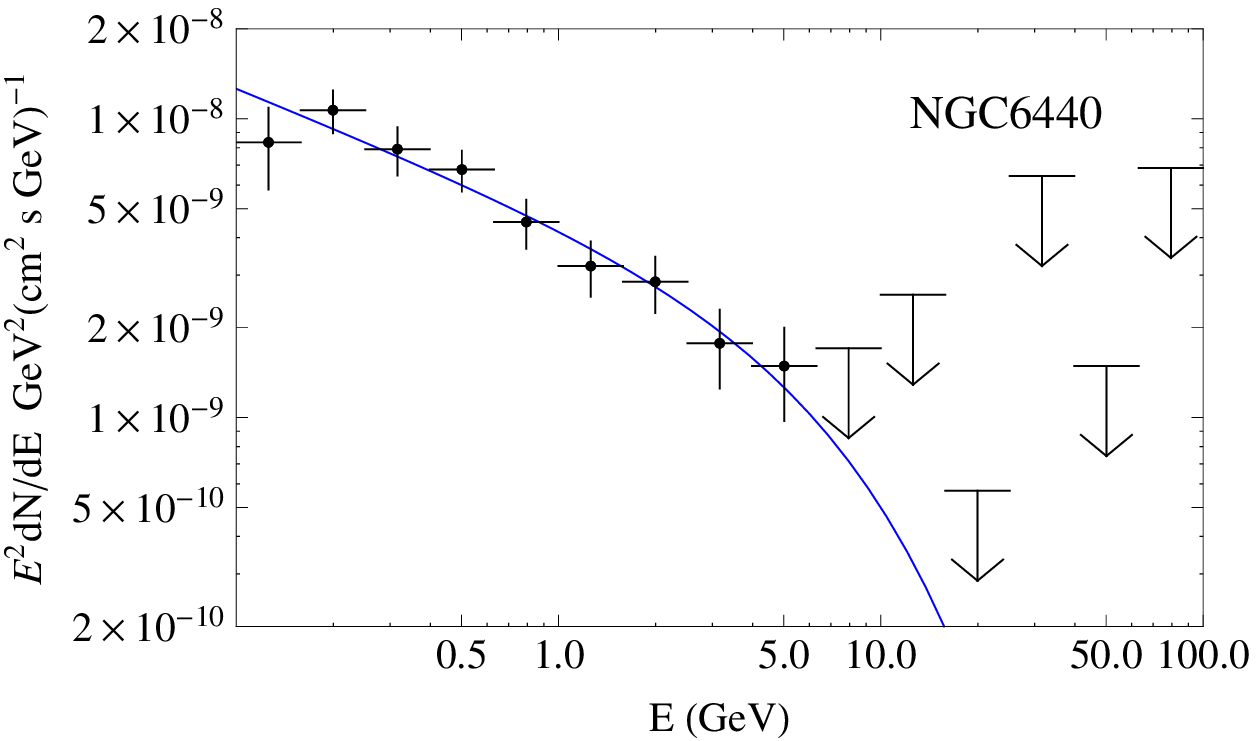} \\
\caption{The gamma-ray spectra of globular clusters along with their best fit model following the spectral form of Eq.~\ref{eq:MSP_Spect}.}
\label{fig:GCs1}
\end{figure*}

\begin{figure*}
\includegraphics[width=3.40in,angle=0]{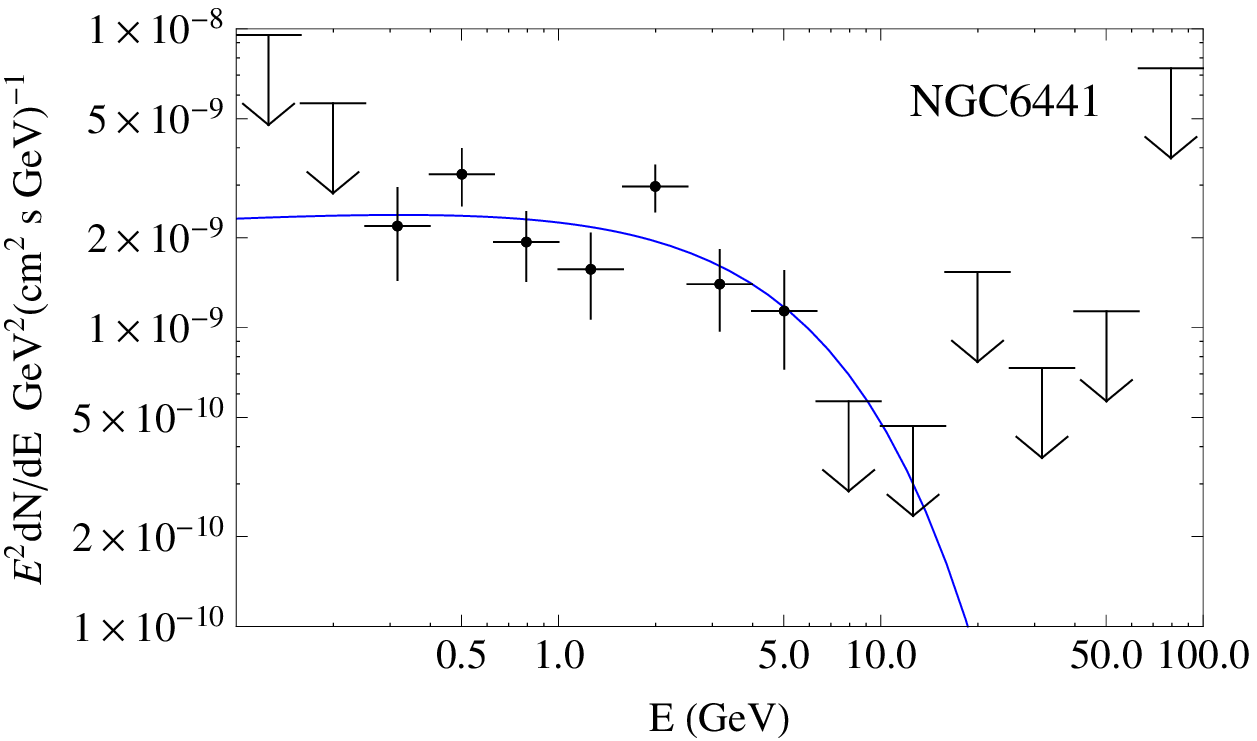} 
\includegraphics[width=3.40in,angle=0]{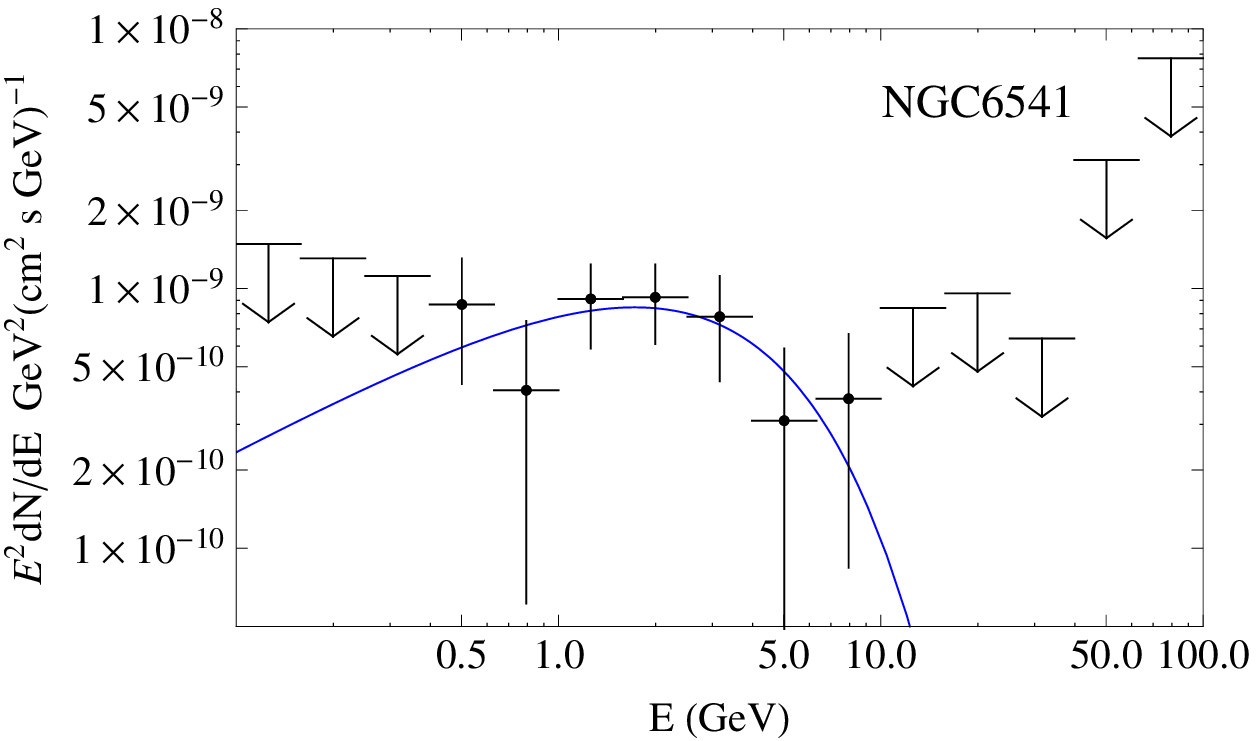} \\
\includegraphics[width=3.40in,angle=0]{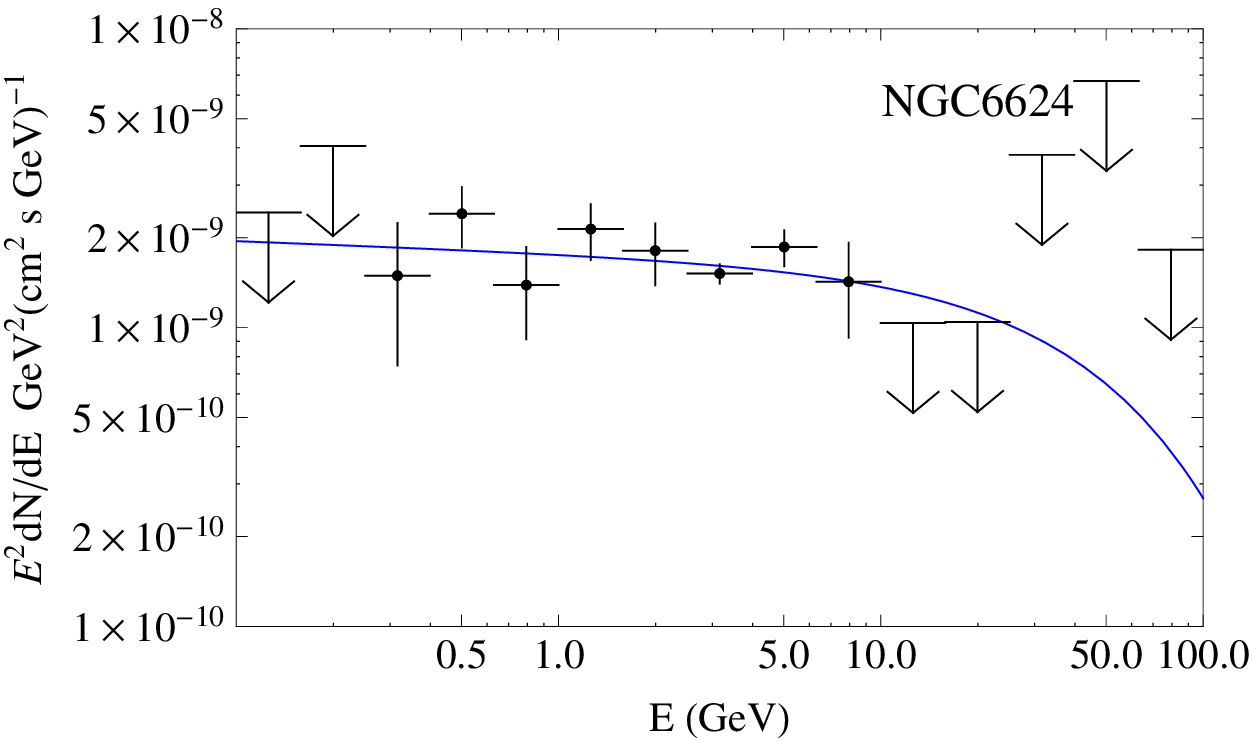} 
\includegraphics[width=3.40in,angle=0]{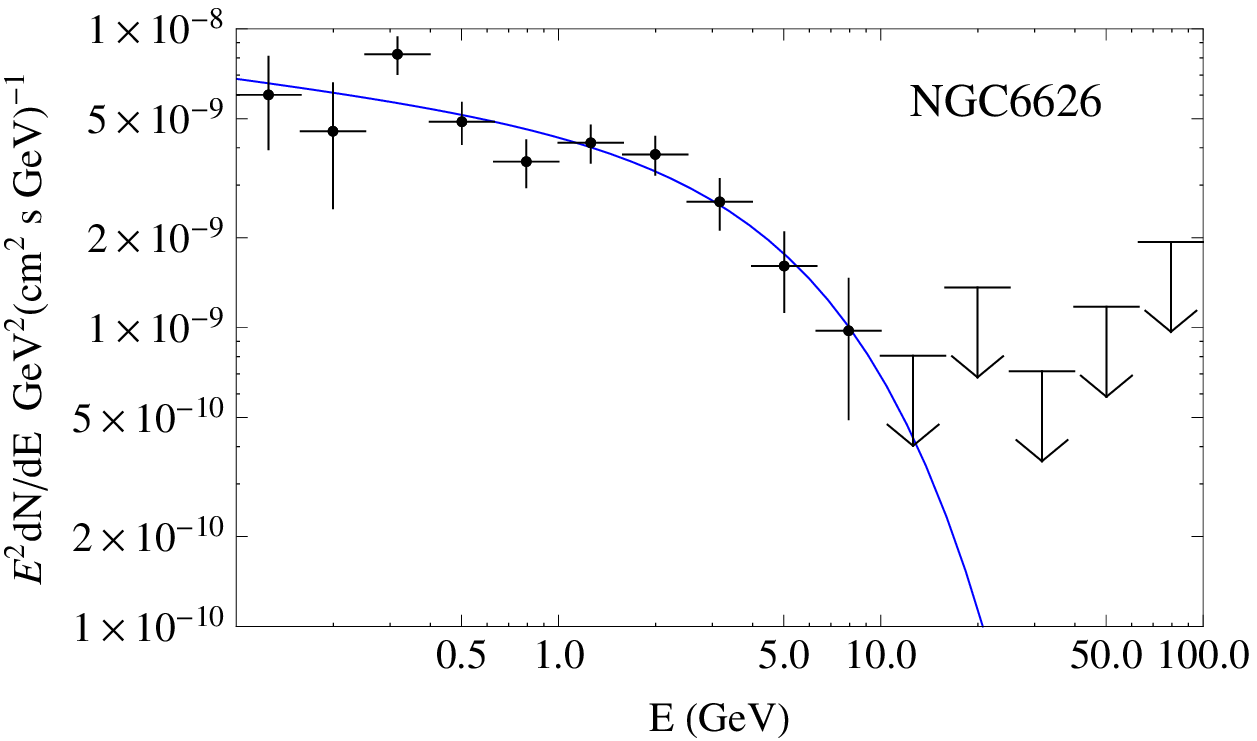}\\ 
\includegraphics[width=3.40in,angle=0]{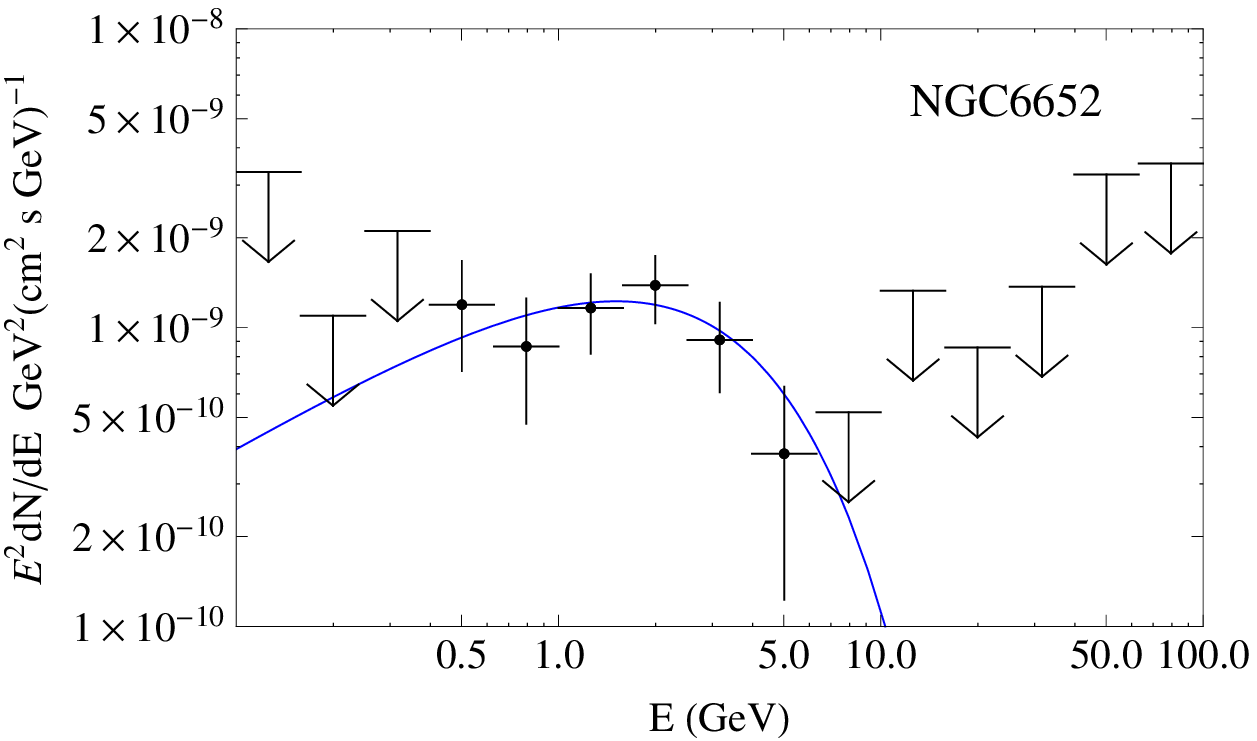} 
\includegraphics[width=3.40in,angle=0]{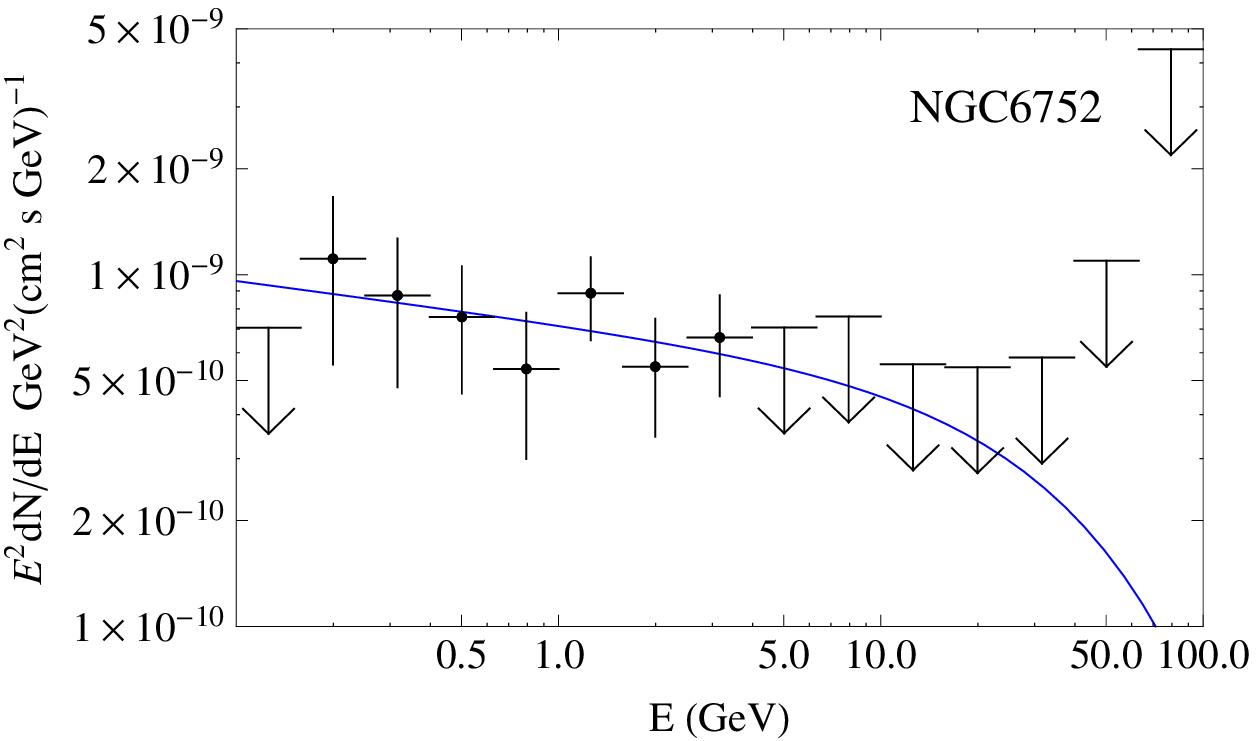} \\
\includegraphics[width=3.40in,angle=0]{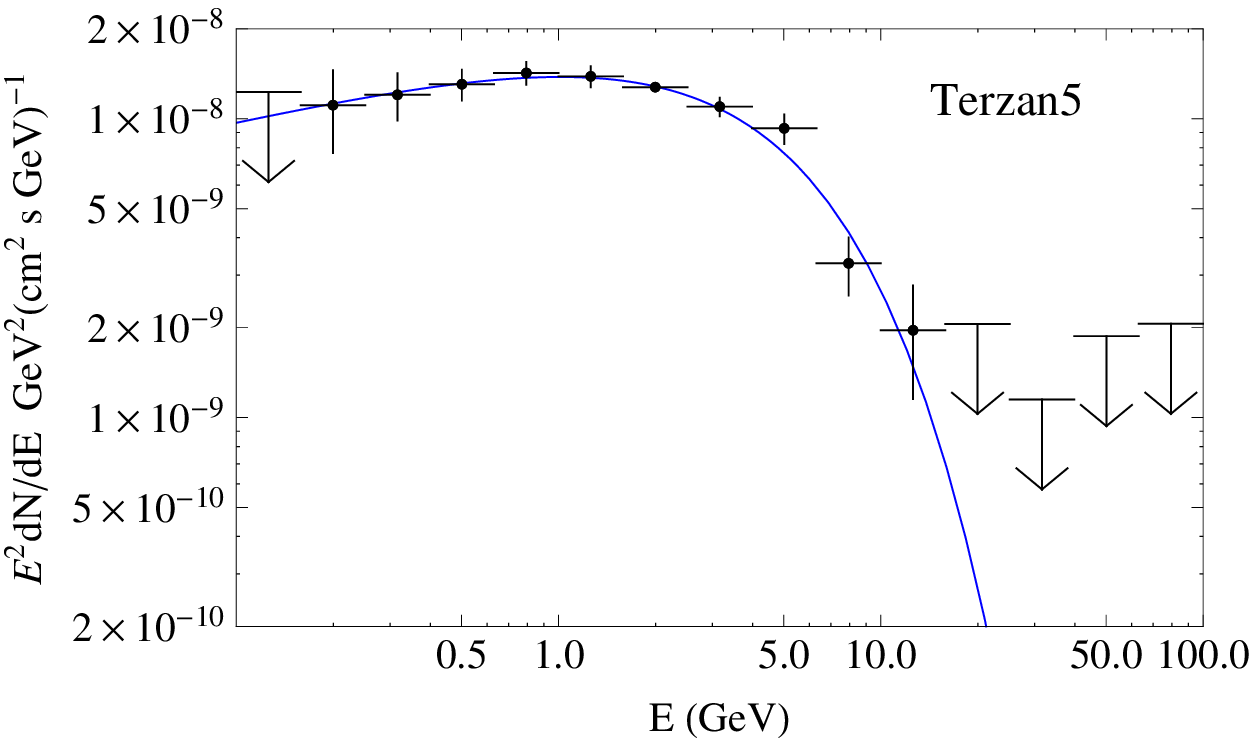}  
\includegraphics[width=3.40in,angle=0]{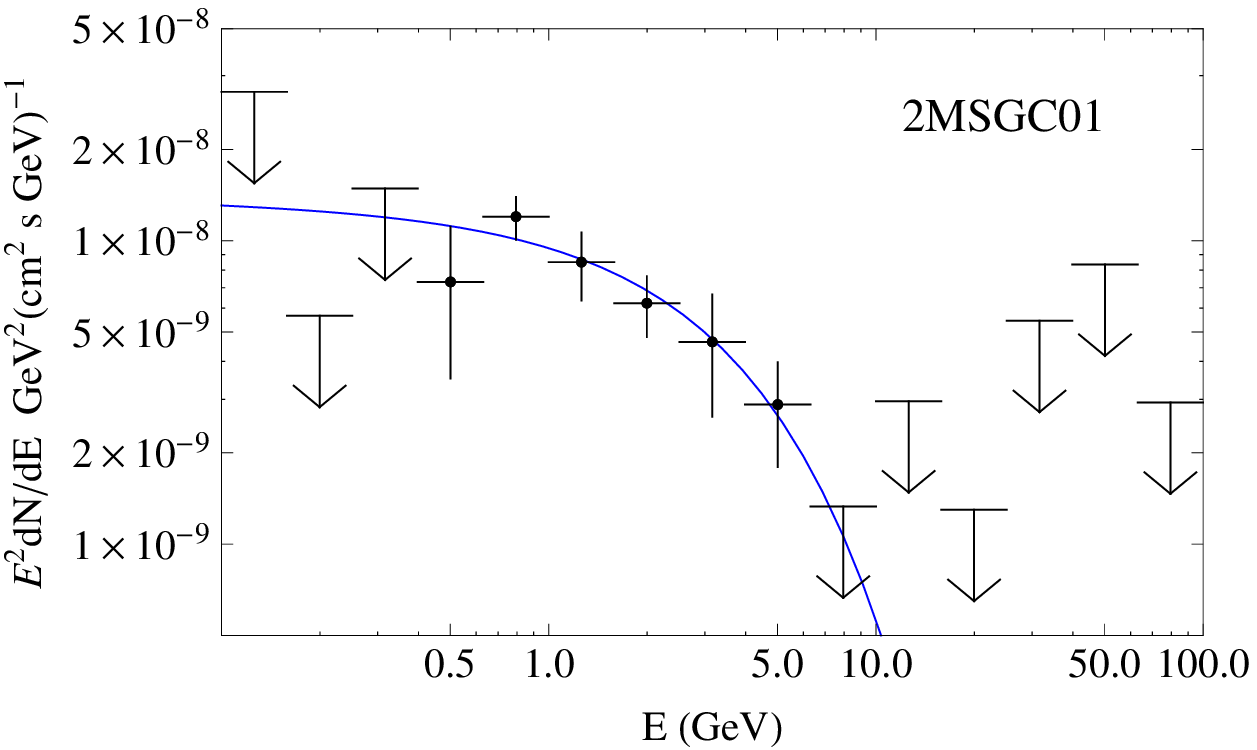}  \\
\caption{The gamma-ray spectra of globular clusters (continued).}
\label{fig:GCs2}
\end{figure*}

The gamma-ray emission observed from globular clusters is generally thought to be dominated by the MSPs contained within such systems. If this is true, then the gamma-ray spectra from globular clusters should closely resemble that observed from MSPs.  In Figs.~\ref{fig:GCs1} and~\ref{fig:GCs2}, and in Table~\ref{GlobClust-table}, we plot and describe the spectra from the 16 globular clusters detected at high-significance by Fermi. After stacking these 16 sources along with 20 other globular clusters either known to contain at least one MSP,
\footnote{%For a catalog of globular clusters containing MSPs, 
See: \url{www.naic.edu/}$\sim$\url{pfreire/GCpsr.html} for a catalog of globular clusters containing MSPs.} or given in Ref.~\cite{2012arXiv1207.7267T}, we arrive at the spectrum shown in the right frame of Fig.~\ref{fig:Stacked_MSPs}. Although the spectra observed from these sources is generally similar to that from MSPs, they tend to be flatter at low-energies ($\alpha \sim-2$ rather than the $\alpha \sim -1.6$ for typical MSPs); see Table~\ref{Stacked-table}. This comparison suggests that MSPs might not produce all of the gamma-ray emission observed from globular clusters (another point source population with a softer spectral index may also contribute). Alternatively, the softer emission observed from globular clusters could be diffuse in nature, such as that arising from inverse Compton scattering. Although one might consider the possibility that this soft gamma-ray emission originates from low-luminosity MSPs, we will show in Secs.~\ref{lumfunc2} and \ref{sec:correlations} that this cannot be the case.

%\begin{figure*}
%\includegraphics[width=5.40in,angle=0]{plots/GlobularClusters_alpha_VS_Ecut_P1_Alternative.eps}
%\caption{The spectral properties of the Globular Clusters as seen by Fermi. Error bars give the 68$\%$ CL range. We also give the stacked %Fermi Globular Clusters spectrum (green diamond). Solid blue: flat, $dN/dE \propto E^{-2}$; red dashed: for the $E^{2}dN/dE$ spectra with%$E_{\rm peak}=0.5$ GeV; brown dots: for the $E^{2}dN/dE$ spectra with $E_{\rm peak}=1.0$ GeV}
%\label{fig:Alpha_VS_Ecut_GCs}
%\end{figure*}

\begin{table*}[t]

\begin{tabular}{|c|c|c|c|c|c|c|c|c|}

\hline

Name    &       $A$ (cm$^{-2}$ s$^{-1}$) & $\alpha$ & $\alpha_{\rm low}$(68$\%$) & $\alpha_{\rm high}$(68$\%$) & $E_{\rm cut}$ (GeV) & $E_{\rm cut_{\rm low}}$(68$\%\

$) & $E_{\rm cut_{\rm high}}$(68$\%$) & $\chi^{2}$/d.o.f. \\

\hline \hline

IC1257            &    1.95$\times 10^{-10}$     &    -1.97      &    -2.37      &    -1.19     &       7.23      &      1.81     &     112     &     0.47  \\

\hline

NGC104 (47 Tuc)     &    5.85$\times 10^{-9}$         &        -1.33   &   -1.43      &    -1.22    &      2.33    &       2.00    &     2.70   &     2.26 \\

\hline

NGC5139        &     1.54$\times 10^{-9}$          &       -1.75   &   -2.25      &    -1.25    &      3.44      &     1.89     &     20.5     &     0.65 \\

\hline

NGC5904 (M5)   &     4.73$\times 10^{-11}$         &       -1.82   &   -2.30      &    -0.26    &       14.9     &     1.40     &      >1000    &    0.29 \\

\hline

NGC6266 (M62)    &     1.33$\times 10^{-10}$       &         -1.99   &   -2.08     &     -1.88      &     27.7   &       8.31     &     554    &    0.88 \\

\hline

NGC6388    &     1.94$\times 10^{-9}$          &       -1.67   &   -1.87     &     -1.47     &     3.43      &     2.40     &    5.76      &     0.48  \\

\hline

NGC6397    &     1.63$\times 10^{-10}$         &       -2.71   &   -2.97     &     -2.44     &      2.56      &     1.43    &     9.73     &     0.82 \\

\hline

NGC6440    &     2.45$\times 10^{-10}$        &        -2.43   &   -2.56     &     -2.30      &     7.94      &     4.29     &    32.6   &     0.72 \\

\hline

NGC6441    &     5.59$\times 10^{-10}$        &        -1.94   &   -2.23      &    -1.61      &     5.36      &     2.79    &     26.8    &     2.60 \\

\hline

NGC6541    &     8.50$\times 10^{-10}$        &        -1.33   &   -2.13      &   -0.21      &      2.55     &    1.15      &     17.3    &     0.71  \\

\hline

NGC6624    &     2.58$\times 10^{-11}$        &        -2.04   &   -2.16      &    -1.92     &      58.6     &     14.7     &     >1000   &    1.33 \\

\hline

NGC6626 (M28)    &     6.97$\times 10^{-10}$        &        -2.13   &   -2.26      &    -2.00     &      5.85      &     3.98     &      10.8    &     1.53 \\

\hline

NGC6652    &     1.31$\times 10^{-9}$        &        -1.36   &   -2.04     &     -0.16     &      2.36     &      0.90     &     17.3   &    1.05 \\

\hline

NGC6752    &     9.46$\times 10^{-12}$          &       -2.12   &   -2.32     &     -1.82     &      48.4     &     4.84    &     >1000    &    0.63 \\

\hline

Terzan5       &  6.19$\times 10^{-9}$            &     -1.75   &   -1.81      &    -1.67      &     4.05     &      3.52     &    4.62     &     0.76 \\

\hline

2MSGC01      &    3.88$\times 10^{-9}$         &    -2.02     &     -2.46  &    -1.53      &     3.24     &     1.68       &      11.2     &      1.12  \\

\hline \hline

\end{tabular}

\caption{The spectral parameters of 16 individual globular clusters (see Eq.~\ref{eq:MSP_Spect}). The ``high'' and ``low'' values encompass the 68$\%$ confidence interval. In the last column, we give the quality of the fit ($\chi^{2}$ per degree-of-freedom) for the best fit values.}

\label{GlobClust-table}

\end{table*}

\begin{table*}[t]

\begin{tabular}{|c|c|c|c|c|c|c|c|c|}

\hline

Name    &       $A$ (cm$^{-2}$s$^{-1}$) & $\alpha$ & $\alpha_{\rm low}$(68$\%$) & $\alpha_{\rm high}$(68$\%$) & $E_{\rm cut}$(GeV) & $E_{\rm cut_{\rm low}}$(68$\%$)\

 & $E_{\rm cut_{\rm high}}$(68$\%$) & $\chi^{2}$/d.o.f. \\

\hline \hline

All MSPs       &            1.33$\times 10^{-7}$       &          -1.57   &    -1.59     &       -1.56     &      3.78     &       3.70    &     3.93    &     7.45 \

\\

\hline

All GCs          &          1.19$\times 10^{-8}$         &        -2.05     &   -2.07      &     -1.69      &     5.76       &     3.05      &   5.93      &   3.19 \

\\

\hline \hline

\end{tabular}

\caption{The spectral parameters of the stacked millisecond pulsars and stacked globular clusters.}

\label{Stacked-table}

\end{table*}

%%%%%%

\newpage

\begin{figure*}[!t]
\includegraphics[width=3.5in,angle=0]{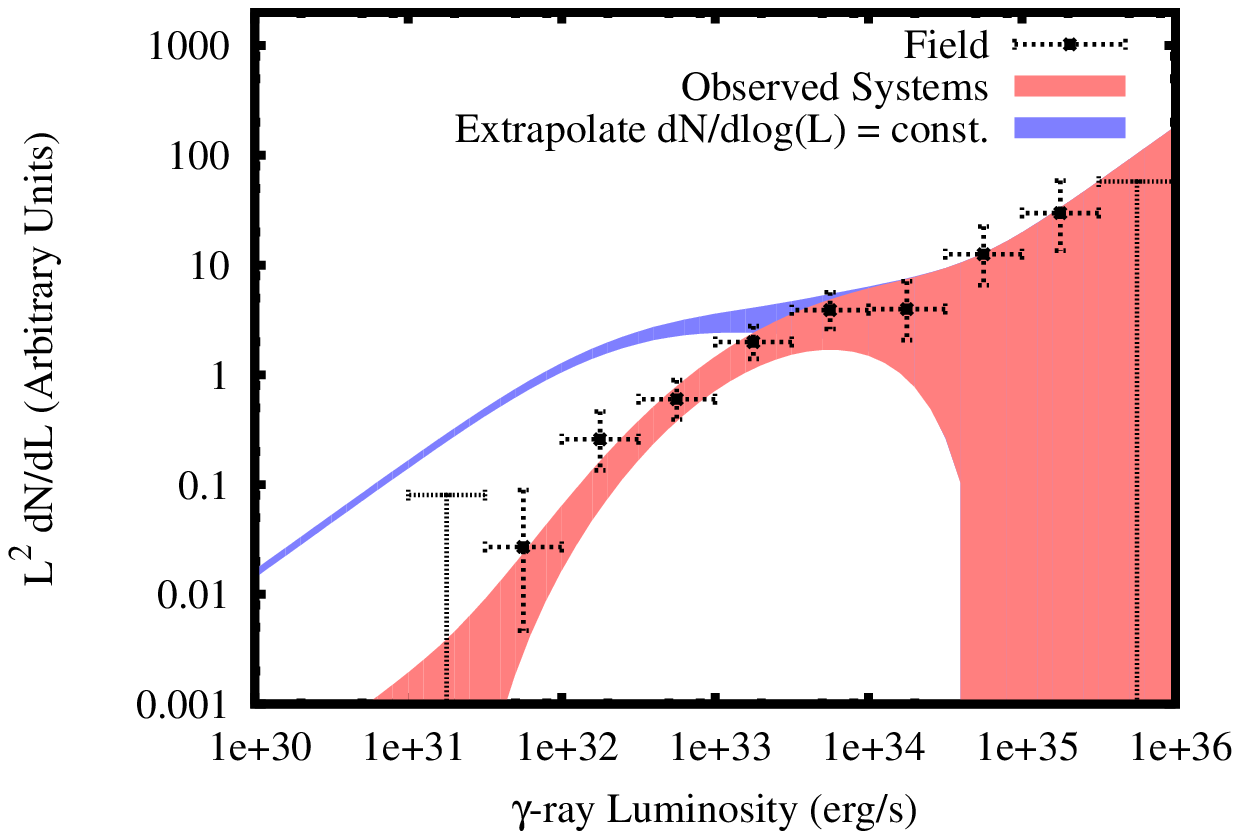}
\includegraphics[width=3.5in,angle=0]{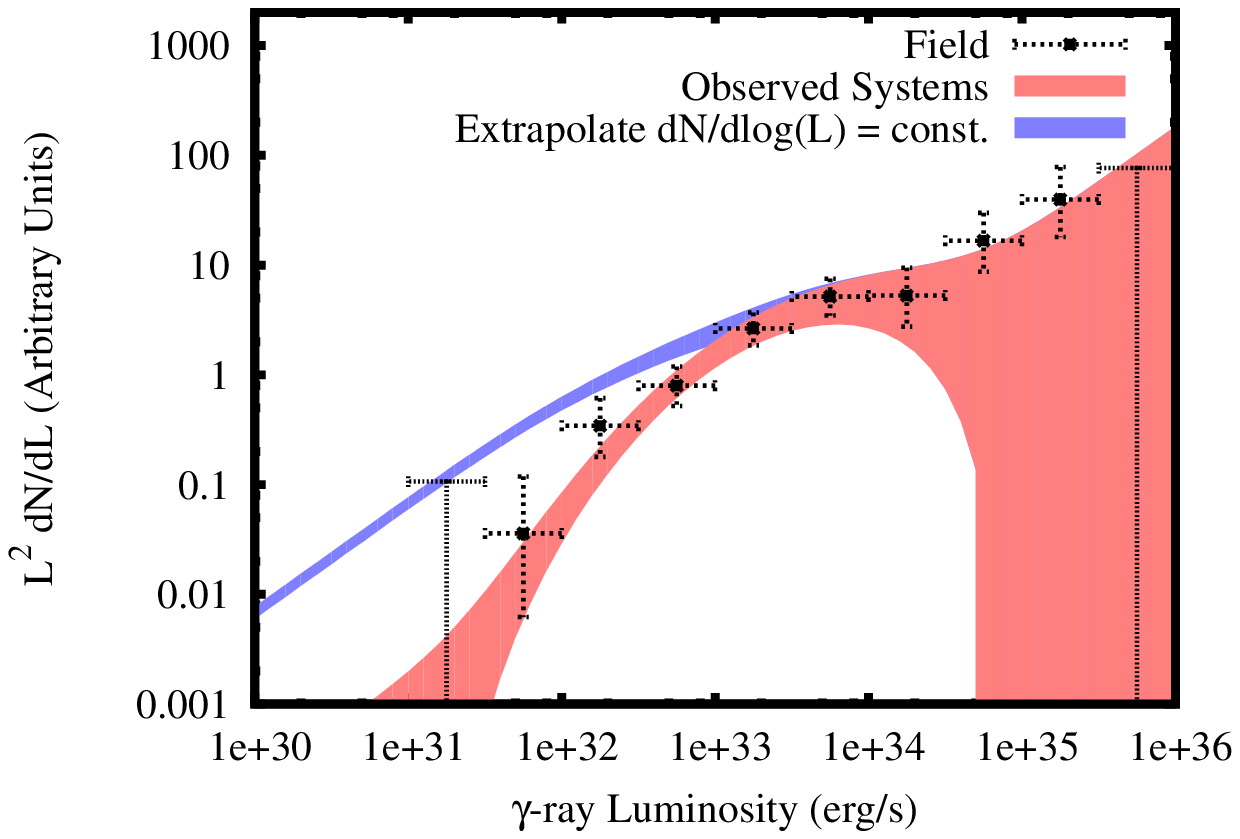}\\
\includegraphics[width=3.5in,angle=0]{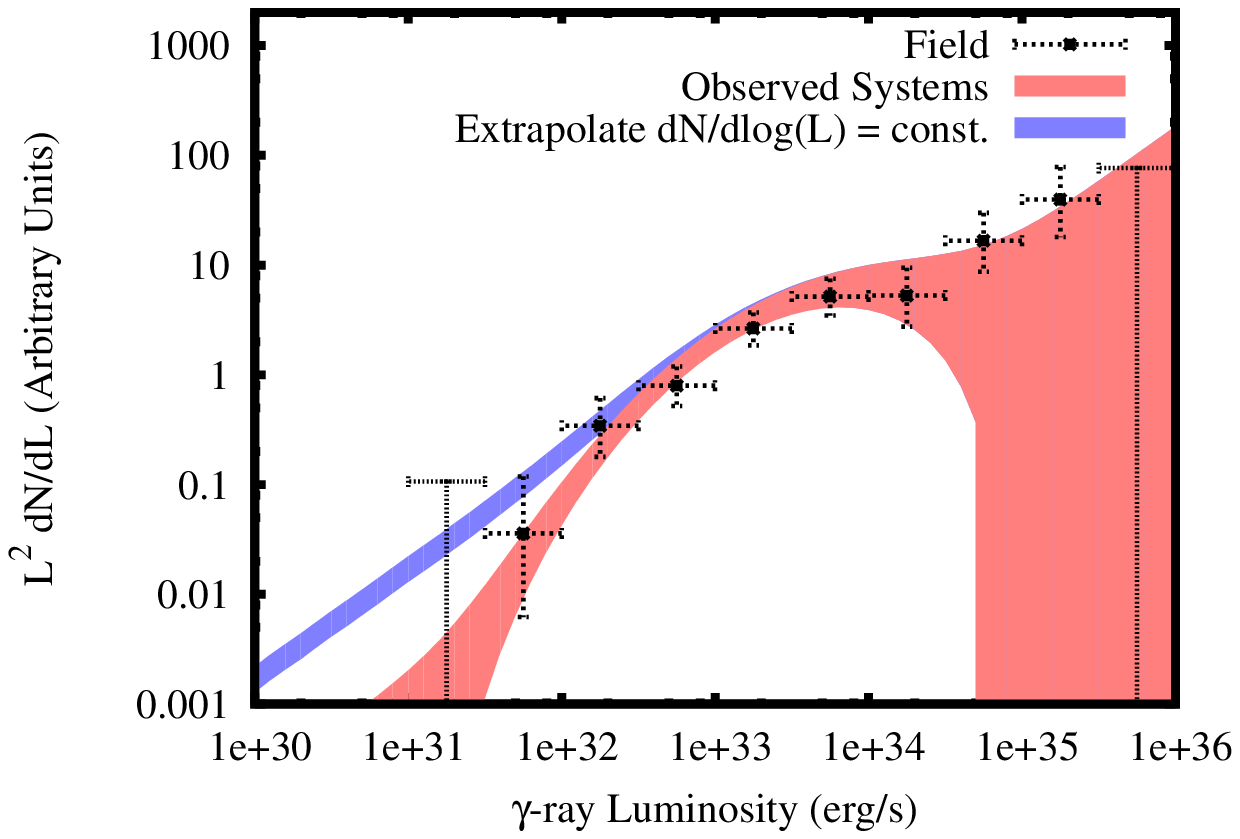}
\includegraphics[width=3.5in,angle=0]{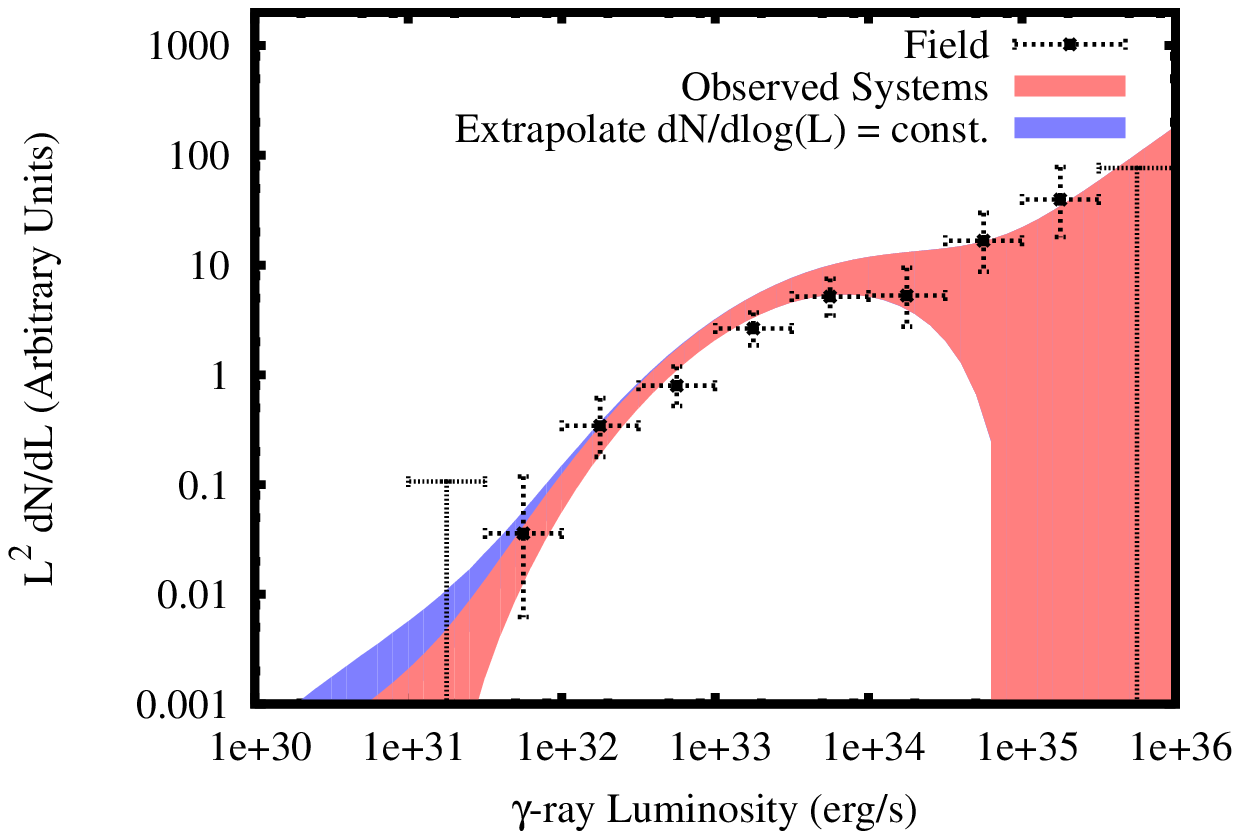}
\caption{The MSP gamma-ray luminosity function as calculated using the two independent and complementary techniques described in the text. The error bars denote the luminosity function determined from the population of MSPs observed by Fermi, excluding those residing in globular clusters (see Sec.~\ref{lumfunc1}). The red and blue bands represent the luminosity function determined using gamma-ray and X-ray observations of the globular cluster 47 Tucanae (see Sec.~\ref{lumfunc2}). In each frame, the red band denotes the luminosity function accounting only for the 17 MSPs observed in X-rays, whereas the blue band accounts for sub-threshold MSPs assuming an extrapolation in which $L\,dN/dL$ is constant at low luminosities (down to a minimum X-ray luminosity of 1.1~$\times$~10$^{26}$~erg/s). The width of the red and blue bands correspond only to the Poisson variance in the number of simulated systems within a decade of each $\gamma$-ray luminosity, and does not represent the statistical and systematic errors on the measured X-ray luminosities of the 47~Tuc MSPs. In the upper left frame, we have assumed an equal degree of beaming at gamma-ray and X-ray wavelengths, while in the other frames we assume that the gamma-ray emission is more isotropic, such that the solid angle of the gamma-ray emission is 1.5 (upper right), 2.0 (lower left), or 2.5 (lower right) times that of the X-ray emission. For a beaming ratio of $\sim$2.0-2.5, these two techniques yield very similar MSP luminosity functions.}
\label{fig:lumfunc}
\end{figure*}

\section{Luminosity Function}
\label{sec:LumFunc}

In this section, we determine the gamma-ray luminosity function of MSPs. We do this using two independent and complementary techniques. First, in the following subsection, we use the sample of nearby MSPs observed by Fermi to directly measure the luminosity distribution of this source population. We then, in Sec.~\ref{lumfunc2}, consider the population of MSPs within the globular cluster 47 Tucanne, and make use of the observed correlation between X-ray and gamma-ray luminosities to infer the gamma-ray luminosity function of the MSPs in that system. 

\subsection{Millisecond Pulsars in the Field of the Milky Way}
\label{lumfunc1}

In this subsection, we derive the gamma-ray luminosity function for the Milky Way's MSP population, excluding those residing within globular clusters. We do this through the following procedure. First, we count the number of MSPs observed over a given range of luminosities (see Table~\ref{lummsptable}). For the most luminous observed MSPs ($L_{\gamma}\sim10^{35}$ erg/s, integrated above 0.1 GeV) located more than $10^{\circ}$ away from the Galactic Plane, Fermi's source catalog should be largely complete, containing the majority of such objects present within the Milky Way. For more typical MSPs ($L_{\gamma} \sim 10^{32}-10^{34}$ erg/s), however, Fermi is only sensitive to those residing within the surrounding $\sim$1-5 kpc. For hard spectrum sources located outside of the Galactic Plane ($|b|>10^{\circ}$), the detection threshold of Fermi's source catalog extends out to distances of approximately $D \simeq$ 1 kpc $\times \, [L_{\gamma}/(1.5\times 10^{32}\,{\rm erg/s})]^{1/2} \,$~\cite{Fermi:2011bm}.  To account for sources that are too distant (and thus too faint) to be detected by Fermi, we multiply the observed number of MSPs in a given luminosity bin by the total number of observed MSPs with greater luminosities divided by the number of those MSPs that reside within the distance to which the detection threshold of Fermi's source catalog extends. For example, 9 MSPs have been observed with luminosities in the range of $\log_{10} (L_{\gamma}/{\rm erg/s})=33.5-34.0$ and at $|b|>10^{\circ}$. After correcting for the fact that 8 out of the 11 MSPs observed with luminosities greater than $10^{34}$ erg/s reside within the threshold distance for this luminosity range (within 4.56 kpc), we arrive at an effective number of 12.38. 

The results of this calculation are shown as error bars in Fig.~\ref{fig:lumfunc} (the meaning of the red and blue regions will be described in the following subsection). In $L^2 dN/dL$ units, the luminosity function rises continuously up to values of $\sim 3 \times 10^{35}$ erg/s, above which no such sources are observed. In $L\,dN/dL$ units, the luminosity function peaks at and is approximately flat between $L\sim 10^{32}-10^{34}$ erg/s, with a significant population extending to above $10^{35}$ erg/s.  Although there is an indication that the luminosity function begins to fall more rapidly below $10^{32}$ erg/s, Fermi is only sensitive to such faint sources if they are very nearby, leading to large statistical errors in this low-luminosity regime. 

\begin{figure}[!t]
\includegraphics[width=3.50in,angle=0]{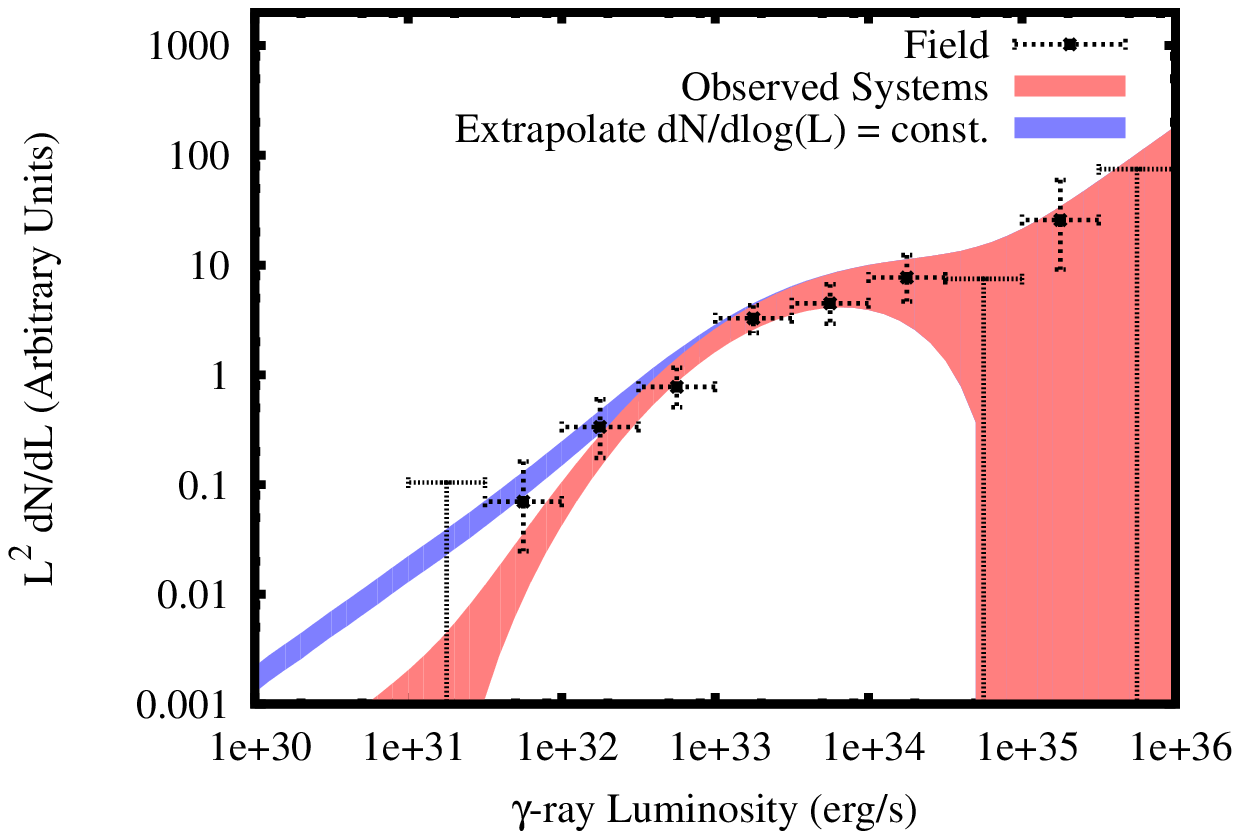}\\
\includegraphics[width=3.5in,angle=0]{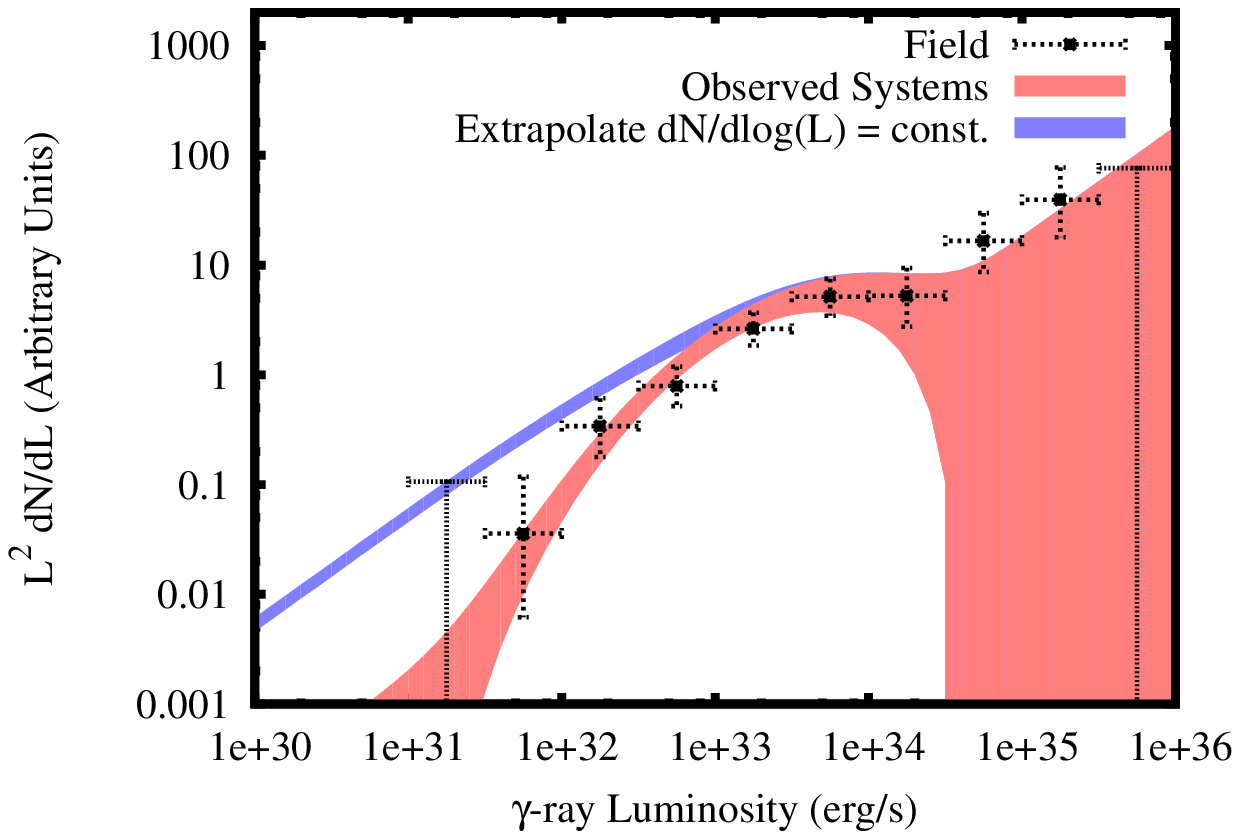}
\caption{Top: The same as in the lower left frame of Fig.~\ref{fig:lumfunc}, but using the distances reported in the Fermi second pulsar catalog~\cite{TheFermi-LAT:2013ssa} when available, instead of those reported in the ATNF catalog~\cite{Manchester:2004bp}. Bottom: The same as in the lower left frame of Fig.~\ref{fig:lumfunc}, but derived after deweighting those simulations in which the total gamma-ray flux exceeds that observed from 47 Tuc.}
\label{fig:lumfunc-alt}
\end{figure}

So far in our determination of the luminosity function, we have calculated the gamma-ray luminosities of individual MSPs using their distances as reported in the Australia Telescope National Facility (ATNF) pulsar catalog~\cite{Manchester:2004bp}.\footnote{For the current ATNF catalog, see \url{http://www.atnf.
csiro.au/people/pulsar/psrcat/.}} We adopt these distances as our default values because of the relative completeness of the ATNF catalog (a large fraction of the MSPs observed by Fermi have distances reported in this catalog). However, some of the distances reported in the Fermi second pulsar catalog are not insignificantly different from those reported in the ATNF catalog (see Table~\ref{lummsptable}). In the upper frame of Fig.~\ref{fig:lumfunc-alt}, we plot the luminosity function as found using the Fermi catalog distances when available (and ATNF values otherwise).

\begin{table*}[t]
\begin{tabular}{|c|c|c|c|c|c|c|c|c|}
\hline
Name    &       $l$ (deg.) & $b$ (deg.) & Flux/($10^{-11}$ erg/cm$^2$s) & Dist.~(kpc) & 2PC Dist.~(kpc) & $L_{\gamma}/(10^{33}$ erg/s)  & $\dot{E}$ (erg/s) & $f_{\gamma}$\\
\hline \hline
J0023+0923 & 111.38 & -52.85 &   0.73 &   0.95 &  0.69 &  0.783 & $1.5\times 10^{34}$& 0.048\\
 \hline
J0030+0451 & 113.13 & -57.61 &   6.11 &   0.28 &  -- &  0.573 &$3.5\times 10^{33}$& 0.16\\
 \hline
J0034-0534 & 111.49 & -68.07 &   1.99 &   0.98 &  0.54  & 2.285 &$2.9\times 10^{34}$& 0.074\\
 \hline
J0101-6422 & 301.19 & -52.72 &   1.33 &   0.73 &  0.55  & 0.849 &$1.1\times 10^{34}$& 0.077\\
 \hline
J0102+4839 & 124.87 & -14.17 &   1.85 &   4.03 & 2.32  & 35.972 &$1.7\times 10^{34}$& 2.1\\
 \hline
J0218+4232 & 139.51 & -17.52 &   4.76 &   5.85 & 2.64 & 194.729 &$2.4\times 10^{35}$&0.81\\
 \hline
J0307+7443 & 131.70 &  14.22 &   1.54 &   0.34 & --  &  0.213 &$2.2\times 10^{34}$&0.0097\\
 \hline
J0340+4130 & 153.79 & -11.02 &   2.07 &   2.67 &  1.73 & 17.624 &$7.5\times 10^{33}$&2.4\\
 \hline
J0437-4715 & 253.39 & -41.96 &   1.89 &   0.16 &  --  & 0.058 &$1.2\times 10^{34}$&0.0048\\
 \hline
J0533+6759 & 144.78 &  18.18 &   0.91 &   6.66 & --  & 48.273 & -- &\\
 \hline
J0605+3757 & 174.19 &   8.02 &   0.55 &   1.16 &  --  & 0.881 & -- &\\
 \hline
J0610-2100 & 227.75 & -18.18 &   0.89 &   5.64 &  3.54 & 33.891 &$8.4\times 10^{33}$&4.0\\
 \hline
J0613-0200 & 210.41 &  -9.31 &   3.34 &   0.90 &  --  & 3.233 &$1.5\times 10^{34}$&0.22\\
 \hline
J0614-3329 & 240.50 & -21.83 &  11.23 &   2.96 &  1.90 & 117.729 &$2.2\times 10^{34}$&5.3\\
 \hline
J0751+1807 & 202.73 &  21.09 &   1.18 &   0.40 & --   & 0.227 & $7.3\times 10^{33}$&0.031\\
 \hline
J1024-0719 & 251.70 &  40.52 &   0.39 &   0.49 &  0.39  & 0.112 &$5.3\times 10^{33}$&0.021\\
 \hline
J1124-3653 & 284.09 &  22.76 &   1.47 &   4.40 & 1.72    &34.061 &$1.7\times 10^{34}$&2.0\\
 \hline
J1125-5825 & 291.89 &   2.60 &   1.06 &   2.98 &  2.62  &11.307 &$8.1\times 10^{34}$&0.13\\
 \hline
J1137+7528 & 295.79 &  -5.17 &   0.21 &  19.53 &  --  &94.570 &--&\\
 \hline
J1142+0119 & 267.54 &  59.40 &   0.63 &   2.04 & --   & 3.120 &--&\\
 \hline
J1231-1411 & 295.54 &  48.38 &  10.32 &   0.45 & --   & 2.500 &$1.8\times 10^{34}$&0.14\\
 \hline
J1301+0833 & 310.81 &  71.29 &   0.76 &   0.91 & --   & 0.754 &--&\\
 \hline
J1302-3258 & 305.59 &  29.83 &   1.67 &   1.86 & --   & 6.931 &--&\\
 \hline
J1311-3430 & 307.68 &  28.18 &   6.27 &   3.72 & --  &103.760 &$4.9\times 10^{34}$&2.1\\
 \hline
J1312+0051 & 314.83 &  63.23 &   1.21 &   1.15 &   -- & 1.910 &--&\\
 \hline
J1446-4701 & 322.50 &  11.43 &   0.76 &   2.03 &  1.46  & 3.753 &$3.8\times 10^{34}$&0.099\\
 \hline
J1514-4946 & 325.25 &   6.81 &   4.02 &   1.27 &   0.94  &7.749 &$1.6\times 10^{34}$&0.48\\
 \hline
J1543-5149 & 327.92 &   2.48 &   2.93 &   1.46 &   -- & 7.477 &$7.3\times 10^{34}$&0.10\\
 \hline
J1544+4937 &  79.17 &  50.17 &   0.29 &   2.30 &  --  & 1.823 &$1.2\times 10^{34}$&0.15\\
 \hline
J1600-3053 & 344.09 &  16.45 &   0.77 &   2.40 &  1.63  & 5.339 &$8.1\times 10^{33}$&0.66\\
 \hline
J1614-2230 & 352.63 &  20.19 &   2.47 &   1.77 &  0.65   &9.270 &$1.3\times 10^{34}$&0.70\\
 \hline
J1630+3734 &  60.25 &  43.22 &   0.86 &   0.85 &  --   &0.745 &--&\\
 \hline
J1640+2224 &  41.05 &  38.27 &   0.18 &   1.15 &  --  & 0.301 &$3.5\times 10^{33}$&0.084\\
 \hline
J1658-5324 & 334.87 &  -6.62 &   2.61 &   1.24 &   0.93  &4.796 &$3.2\times 10^{34}$&0.14\\
 \hline
J1713+0747 &  28.75 &  25.22 &   1.05 &   1.05 &  --  & 1.379 &$3.5\times 10^{33}$&0.39\\
 \hline
J1732-5049 & 340.03 &  -9.46 &   0.66 &   1.81 &  --   &2.577 &$3.7\times 10^{33}$&0.62\\
 \hline
J1741+1351 &  37.88 &  21.64 &   0.55 &   1.41 &  1.08   &1.309 &$2.2\times 10^{34}$&0.057\\
 \hline
J1744-1134 &  14.79 &   9.18 &   4.15 &   0.42 &  --   &0.876 &$5.2\times 10^{33}$&0.17\\
 \hline
J1745+1017 &  34.87 &  19.26 &   1.53 &   1.36 & --   & 3.388 &$5.3\times 10^{33}$&0.63\\
 \hline
J1747-4036 & 350.20 &  -6.41 &   1.05 &   5.74 &  3.39  &41.352 &$1.2\times 10^{35}$&0.34\\
 \hline
J1810+1744 &  44.64 &  16.80 &   2.30 &   2.49 &  2.00  &17.084 &$4.0\times 10^{34}$&0.43\\
 \hline
J1811-2405 &   7.07 &  -2.56 &   4.24 &   1.70 &  --  &14.669 &$2.8\times 10^{34}$&0.23\\
 \hline
J1816+4510 &  72.83 &  24.74 &   1.16 &   4.20 &  --  &24.459 &$5.2\times 10^{34}$&0.45\\
 \hline
J1843-1113 &  22.05 &  -3.40 &   2.56 &   1.97 &--    &11.881 &$6.0\times 10^{34}$&0.20\\
 \hline
J1858-2216 &  13.58 & -11.39 &   1.10 &   1.35 &  0.94  & 2.400 &$1.1\times 10^{34}$&0.21\\
 \hline
J1902-5105 & 345.65 & -22.38 &   2.15 &   2.11 &  1.18  &11.473 &$6.8\times 10^{34}$&0.16\\
 \hline
J1939+2134 &  57.51 &  -0.29 &   0.44 &   5.00 & 3.56   &13.167 &$1.1\times 10^{36}$&0.011\\
 \hline
J1959+2048 &  59.19 &  -4.70 &   1.84 &   1.53 &   2.49  &5.157 &$1.6\times 10^{35}$&0.032\\
 \hline
J2017+0603 &  48.62 & -16.02 &   3.71 &   1.32 &  1.57   &7.734 &$1.3\times 10^{34}$&0.59\\
 \hline
J2043+1711 &  61.92 & -15.32 &   2.72 &   1.13 &  1.76   &4.158 &$1.5\times 10^{34}$&0.28\\
 \hline
J2047+1053 &  57.05 & -19.67 &   0.58 &   2.23 & 2.05   & 3.475 &$1.1\times 10^{34}$&0.30\\
 \hline
J2051-0827 &  39.19 & -30.41 &   0.24 &   1.28 &  1.04  & 0.480 &$5.5\times 10^{33}$&0.080\\
 \hline
J2124-3358 &  10.92 & -45.44 &   4.08 &   0.30 &  --  & 0.440 &$6.9\times 10^{33}$&0.063\\
 \hline
J2129-0429 &  48.91 & -36.94 &   0.83 &   1.03 &  --  & 1.057 &--&\\
 \hline
J2214+3000 &  86.85 & -21.66 &   3.39 &   1.32 & 1.54    &7.069 &$1.8\times 10^{34}$&0.39\\
 \hline
J2215+5135 &  99.87 &  -4.16 &   2.58 &   3.30 &  3.01 & 33.562 &$5.2\times 10^{34}$&0.65\\
 \hline
J2241-5236 & 337.45 & -54.93 &   3.35 &   0.68 &   0.51  &1.853 &$2.5\times 10^{34}$&0.073\\
 \hline
J2256-1024 &  59.22 & -58.29 &   0.56 &   0.91 & --   & 0.559 &--&\\
 \hline
J2302+4442 & 103.40 & -14.00 &   3.78 &   0.75 & 1.19   & 2.544 &$3.7\times 10^{33}$&0.69\\
\hline \hline
\end{tabular}
\caption{Gamma-ray fluxes, luminosities, and distances for MSPs used in our luminosity function determination, as described in Sec.~\ref{lumfunc1}. Fluxes and luminosities are integrated above 0.1 GeV. In our default calculation, we use the distances reported in the ATNF catalog (column 5)~\cite{Manchester:2004bp}. In the sixth column, we give the distances reported in Fermi's second pulsar catalog~\cite{TheFermi-LAT:2013ssa}, when different from those reported by ATNF. Note that J0737-3039A is not included, as it resides within a globular cluster.}
\label{lummsptable}
\end{table*}

%\begin{figure*}[!t]
%%\includegraphics[width=3.5in,angle=0]{plots/out7_L2dNdL.eps}
%\includegraphics[width=3.5in,angle=0]{plots/out7_beaming_L2dNdL.eps}\\
%\caption{The same as in the right frame of Fig.~\ref{fig:lumfunc}, but derived after discarding those simulations in which the total gamma-ray flux exceeds that observed from 47 Tuc.}
%\label{fig:lumfunc2}
%\end{figure*}

%\begin{figure*}
%\includegraphics[width=4.40in,angle=0]{lumfunc-finalcorrect.eps}
%\caption{Luminosity funciton...}
%\label{lumfunc1}
%\end{figure*}

\subsection{Millisecond Pulsars in Globular Clusters}
\label{lumfunc2}

The gamma-ray emission from globular clusters is generally believed to be dominated by the MSPs they contain, and observations of such systems can be used to constrain the gamma-ray emission from MSPs which individually lie below the Fermi detection threshold. The Fermi second source catalog includes 11 globular clusters detected as gamma-ray point sources~\cite{Fermi:2011bm} (see also Ref.~\cite{collaboration:2010bb}). These systems are each thought to contain $\sim$10--100 MSPs, which collectively produce the majority of their gamma-ray emission. Due to the significant distances to even the nearest globular clusters,\footnote{The two nearest globular clusters, M4 and NGC 6397 (each of which are 2.2 kpc from the Solar System), have not yet been detected at gamma-ray wavelengths.} and to the relatively poor point spread function of gamma-ray instruments, Fermi has resolved as individual sources only two MSPs within globular clusters (PSR J1823-3021A in NGC 6624 and PSR J1824-2452A in M28).  MSPs, however, have been detected at radio and X-ray wavelengths in the majority of globular clusters detected by Fermi. And although, on an individual pulsar basis, there is only a weak correlation observed between the emission at radio, X-ray, and gamma-ray wavelengths, the fairly large sample of MSPs present within the largest globular clusters allows us to reduce the statistical uncertainty involved, making it possible to deduce the underlying gamma-ray luminosity function of the MSPs contained within such a system.

%Since nearly all known MSPs were first detected in radio observations, there could be a significant selection effect in the ratio of radio intensity to gamma-ray emission for observed MSPs. In this work, we will examine the correlation between the X-ray emission and gamma-ray emission in radio selected MSPs observed in all three wavelengths. 

In studying the gamma-ray flux observed by Fermi at energies above 0.1 GeV and the non-thermal X-ray flux observed between 0.3 and 10~keV,  \citet{TheFermi-LAT:2013ssa} found that a correlation exists between these two fluxes. They fit this correlation to a normal distribution following $\log_{10}(F_\gamma$/F$_X$)~=~2.31~$\pm$~0.48~\citep{TheFermi-LAT:2013ssa}. This correlation is consistent with the observation that MSPs convert (on average) approximately 10\% of their spin-down
energy into gamma-rays~\cite{Fermi:2011bm,TheFermi-LAT:2013ssa}, and approximately
0.06\% of their spin-down energy into X-rays~\cite{2009ASSL..357.....B}. In attempting to apply this correlation to the population of globular cluster MSPs detected in X-rays, but not in gamma-rays, we note that the correlation could be biased in several ways. Most importantly, a large population of X-ray bright, but gamma-ray dim MSPs could exist, which would not be absent in the field sample. Upon examining the list of X-ray detected rotationally-powered MSPs from Ref.~\cite{2009ASSL..357.....B} (see Table 6.7, p. 132), however, we find that Fermi has successfully detected gamma-ray pulsations from 10 of the 13 field MSPs in this catalog.\footnote{The three systems currently missing in gamma-ray observations are B1257+12, J1012+5307 and B1534+12} This argues against there being a strong selection effect in the X-ray/gamma-ray correlation.

Another difficulty in translating the correlation from \citet{TheFermi-LAT:2013ssa} into a constraint on the population of globular cluster MSPs is the breakdown of the X-ray flux into thermal and non-thermal components. This division, employed by the Fermi analysis, makes physical sense due to the fact that only the non-thermal flux is thought to originate from the synchrotron radiation of the relativistic electron population that also produces the gamma-ray emission. However, this makes it difficult to compare the results of different groups who have studied globular cluster MSPs, since they have used varying classifications in order to separate the X-ray flux into thermal and non-thermal components. We note that work by \citet{Marelli:2012sf}, which formed the basis for the Fermi analysis, additionally calculated the thermal X-ray component of rotationally powered MSPs. Using these results, we find that MSPs which are modeled with only thermal fluxes actually have a relative gamma-ray flux that is even larger compared to the observed X-ray flux than MSPs with a measured non-thermal flux. In these systems, the correlation follows a normal distribution centered at  $\log_{10}(F_\gamma$/F$_X$)~=~3.03. In what follows, we assume that the correlation reported by \citet{TheFermi-LAT:2013ssa} holds for all X-ray detected MSPs, regardless of the spectrum of the X-ray emission.

Using this model, we can calculate the gamma-ray flux expected from the X-ray bright MSP population of any given globular cluster. Due to its large number (23) of MSPs detected at radio wavelengths~\citep{msps_47Tuc, msps_10_47Tuc, msps_more_47Tuc, Camilo:1999fc}, we concentrate on 47~Tuc. Work by \citet{Bogdanov:2006ap} found X-ray sources coincident to 19 of these systems, and provide an X-ray flux from 0.3-8~keV for 17 of the observed MSPs. These systems are all radio selected, and the high fraction of radio MSPs in 47~Tuc that are detected in X-rays argues against the possibility that these systems are systematically over-luminous in X-rays. We take the X-ray flux for each MSP to follow the blackbody power-law flux calculation given by \citet{Bogdanov:2006ap}. This produces a modeled X-ray flux that is roughly 10\% smaller than the neutron star hydrogen atmosphere model alternatively presented in the paper. We also conservatively assume the flux at 8-10~keV to be negligible, which is true in most circumstances. 

In order to calculate the expected gamma-ray luminosity, we produce a Monte Carlo algorithm for the luminosity of each detected X-ray pulsar, drawing the gamma-ray to X-ray flux ratio for each source from the distribution of Ref.~\citep{TheFermi-LAT:2013ssa}. In Fig.~\ref{fig:lumfunc} (and also in Fig.~\ref{fig:lumfunc-alt}), we show the luminosity function of the MSP population within 47 Tuc, as determined using this method. In each frame, the red band denotes the luminosity function accounting only for the 17 MSPs with reported X-ray fluxes. The blue bands, in contrast, account for sub-threshold MSPs, assuming an extrapolation in which $L\,dN/dL$ is constant at low luminosities (down to a minimum X-ray luminosity of 1.1~$\times$~10$^{26}$~erg/s). The size of the statistical error bars corresponds to the Poisson variance in the number of simulated systems within a decade of each gamma-ray luminosity, and does not represent statistical and systematic errors on the measured X-ray luminosities of the 47 Tuc MSPs.

\begin{figure}[!t]
\includegraphics[width=3.40in,angle=0]{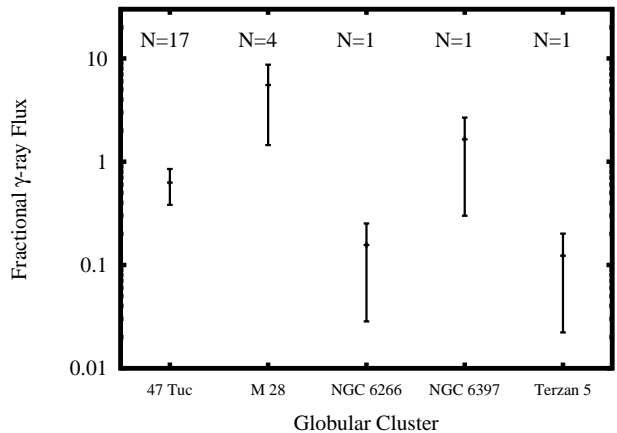}
\caption{The fraction of the total observed gamma-ray emission from five globular clusters that is predicted by our simulation (based on the observed correlation between X-ray and gamma-ray emission from MSPs). Each error bar represents the range covered by 68\% of our simulations. The quantities $N=17$, 4, 1, 1, 1 denote the number of MSPs detected as X-ray sources in each globular cluster.} 
%Note that one of the four X-ray MSPs detected in M28 supplies (in an average simulation) more than
%90\% of the predicted flux and thus can account for the entire gamma-ray emission observed from this %system.
\label{fig:averages}
\end{figure}

In the upper left frame of Fig.~\ref{fig:lumfunc}, we have assumed that the X-ray and gamma-ray emission from MSPs are anisotropic to the same degree (are equally beamed). This is not expected to be the case, however, as models of MSP emission generally predict a gamma-ray beam that significantly exceeds the size of the radio or X-ray beams~\cite{Guillemot:2014kta,Venter:2009eu,Watters:2008bp,Watters:2010jb} (for a recent review, see Ref.~\cite{Pierbattista:2014ona}). In the upper right, lower left, and lower right frames of Fig.~\ref{fig:lumfunc}, we adopt a solid angle for the gamma-ray emission that is 1.5, 2.0, and 2.5 times larger that of the X-ray emission, respectively.
% (for the case of a ratio of 2, for example, we calculate this result by assuming that for each detected X-ray system there is a second undetected system with the same characteristics). 
%
For reasonable values of $\sim$2.0-2.5 for this beaming ratio, we find good agreement between the luminosity functions found using the techniques of this subsection and the previous subsection.

For the case in which the X-ray emission is beamed twice as strongly as the gamma-rays, 70\% the simulations predict a gamma-ray flux from the 17 X-ray measured MSPs that exceeds the total observed emission, and thus is unphysical. In the lower frame of Fig.~\ref{fig:lumfunc-alt}, the luminosity function was calculated after deweighting these simulations. After doing this, our median simulation predicts that the 17 X-ray measured pulsars in 47 Tuc produce a gamma-ray luminosity of 2.27~$\times$~10$^{-11}$~erg~cm$^{-2}$s$^{-1}$, or approximately 86\% of the total gamma-ray luminosity observed by Fermi (2.65~$\times$~10$^{-11}$~erg~cm$^{-2}$s$^{-1}$).

%For the case of equal beaming at X-ray and gamma-ray wavelengths, our median simulation predicts that the 17 X-ray measured pulsars in 47 Tuc produce a gamma-ray luminosity of 1.47~$\times$~10$^{-11}$~erg~cm$^{-2}$s$^{-1}$, or approximately 55\% of the total gamma-ray luminosity observed by Fermi (2.65~$\times$~10$^{-11}$~erg~cm$^{-2}$s$^{-1}$). In approximately 9\% of simulations, the gamma-ray flux from these 17 MSPs exceeds the total observed emission, and thus is unphysical. In the lower frame of Figs.~\ref{fig:lumfunc-alt}, the luminosity function was calculated after discarding these simulations. 

%If we assume a gamma-ray beam to X-ray beam ratio of 2, we find that the emission from the bright MSPs overproduces the total gamma-ray luminosity observed rom 47~Tuc in 71\% of simulations, and the remaining simulations on average produce 85\% of the total gamma-ray emission.

So far in this subsection, we have considered only the MSPs in the globular cluster 47~Tuc. In Fig.~\ref{fig:averages}, we show the fraction of the total observed gamma-ray luminosity that is predicted in our model from the X-ray detected MSPs in five globular clusters, assuming an equal degree of beaming at gamma-ray and X-ray wavelengths. The error bars shown in this figure represent the range predicted in 68\% of our simulations (the data-point is the mean predicted value). Our ability to use these systems to draw meaningful conclusions, however, is limited by the very small number of MSPs detected in X-rays (only 4 in M28, and one 1 in each of NGC 6266, NGC 6397, and Terzan 5).  In M28, a single X-ray source is predicted by our (median) simulation to produce more than 90\% of the observed gamma-ray emission from this globular cluster. Despite its statistical limitations, this plot demonstrates that in these five systems, all or most of the observed gamma-ray emission could plausibly be produced by the handful of X-ray detected MSPs (this conclusion was also reached in Ref.~\cite{Johnson:2013uza} for the case of M28).

By comparing the luminosity function calculated in subsection~\ref{lumfunc1} using the MSPs individually resolved by Fermi in the Galactic field to that that calculated in this subsection using the MSP population of 47~Tuc, we find good agreement between these two techniques. In particular, for a gamma-ray to X-ray beam ratio of $\sim$2.0-2.5, these two independent techniques yield luminosity functions that are in excellent agreement with each other (see the right frames of Fig.~\ref{fig:lumfunc}). In contrast, this comparison seems to further disfavor models in which the degree of gamma-ray beaming and X-ray beaming are equal. 

These results support the conclusion that the total gamma-ray emission from MSPs is dominated by the brightest members of their population, and appears to exclude the possibility that there exists a sizable sub-population of gamma-ray faint MSPs which dominates the total gamma-ray emission from globular clusters.\footnote{The luminosity function derived here is similar to that used in Ref.~\cite{Hooper:2013nhl}, but quite different from those adopted in Ref.~\cite{Yuan:2014rca}.}  Quantitatively, we can rule out at the 95\% confidence level that the 17 MSPs detected in X-rays produce less than 27\% of the total observed gamma-ray emission from 47~Tuc. In more typical simulations, nearly all of the observed gamma-ray emission from this system is generated by these 17 sources. We also note that the dimmest MSP in 47~Tuc observed by \emph{Chandra} has an X-ray flux of 7.1~$\times$~10$^{-16}$~erg~cm$^{-2}$s$^{-1}$, which translates (in a typical simulation) into a gamma-ray luminosity of 2.4~$\times$~10$^{32}$~erg~s$^{-1}$~\cite{1985IAUS..113..541W}. While such a system is unlikely to be detected in gamma-rays at the distance of 47~Tuc, the number of systems of this or greater luminosity is constrained by by the paucity of such X-ray sources.

%\begin{figure*}[!t]
%\includegraphics[width=3.40in,angle=0]{plots/47Tuc_beaming_LdNdL.eps}
%\includegraphics[width=3.40in,angle=0]{plots/47Tuc_beaming_L2dNdL.eps}
%\caption{Same as Figure~\ref{fig:47Tuc_nobeam} but assuming a relative beaming fraction between the $\gamma$-ray and radio beam of 2.}
%\label{fig:47Tuc_beam}
%\end{figure*}

%In all cases, we compare the observed $\gamma$-ray luminosity function in our simulations to the measured MSP $\gamma$-ray luminosity function in Figure~{\bf get figure number}. Broadly speaking, the simulated and observed $\gamma$-ray luminosity functions are in good agreement, giving credence to both measurements. On a more detailed level, we find much better agreement for a beaming fraction of approximately 2, rather than a beaming fraction of unity. While this analysis suffers from large statistical and systematic uncertainties at the present time, we note that this method may some-day provide an additional observational constraint on the $\gamma$-ray and radio beaming fractions.

\begin{figure*}
\includegraphics[width=3.40in,angle=0]{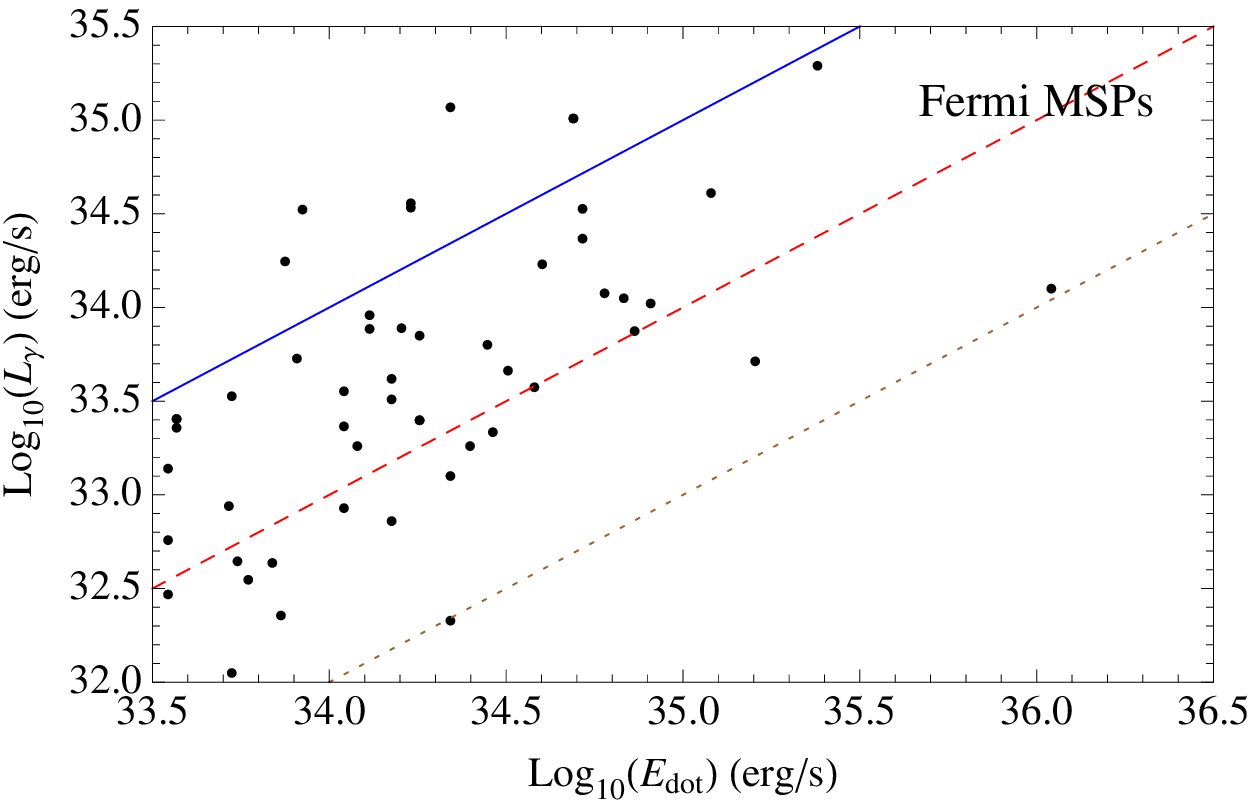} 
\includegraphics[width=3.40in,angle=0]{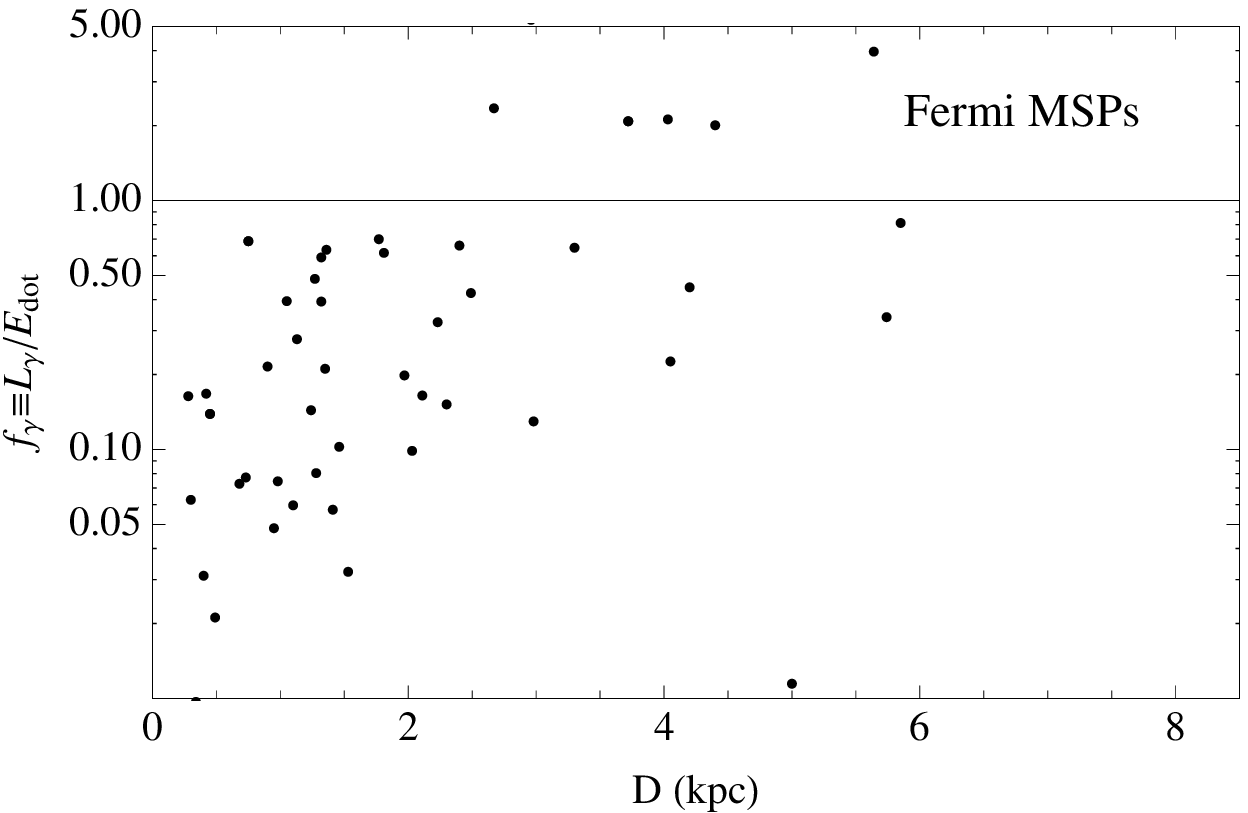} \\
\includegraphics[width=3.40in,angle=0]{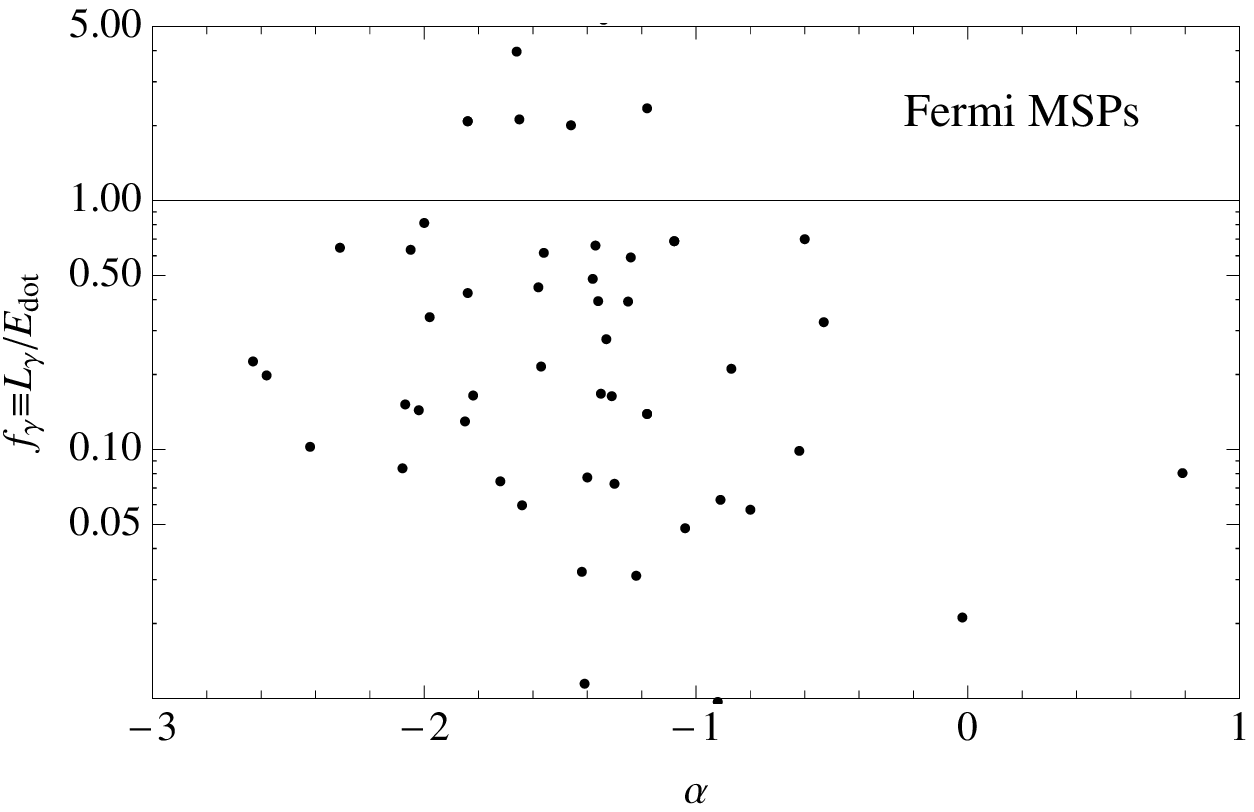} 
\includegraphics[width=3.40in,angle=0]{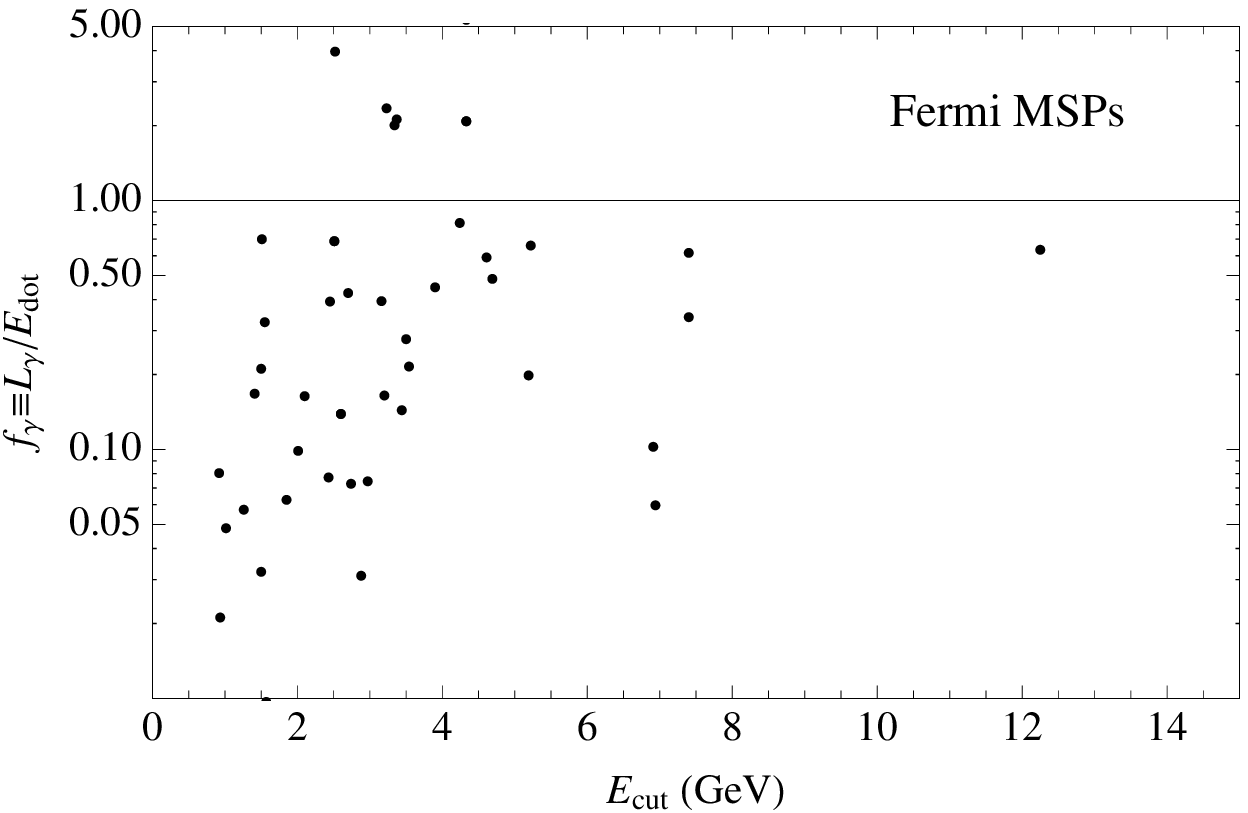} 
\caption{Top left: The isotropic luminosity in gamma rays vs the spin-down power. The solid blue, red dashed, and brown dotted lines represent $f_{\gamma}$=1.0, 0.1, and 0.01, respectively. Top right: the ratio $f_{\gamma}$ vs the distance of MSPs from Earth; showing clear evidence of a selection effect. Bottom left and right: the ratio $f_{\gamma}$ vs the spectral index and cut-off of MSPs in gamma-rays. Very soft and dim MSPs in gamma-rays may also be missing from the catalogue (see the bottom left corner of the bottom left plot).}
\label{fig:Correlations1}
\end{figure*}

\section{Implications for the Physics of Millisecond Pulsars}
\label{sec:correlations}

Gamma-ray emission from MSPs has been suggested to originate either from regions where 
magnetospheric charge depletion occurs (outer gaps models), from regions close to the neutron 
star magnetic poles (polar cap models), or from regions that start close to the neutron star magnetic poles
and extend almost as far as the light cylinder (slot gap models). In this section, we discuss the implications of our results within the context of these classes of emission models.

In outer gap models~\cite{Cheng:1986qt}, large electric fields exist along the magnetic lines within the gap regions ($\overrightarrow{B} \cdot \overrightarrow{E} \neq 0$). Given the geometry of the gap regions and that the observed gamma rays are mainly emitted close to the gaps~\cite{Cheng:1986qt, 1994ApJ...436..754C, Romani:1994sc}, the gamma-ray beaming is highly irregular and a large gamma-ray flux can be observed if the line-of-sight crosses the edges of the outer gaps where the enhancement is maximal. Yet, even between the peaks of the emission, large fluxes can be observed.     
In polar cap models~\cite{Daugherty:1995zy, Dyks:2002nc, Grenier:2006up}, hollow cones close to the poles are produced as a result of cascade radiation. The opening angles are typically very small and 
the observed gamma-ray emission strongly depends on the observer's line-of-sight. Given the small opening angles for a given pulsar, an observer has only a small probability of observing a large gamma-ray flux, and a much greater likelihood of observing only a small fraction of its angle-average emission. In this model, the apparent gamma-ray efficiencies of MSPs will tend to be very high, or very low, depending on the orientation of the beam. Finally, in slot gap models the gamma-ray emission is dominated along caustics that originate close to the polar caps but extend out to the light cylinder \cite{Grenier:2006up, 1983MNRAS.202..495M}. Thus, for an observer seeing many MSPs at different viewing angles, some of the pulsars are going to appear as very efficient gamma-ray emitters, when the line-of-sight crosses along a significant portion of the caustics. Moreover, due to the more extended size of the caustics  (compared to polar cap models) and the fact that the photons are radiated tangentially to the closed 
magnetic field lines, which are curved, many MSPs will also appear as fairly efficient gamma-ray emitters. The total population of MSPs should thus have a wide range of efficiencies in producing gamma-rays.      
 
Given that we are not able to measure the beaming angle relative to the line-of-sight for any given MSP, we can only compare the predictions of these models to that observed from an overall population or ensemble of such sources.  For each MSP observed by Fermi, we calculate the ratio of the total isotropic equivalent gamma-ray luminosity above 0.1 GeV (using the distances quoted in Table~\ref{MSP-table}) to the total spin-down power, $\dot{E}$:
\begin{equation}
f_{\gamma} \equiv \frac{L_{\gamma}}{\dot{E}} =  \frac{4 \pi^{2} I \dot{P}}{P^{3}}.
\label{eq:fgamma}
\end{equation}
For the moment of inertia, $I$, we assume that pulsars have a radius of 10 km and a mass of 1.4$M_{\odot}$. For the period, $P$, and its derivative $\dot{P}$, we take the values reported in the ATNF pulsar catalog~\cite{Manchester:2004bp}.                  

In the upper left frame of Fig.~\ref{fig:Correlations1}, we plot the isotropic equivalent gamma-ray luminosity and spin-down power of our MSP sample. The diagonal lines refer to $f_{\gamma}=$1, 0.1, and 0.01, from top-to-bottom. There are 6 MSPs for which  $f_{\gamma}>1$, indicating a degree of beaming toward the Earth. The appearance of such high values of $f_{\gamma}$ is expected given that in all models the gamma-ray emissivity is predicted to be highly anisotropic. We also note the wide variation in values of $f_{\gamma}$, ranging from $\sim$0.005 to $\sim$5, with many of these objects exhibiting $f_{\gamma} \simeq 0.1$ (see also Ref.~\cite{Calore:2014oga}). As discussed in Sec.~\ref{sec:LumFunc}, the current Fermi catalogue is complete out to
distances of $D \sim 1$ kpc for MSPs with luminosities larger than $10^{32}$ erg/s. Thus the 14 objects 
that are within a kpc from us and for which we can calculate the ratio $f_{\gamma}$ provide a small but representative subset regarding the ratio $f_{\gamma}$ (at least down to $f_{\gamma} \sim 0.01$). Within this subset of 14 MSPs, $f_{\gamma}$ varies between 0.005 and 0.69, with 8 objects having $f_{\gamma}$ between 0.05 and 0.25.\footnote{If we use 2PC distances when available, instead of those reported in the ATNF catalog, we find 17 MSPs within 1 kpc with $f_{\gamma}> 0.005$, of which 8 have $f_{\gamma}$ between 0.05 and 0.25.} This observation appears to favor the slot gap and the outer gap models over polar cap models, which do not predict a large fraction of MSPs with intermediate values of $f_{\gamma} \simeq$ 0.1 (see also, Refs.~\cite{Guillemot:2014kta,Venter:2009eu,Watters:2008bp,Watters:2010jb}). For MSPs more distant than $\sim$1 kpc, there are selection effects that prevent us from detecting MSPs with low values of $f_{\gamma}$. This effect is apparent in the upper right frame of Fig.~\ref{fig:Correlations1}. On the contrary, the observed MSP sample shows no discernable correlation between $f_{\gamma}$ and the spectral index $\alpha$ or $E_{\rm cut}$ (see the lower frames of Fig.~\ref{fig:Correlations1}). 

%Also, as expected, MSPs with high values of $E_{\rm cut}$ tend to exhibit high values of $f_{\gamma}$.

\begin{figure*}
\includegraphics[width=3.40in,angle=0]{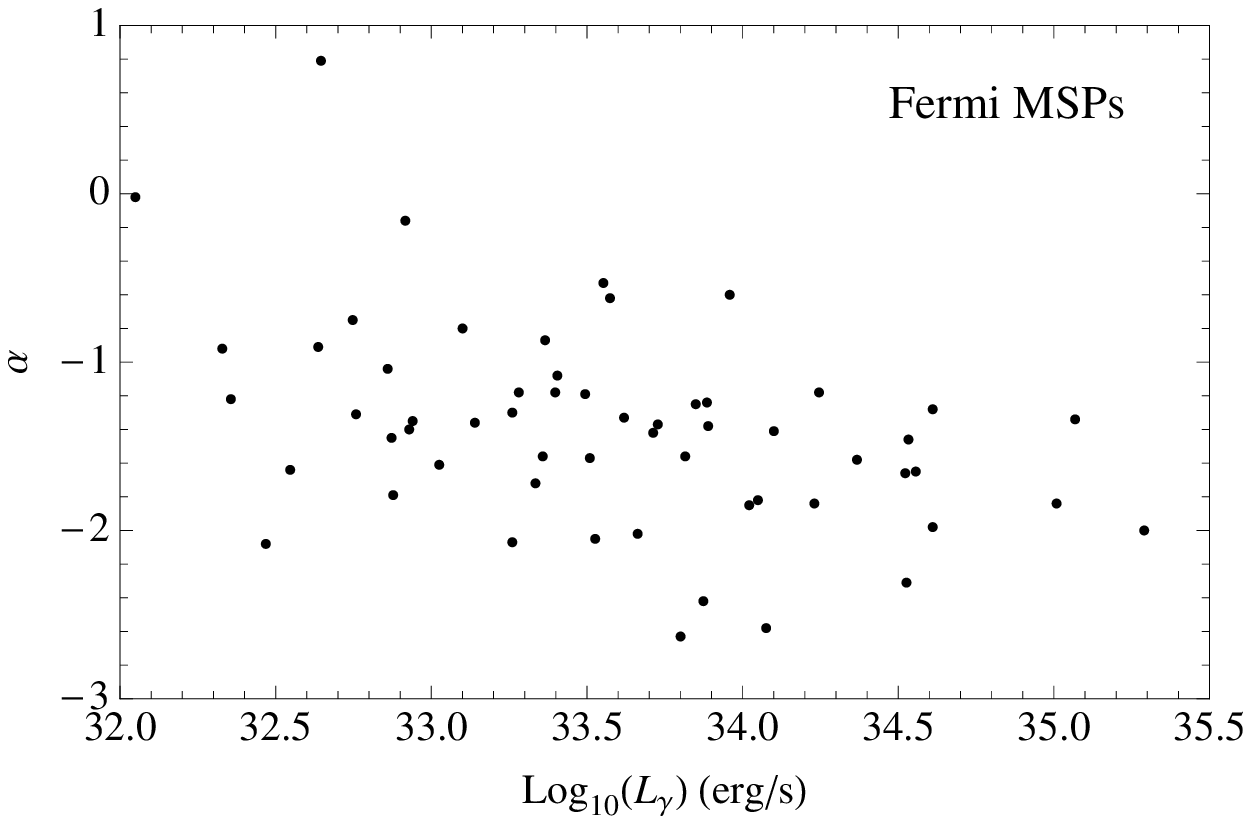}
\includegraphics[width=3.40in,angle=0]{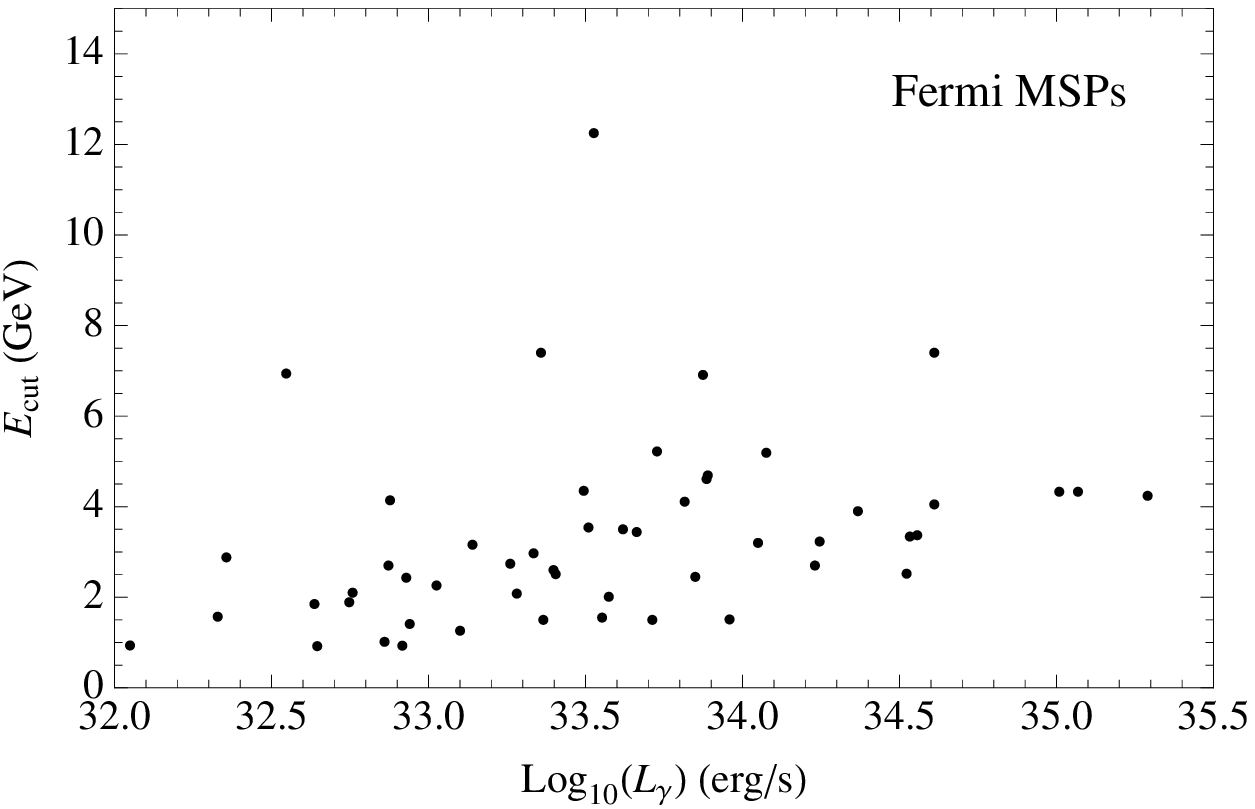} \\
\caption{The spectral power law and cut-off vs the isotropic gamma-ray luminosity in the energy range of 0.1 to 100 GeV.}
\label{fig:Correlations2}
\end{figure*}

In Fig.~\ref{fig:Correlations2}, we plot the fitted values of the spectral parameters ($\alpha$ and $E_{\rm cut}$) versus the gamma-ray luminosity of Fermi's MSPs. The first evident observation is that there is a lack of bright MSPs with hard spectral indices. Such objects are the easiest to discriminate over the diffuse backgrounds out to distances of a few kpc. Thus we find no evidence for a large population of MSPs with $\alpha > -0.5$. Objects with $L_{\gamma} < 10^{33.5}$ erg/s and $\alpha < -2.2$ also appear to be absent from the observed catalog.  

\begin{figure*}
\includegraphics[width=3.40in,angle=0]{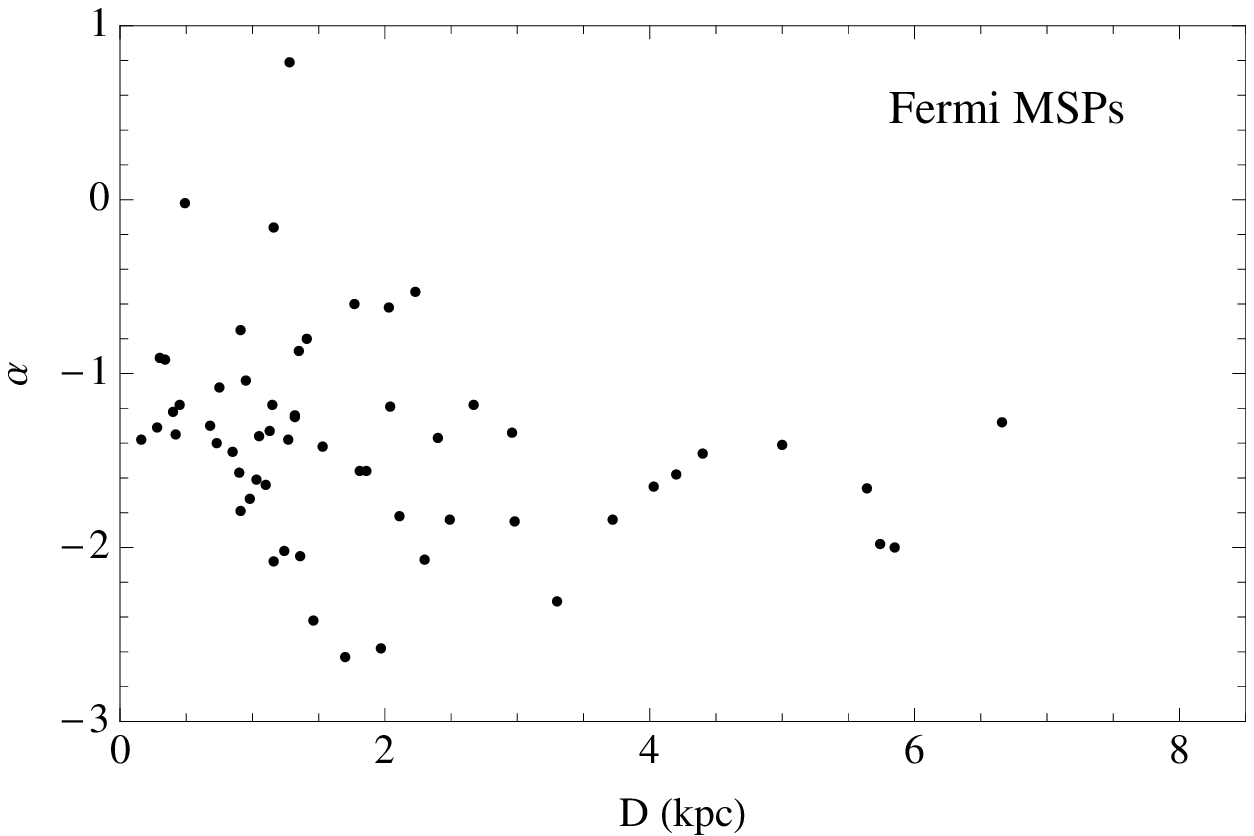} 
\includegraphics[width=3.40in,angle=0]{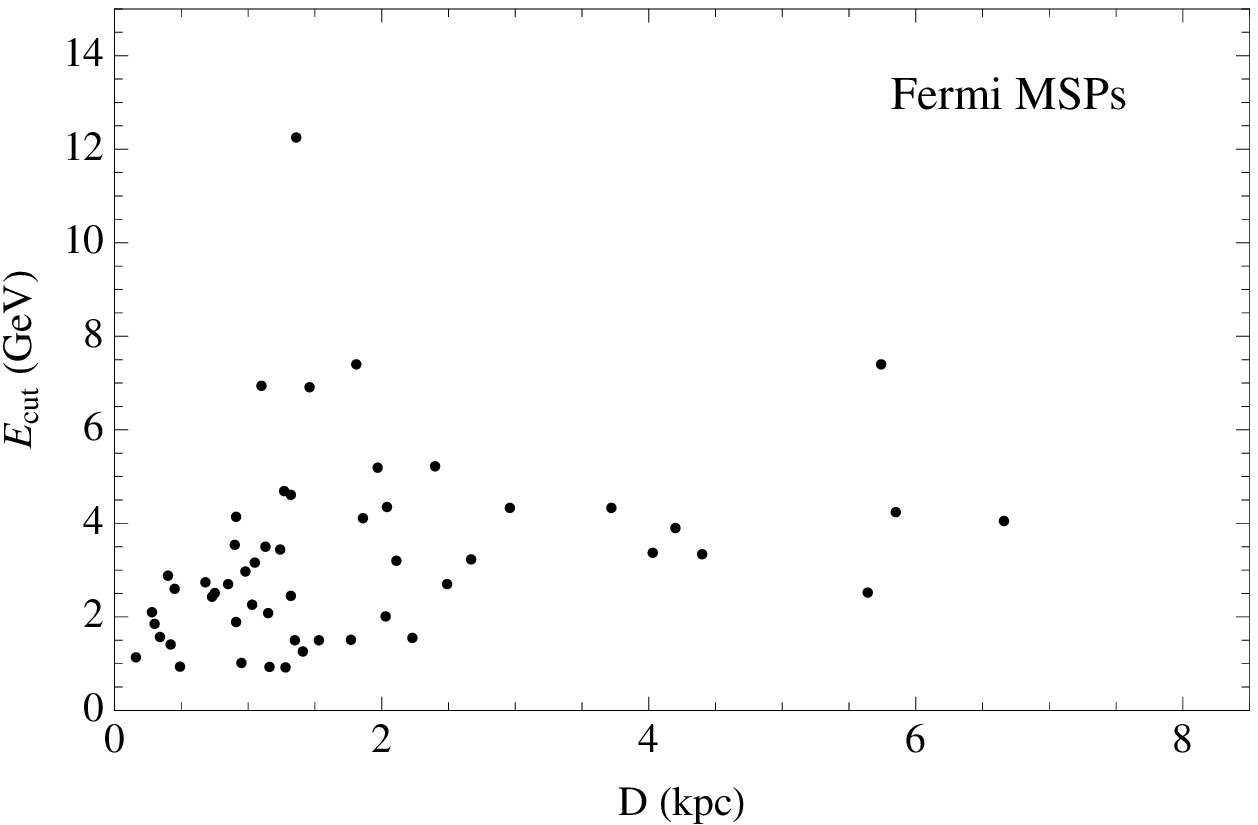} \\
\includegraphics[width=3.40in,angle=0]{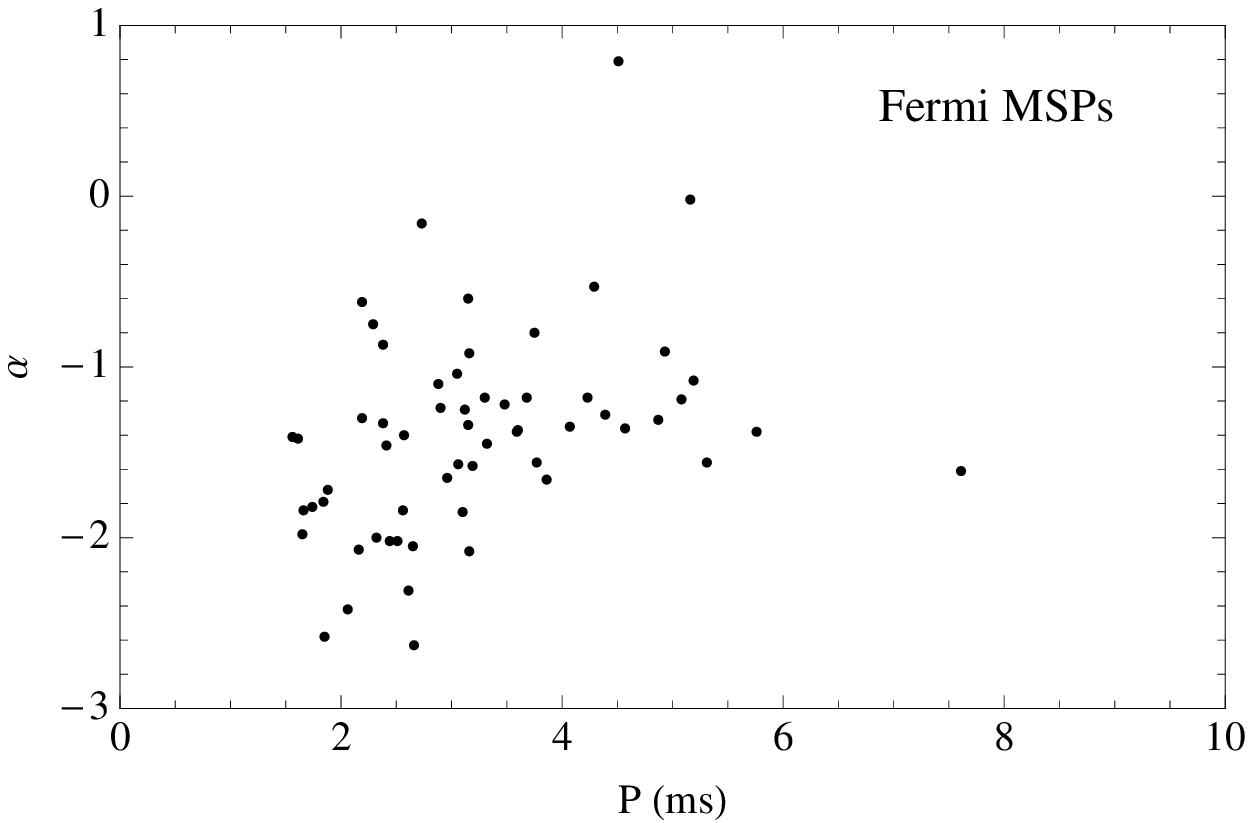} 
\includegraphics[width=3.40in,angle=0]{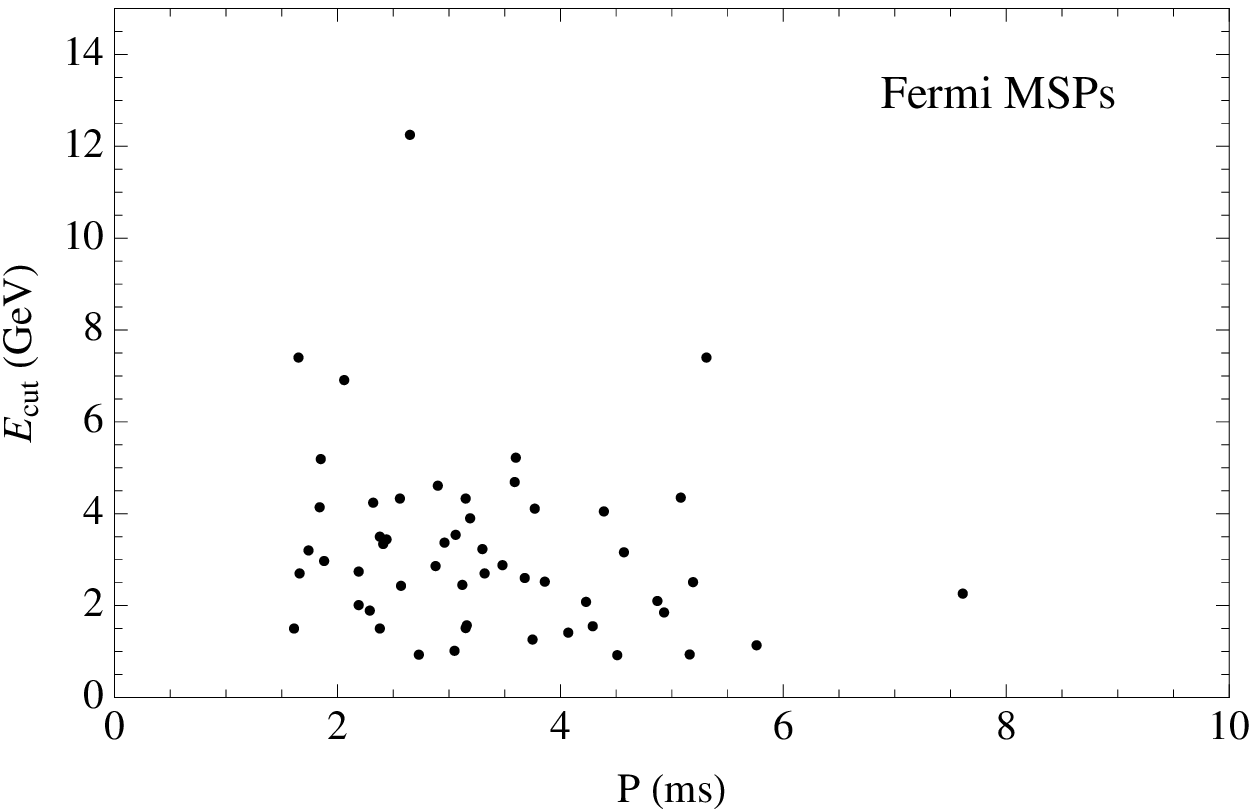} \\
\includegraphics[width=3.40in,angle=0]{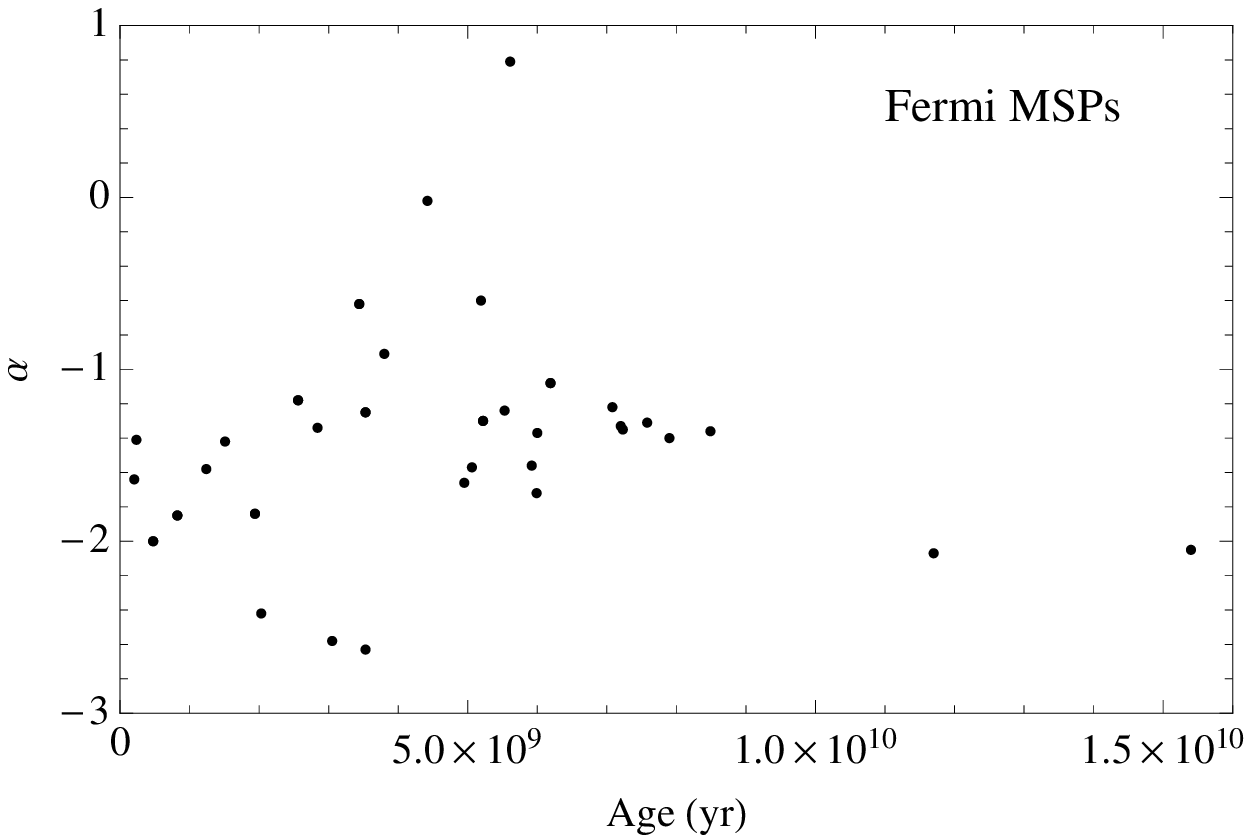} 
\includegraphics[width=3.40in,angle=0]{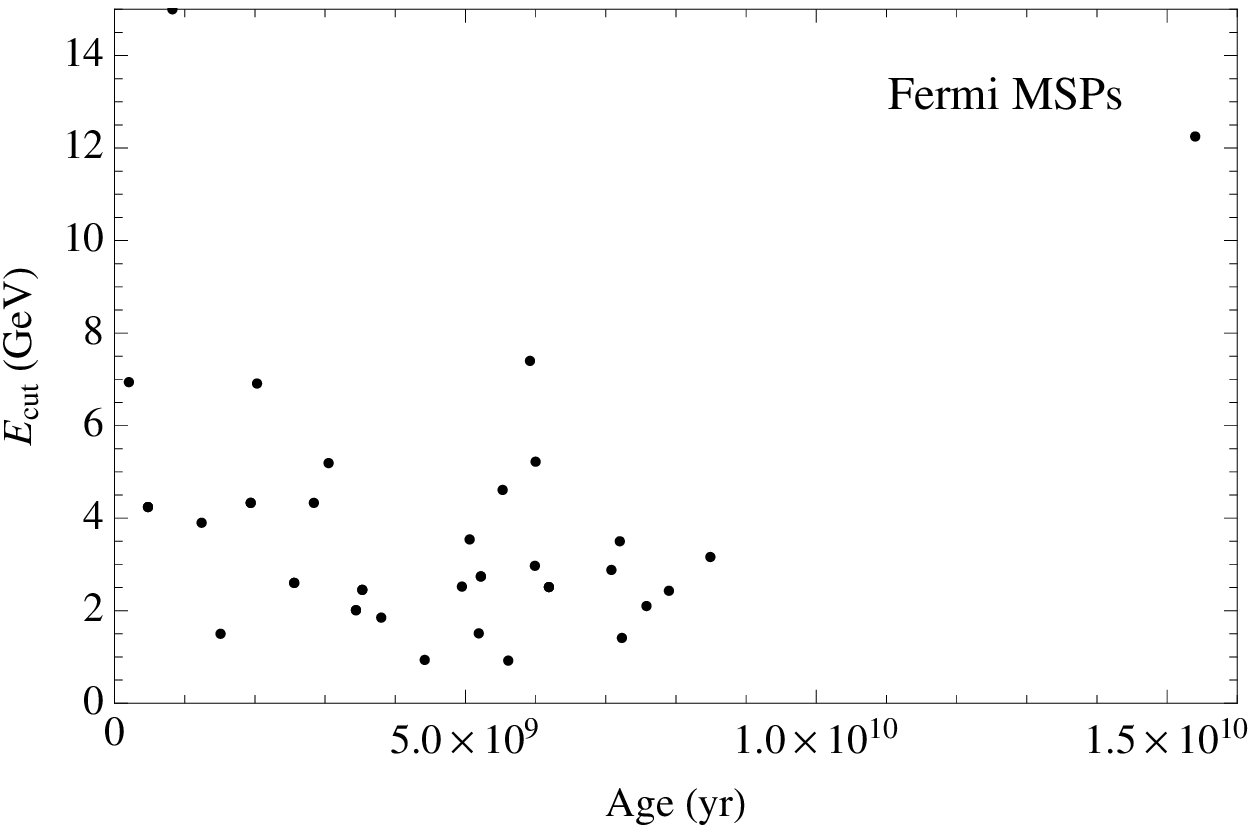}\\
\includegraphics[width=3.40in,angle=0]{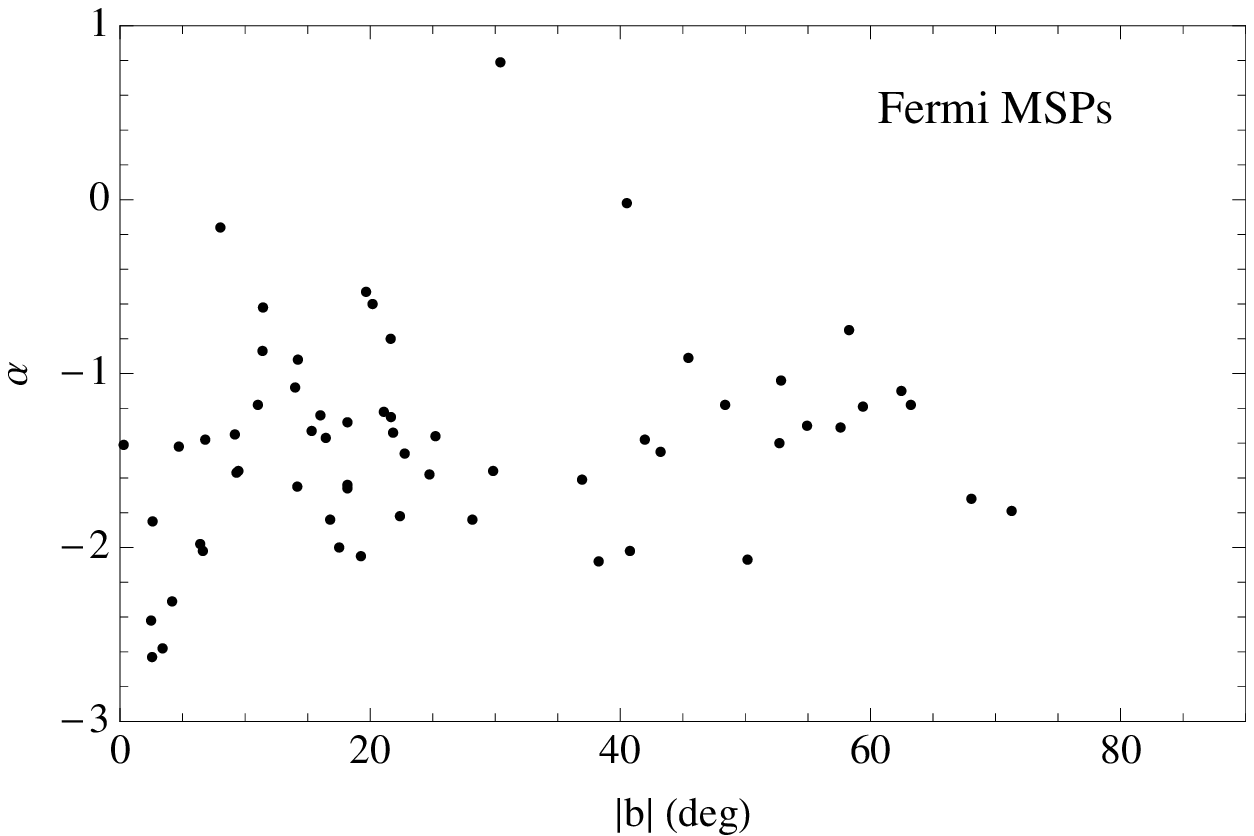} 
\includegraphics[width=3.40in,angle=0]{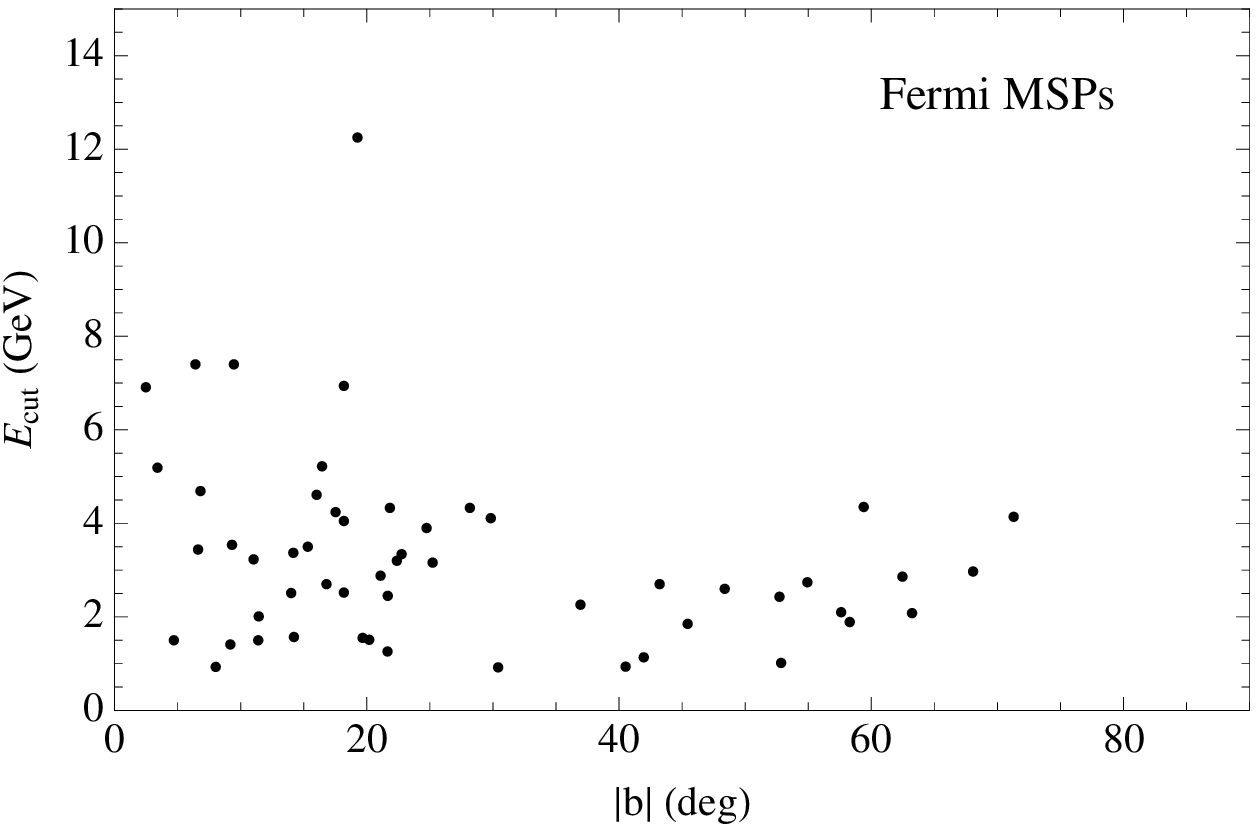} \\
\caption{The spectral power law and cut-off vs the distance, period, characteristic age ($\tau=P/2\dot{P}$), and Galactic Latitude of our MSP sample.}
\label{fig:Correlations3}
\end{figure*}

In Fig.~\ref{fig:Correlations3}, we show the correlations between the fitted spectral parameters ($\alpha$ and $E_{\rm cut}$) and distance, period, characteristic age ($\tau=P/2\dot{P}$), and Galactic Latitude.

%. Finally, we show in Fig.~\ref{fig:Correlations4} the fitted spectral parameters versus Galactic Latitude. 

\section{Discussion and Summary}
\label{sec:Conclusions}

The Fermi Gamma-Ray Space Telescope has provided us with a wealth of new information regarding pulsars. In this study, we have used the current Fermi data set to determine the gamma-ray spectra and gamma-ray luminosity function of pulsars with millisecond-scale periods.  Our primary findings can be summarized as follows:
\begin{itemize}
\item{We have presented the gamma-ray spectra of 61 millisecond pulsars. Although some source-to-source variation is observed, most peak at energies near $\sim$1-2 GeV. The stacked spectrum of this population is well described by $dN/dE \propto E^{-1.57} \exp(-E/3.78\,{\rm GeV})$.}
\item{Although the majority of the 16 high-significance globular clusters studied in our analysis exhibit spectral shapes that are similar to those from millisecond pulsars, several produce significantly softer spectra. This could be indicative of an additional gamma-ray source population present in these systems. The stacked spectrum of the 36 globular clusters studied here is well described by $dN/dE \propto E^{-2.05} \exp(-E/5.76\,{\rm GeV})$.}
\item{We have determined the gamma-ray luminosity function of millisecond pulsars in the field of the Milky Way, and in the globular cluster 47 Tucanae. These two independent and complementary determinations indicate that the luminosity function of the millisecond pulsar population peaks (in $L dN/dL$ units) near $L\sim 10^{32}-10^{34}$ erg/s (integrated above 0.1 GeV), and extends to above $10^{35}$ erg/s. Most of total gamma-ray emission from this population comes from a relatively small number of bright sources.}
\item{Multi-wavelength observations of the millisecond pulsar population suggest that their gamma-ray emission is generally less strongly beamed than the corresponding emission at X-ray wavelengths.}
\item{The distribution of the apparent gamma-ray efficiencies of the millisecond pulsars observed by Fermi provides support in favor of slot gap or outer gap models, relative to polar cap scenarios.}
\end{itemize}

\vskip 0.2 in
\section*{Acknowledgments}  
We would like to thank Craig Heinke and Alex Drlica-Wagner for valuable discussions. This work has been supported by the US Department of Energy. TL is supported by the National Aeronautics and Space Administration through Einstein Postdoctoral Fellowship Award Number PF3-140110.
\vskip 0.05in

\appendix

\section{Gamma-Ray Spectra of Millisecond and Young Pulsars}
\label{app:MSPspectra}

In this appendix, we show the gamma-ray spectra of the millisecond and young pulsars considered in our study. For each case we fit the gamma-ray data with a power-law and an exponential cut-off (Eq.~\ref{eq:MSP_Spect}). The spectral properties of all MSPs are given in Table~\ref{MSP-table}. For most of the young, gamma-ray bright pulsars, we find that a sub-exponential parameterization is clearly favored over the standard exponential form.

\begin{figure*}
\includegraphics[width=3.40in,angle=0]{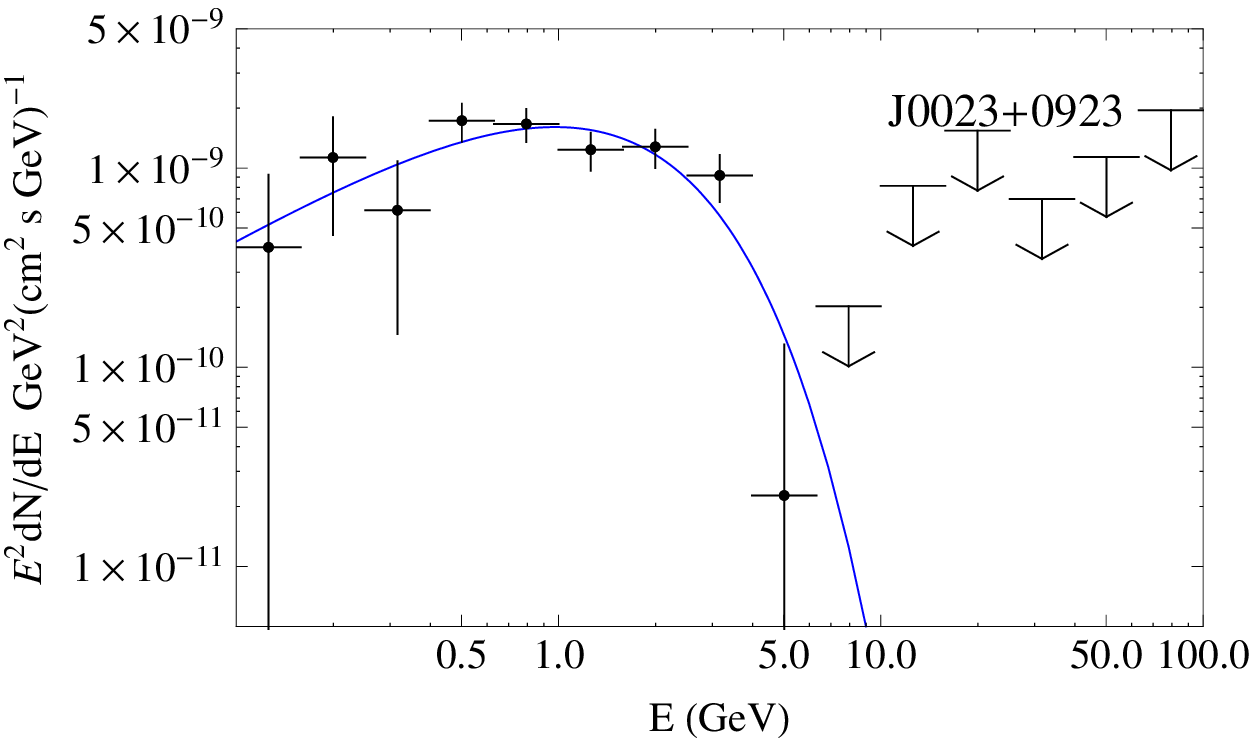} 
\includegraphics[width=3.40in,angle=0]{plots/MSP/J0030p0451.eps} \\
\includegraphics[width=3.40in,angle=0]{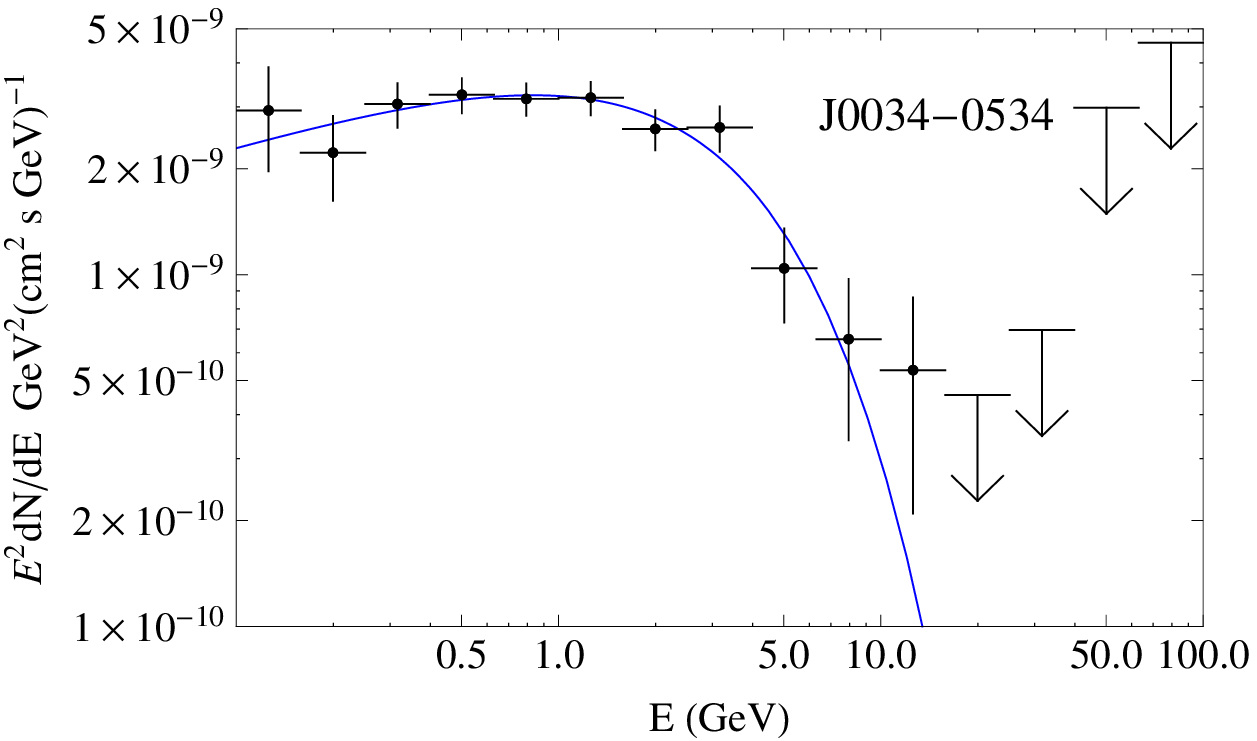} 
\includegraphics[width=3.40in,angle=0]{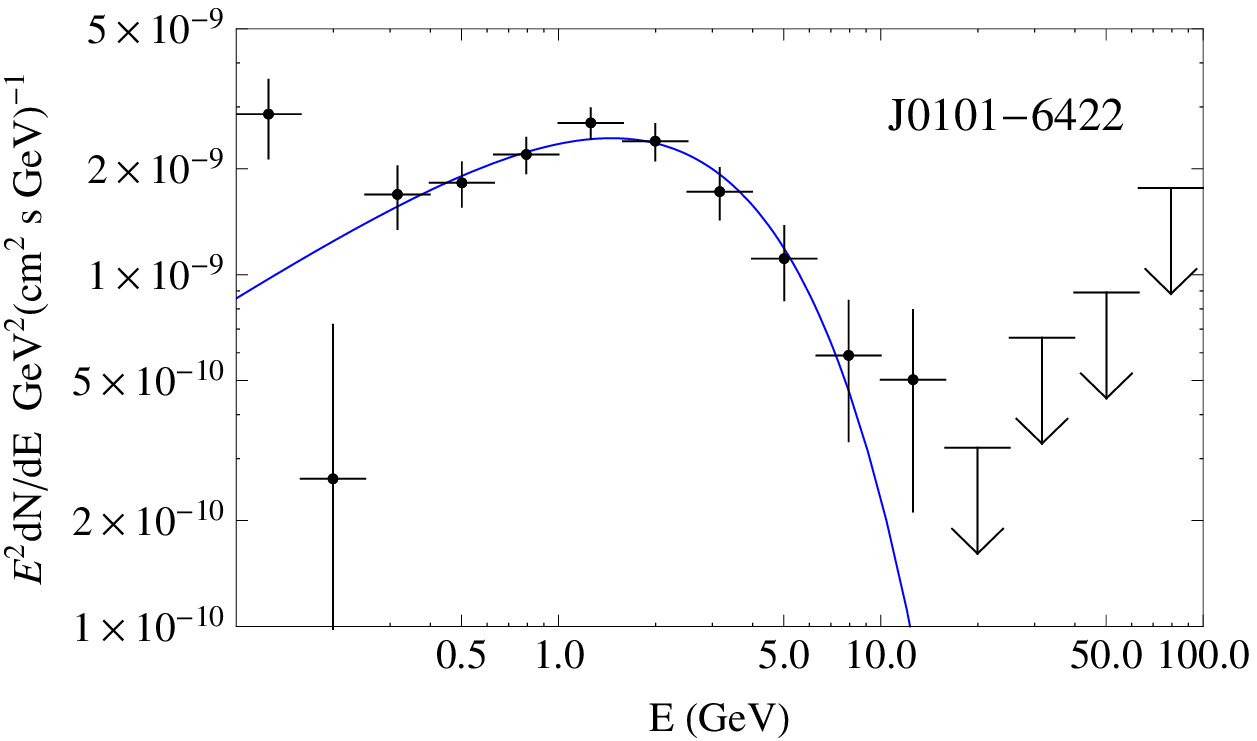} \\
\includegraphics[width=3.40in,angle=0]{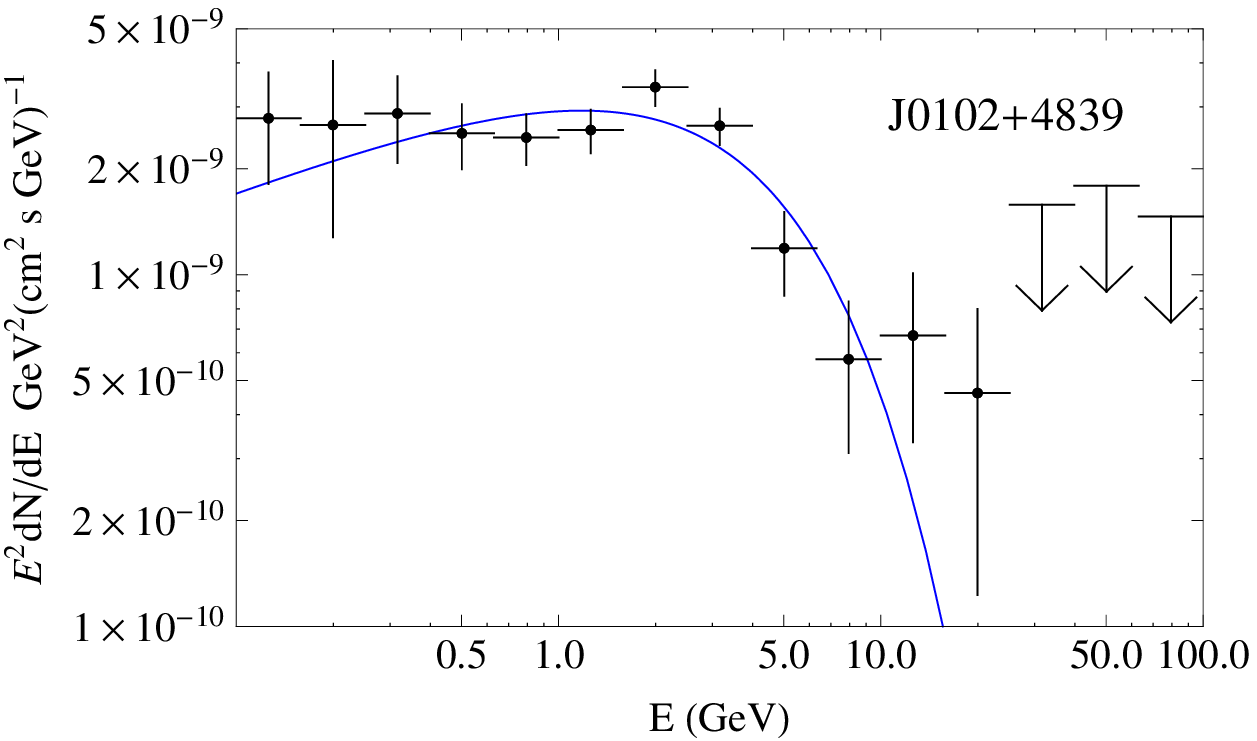} 
\includegraphics[width=3.40in,angle=0]{plots/MSP/J0218p4232.eps} \\
\includegraphics[width=3.40in,angle=0]{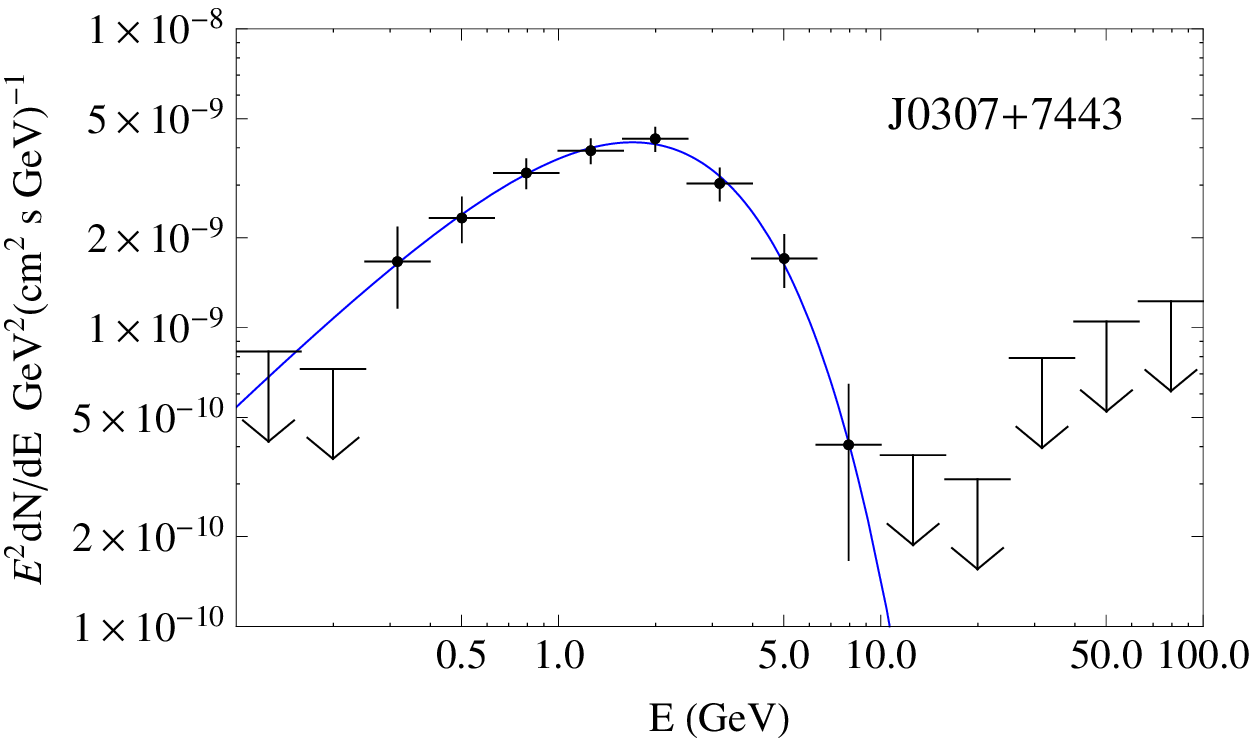} 
\includegraphics[width=3.40in,angle=0]{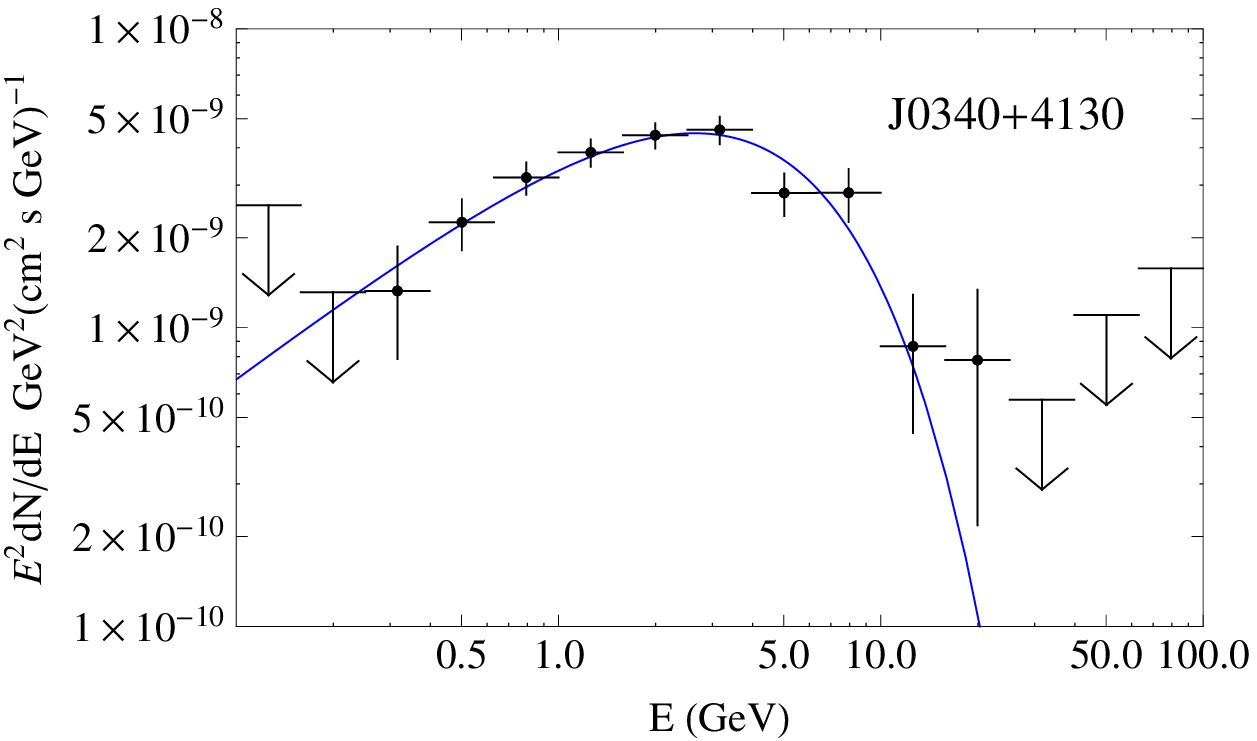}
\caption{The gamma-ray spectra of MSPs.}
\label{fig:MSPs1}
\end{figure*}

\begin{figure*}
\includegraphics[width=3.40in,angle=0]{plots/MSP/J0437m4715.eps} 
\includegraphics[width=3.40in,angle=0]{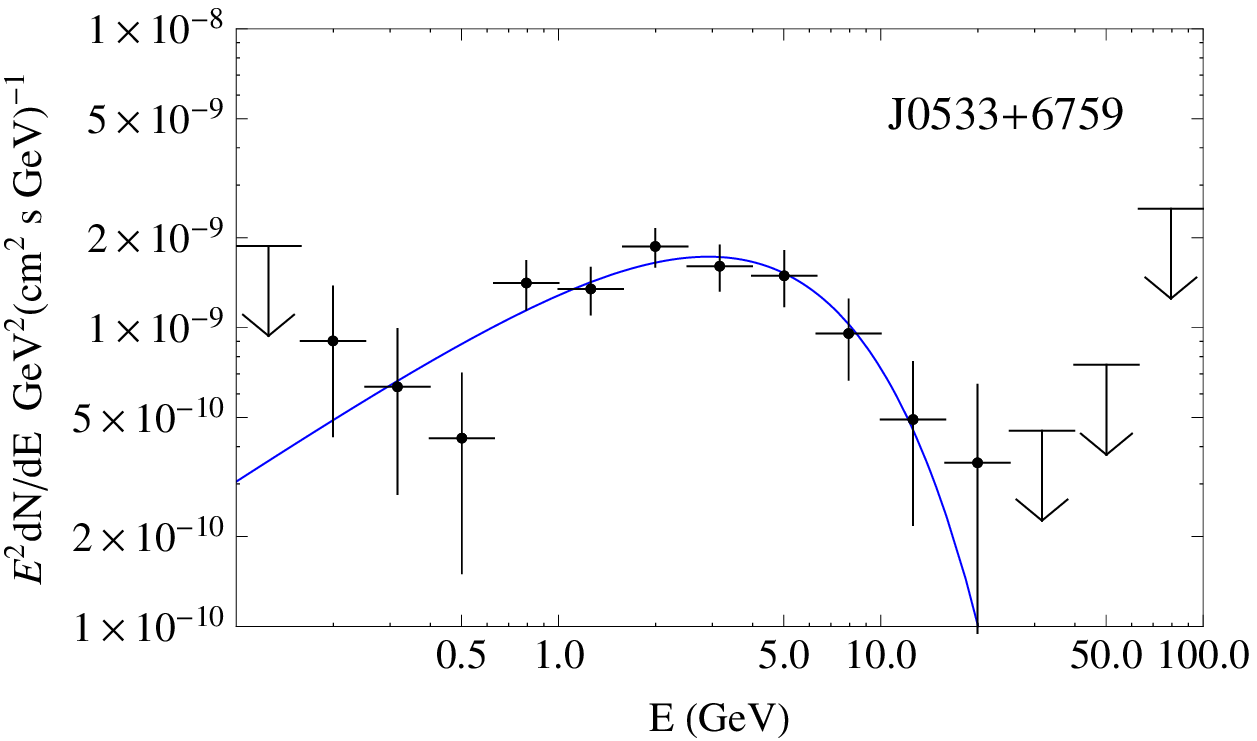} \\
\includegraphics[width=3.40in,angle=0]{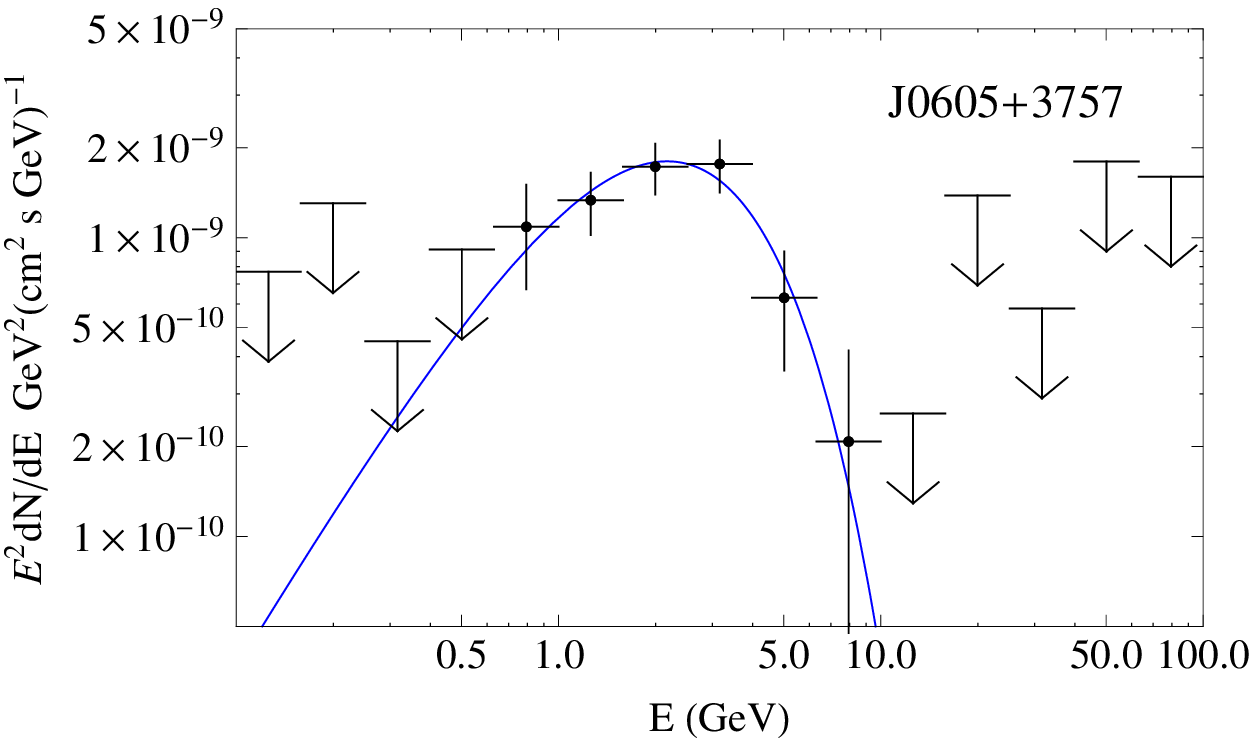} 
\includegraphics[width=3.40in,angle=0]{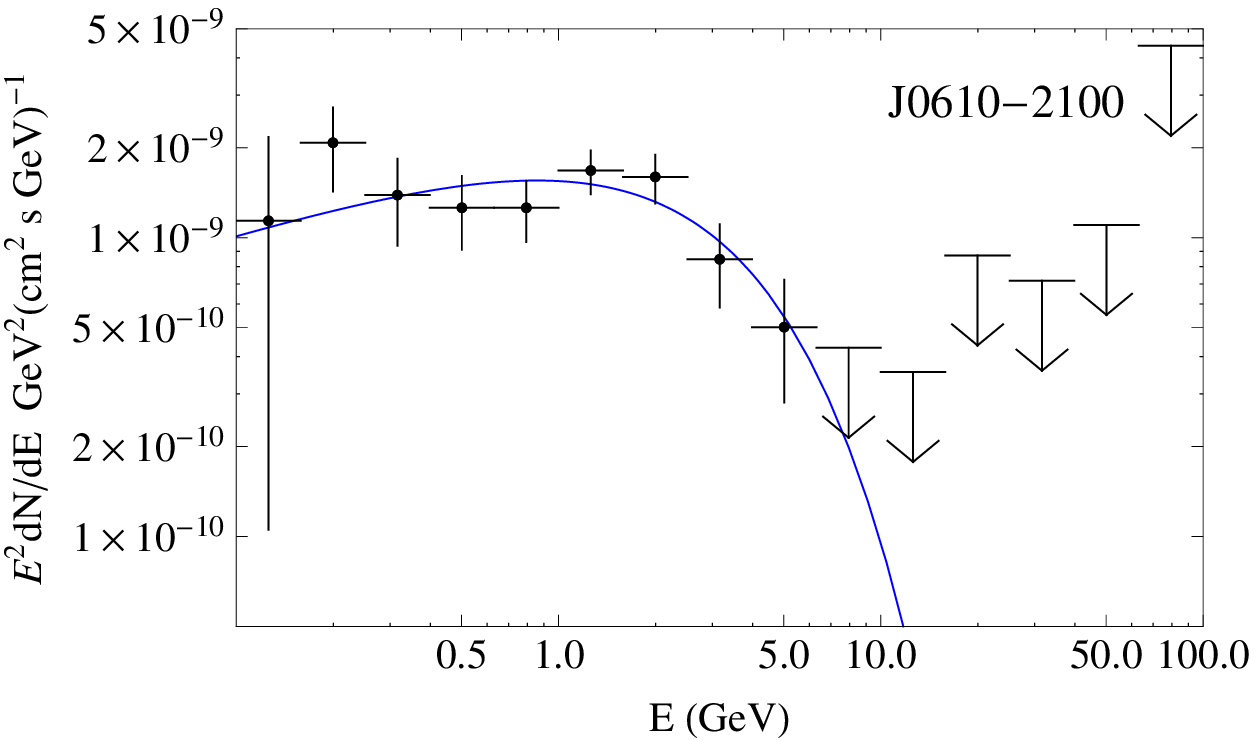} \\
\includegraphics[width=3.40in,angle=0]{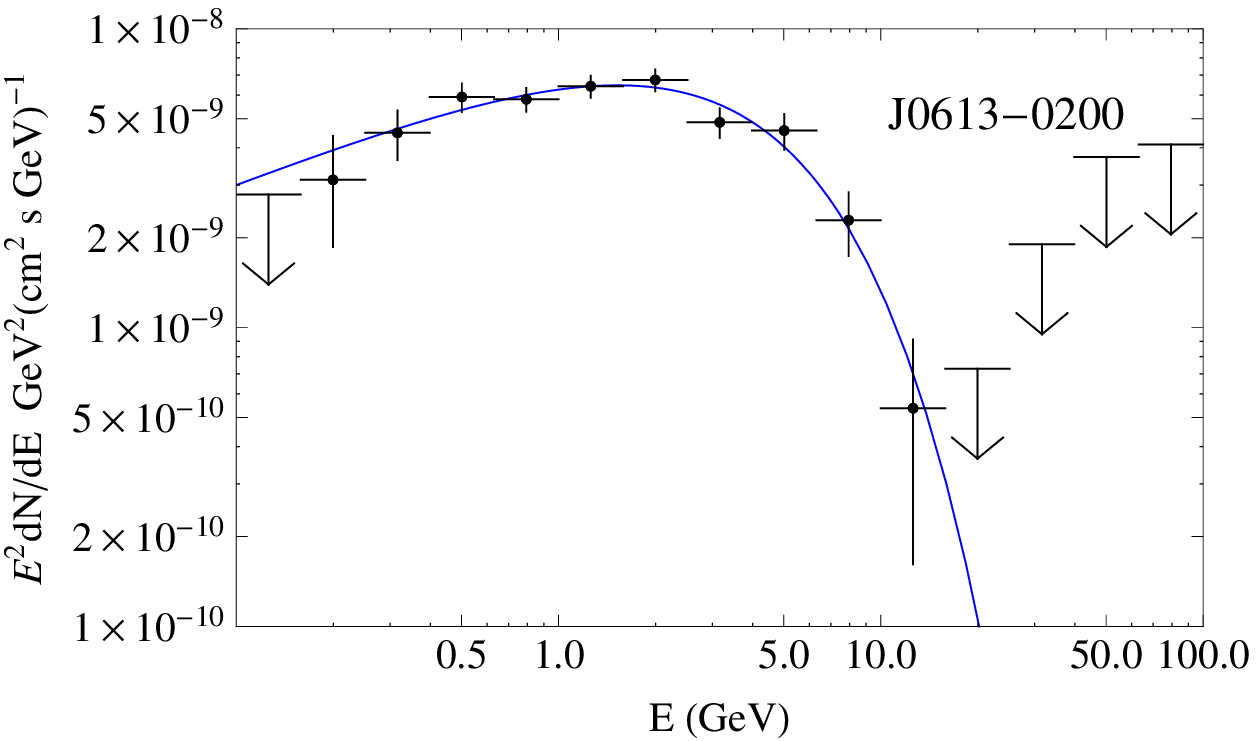} 
\includegraphics[width=3.40in,angle=0]{plots/MSP/J0614m3329.eps} \\
\includegraphics[width=3.40in,angle=0]{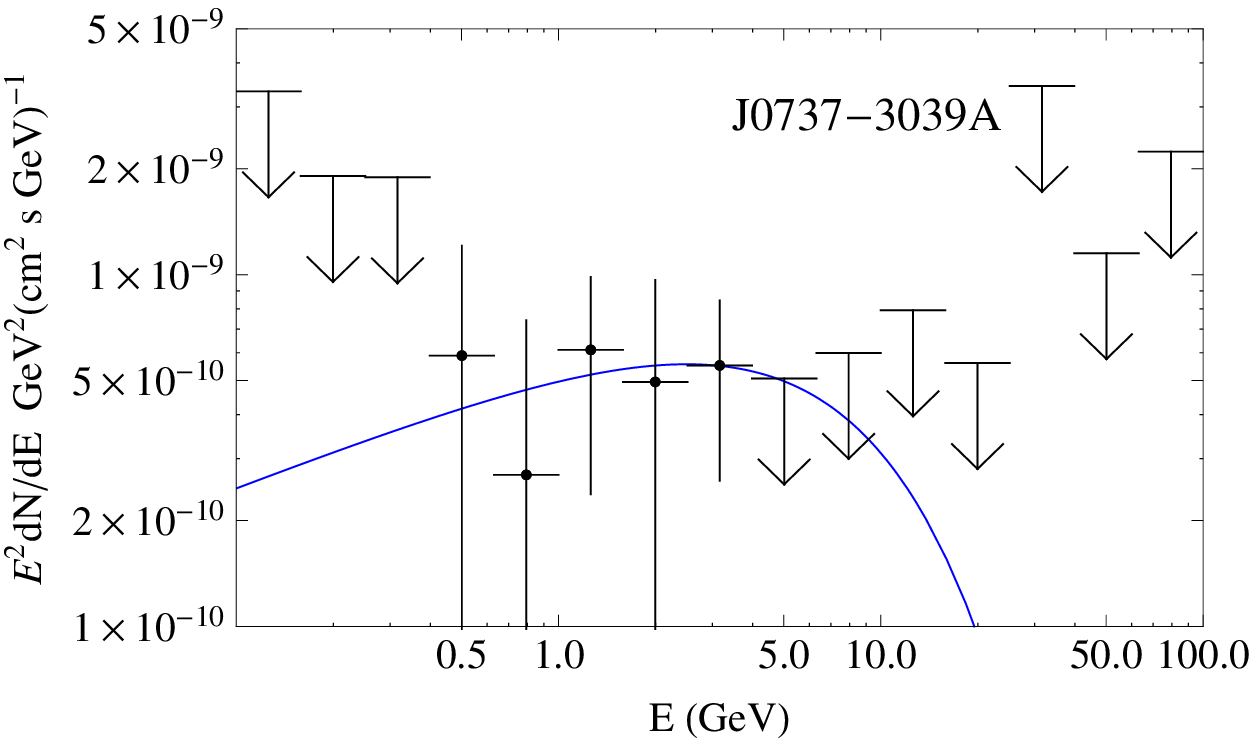} 
\includegraphics[width=3.40in,angle=0]{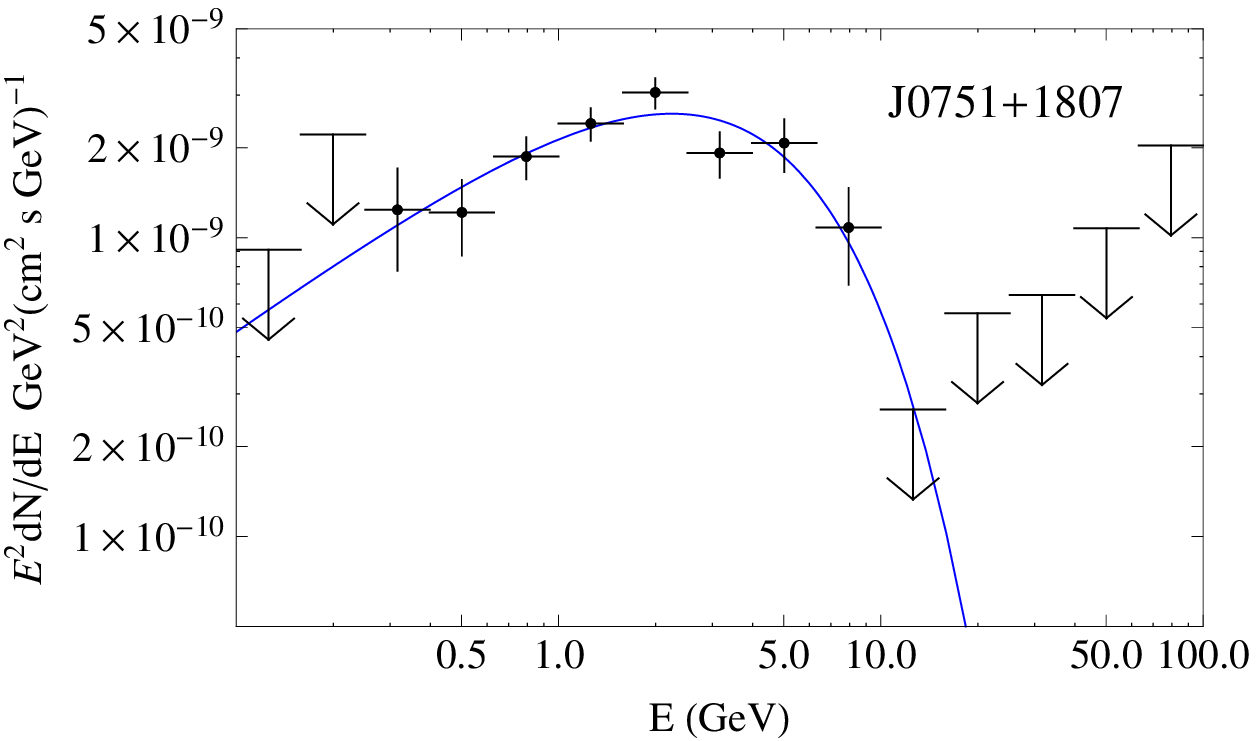} 
\caption{The gamma-ray spectra of MSPs (continued).}
\label{fig:MSPs2}
\end{figure*}

\begin{figure*}
\includegraphics[width=3.40in,angle=0]{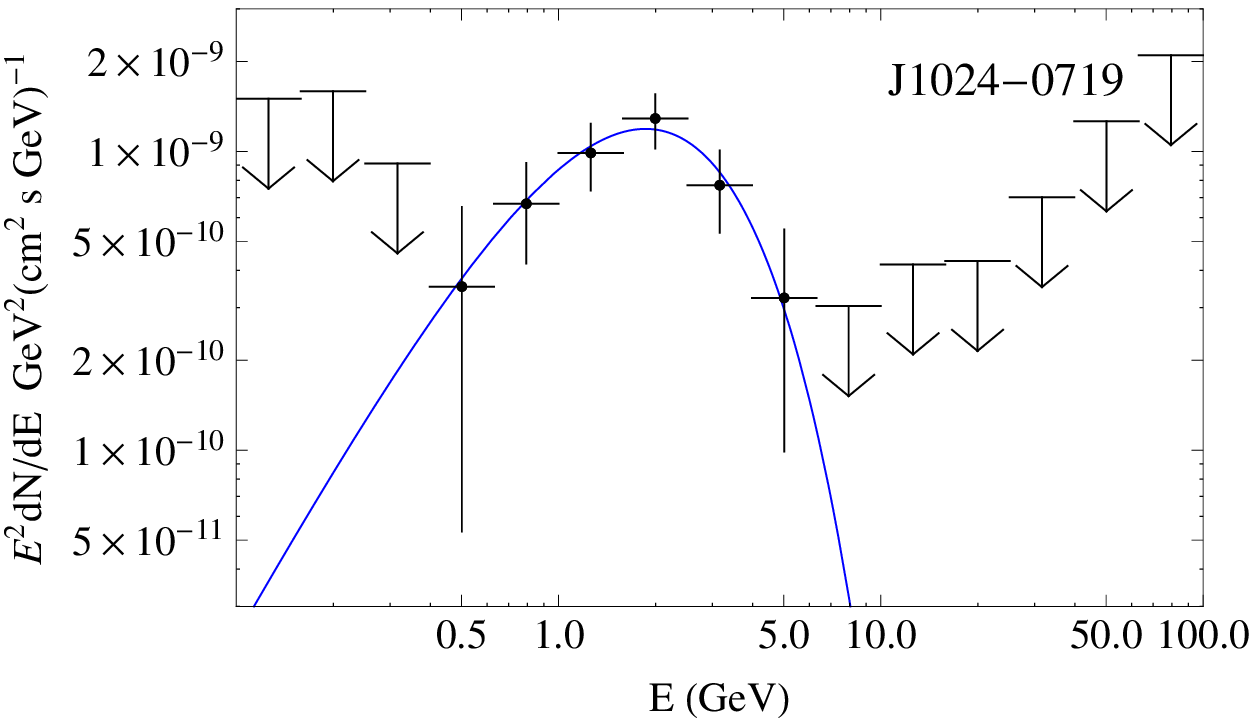} 
\includegraphics[width=3.40in,angle=0]{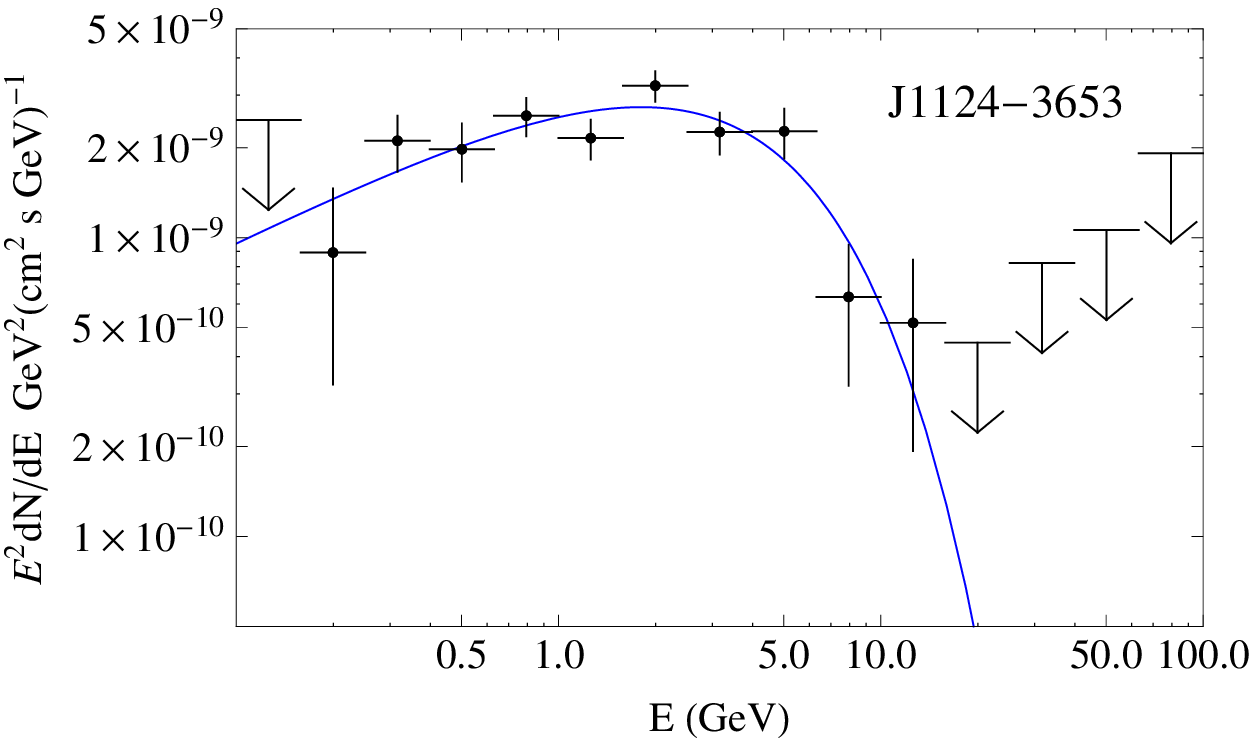} \\
\includegraphics[width=3.40in,angle=0]{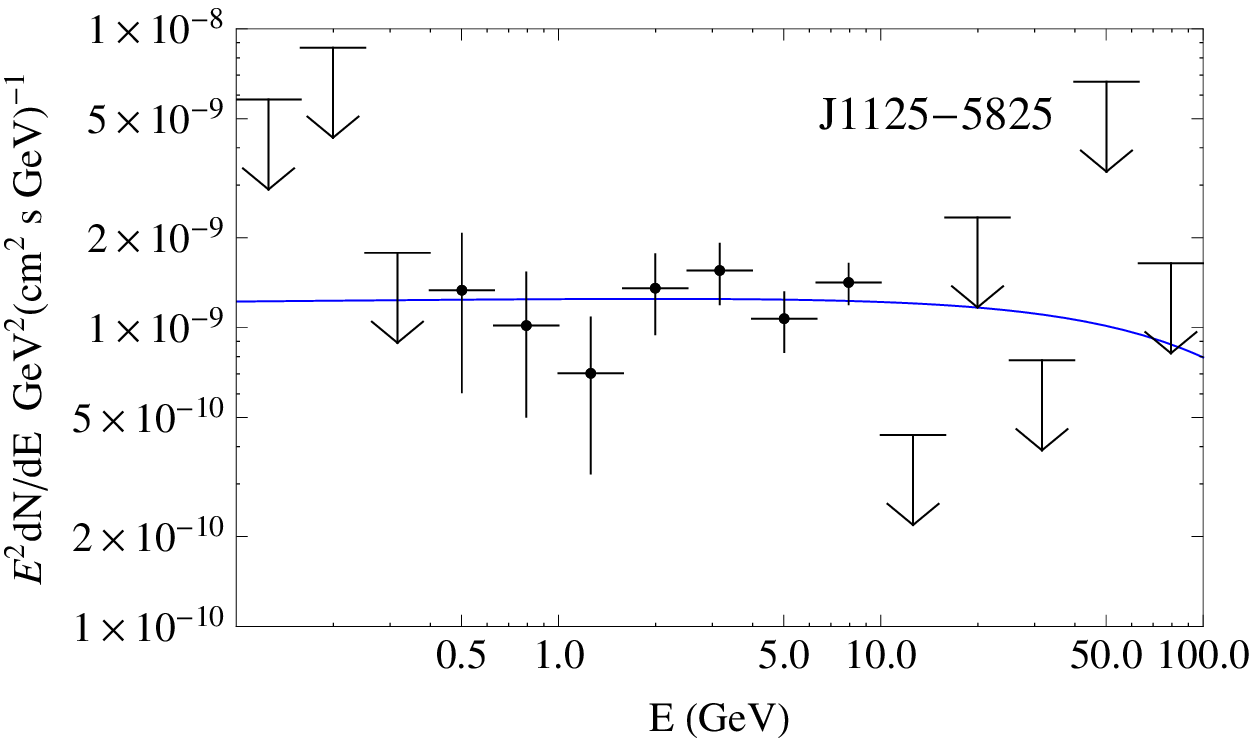} 
\includegraphics[width=3.40in,angle=0]{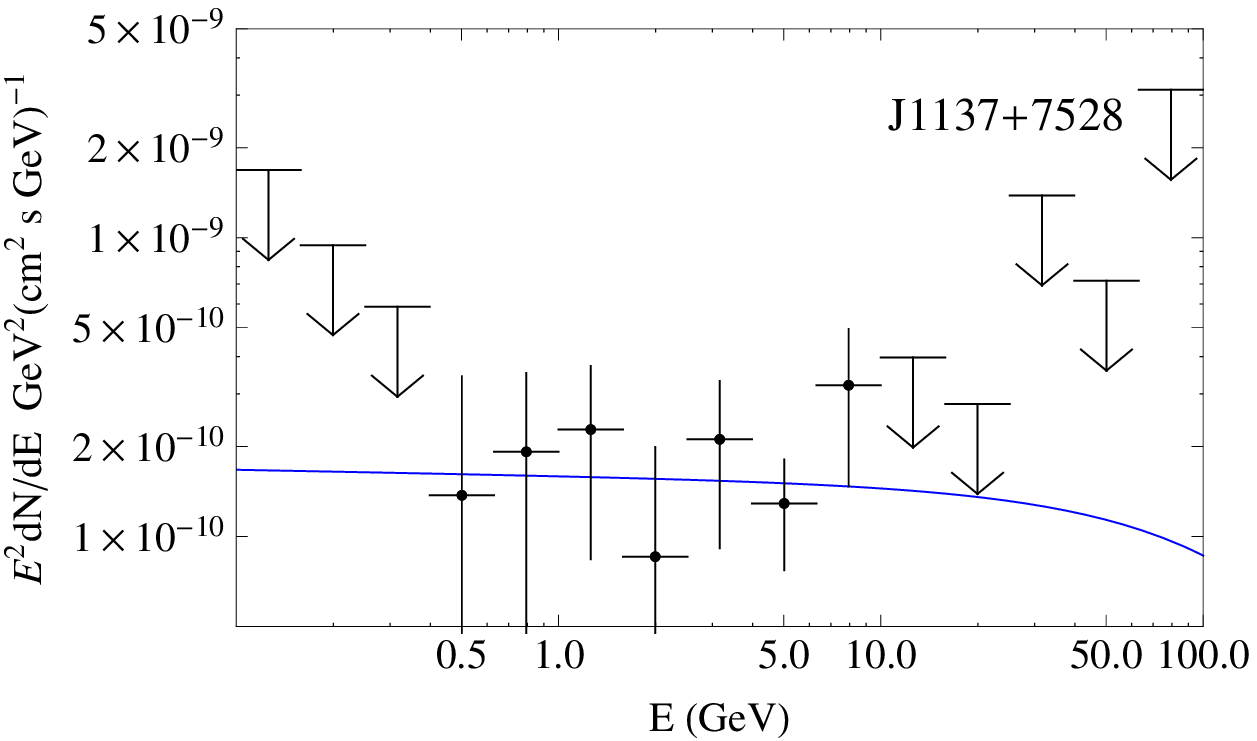} \\
\includegraphics[width=3.40in,angle=0]{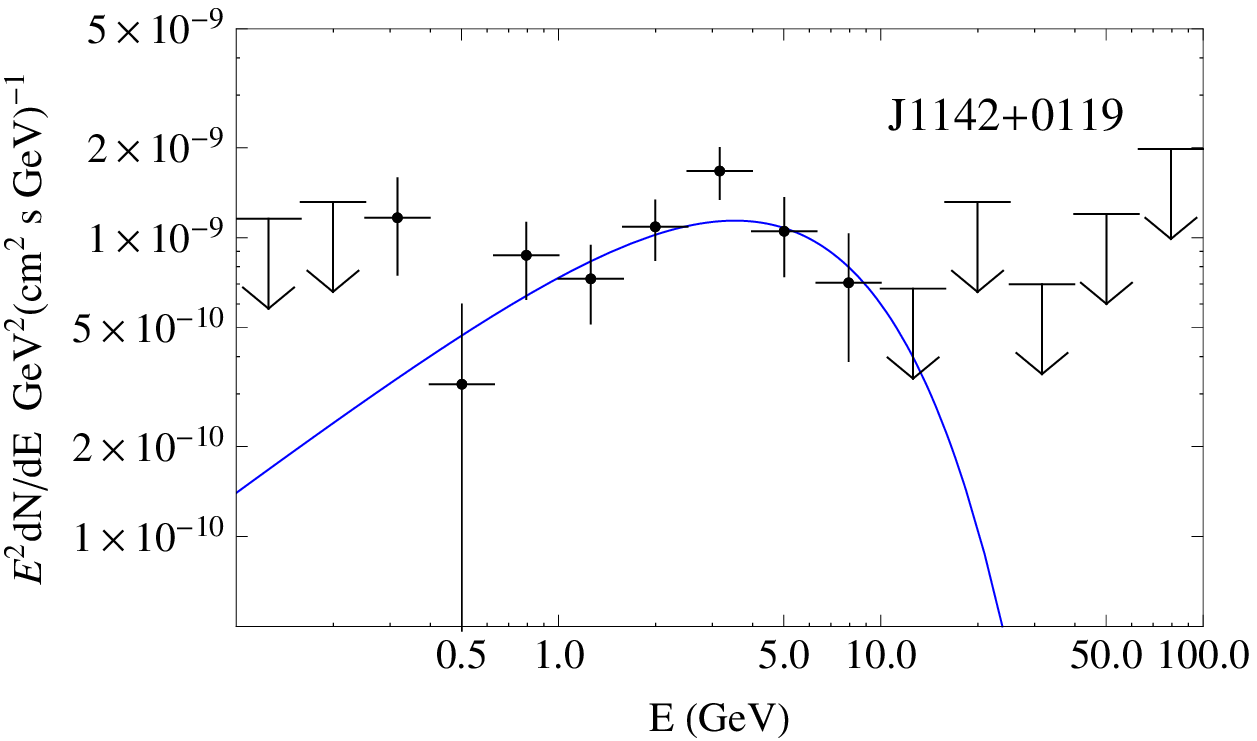} 
\includegraphics[width=3.40in,angle=0]{plots/MSP/J1231m1411.eps} \\
\includegraphics[width=3.40in,angle=0]{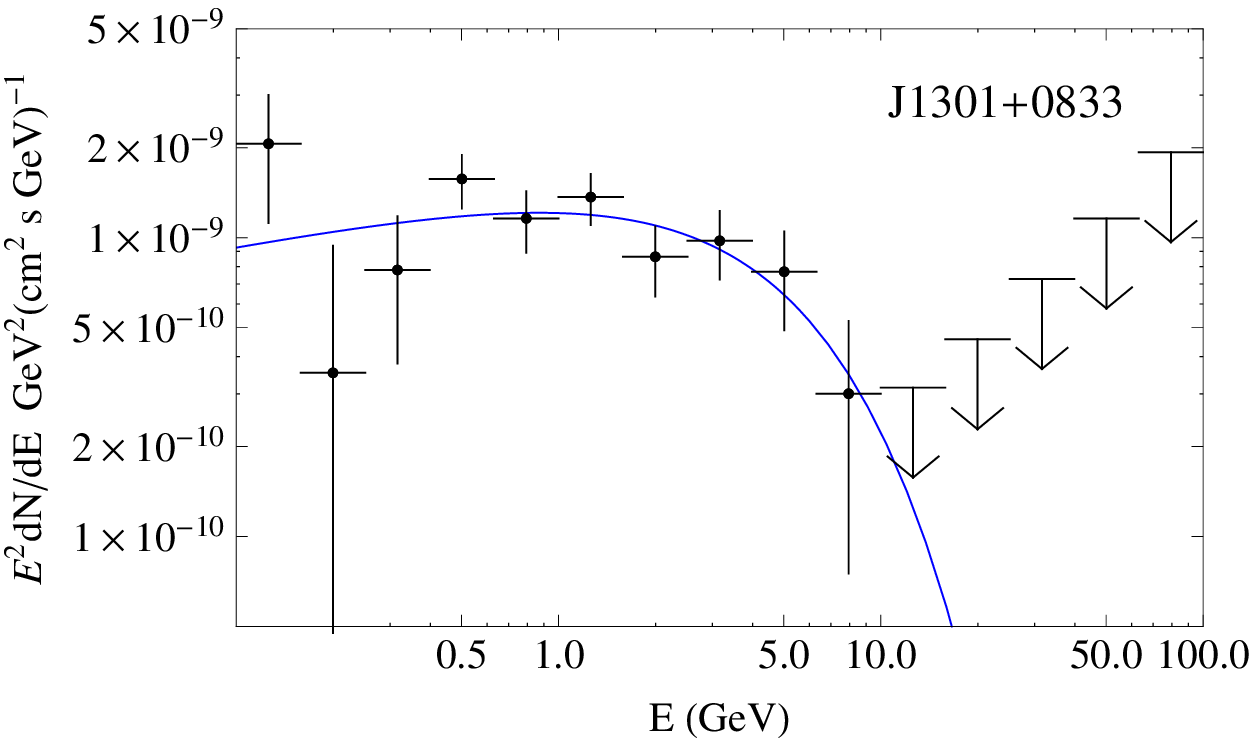} 
\includegraphics[width=3.40in,angle=0]{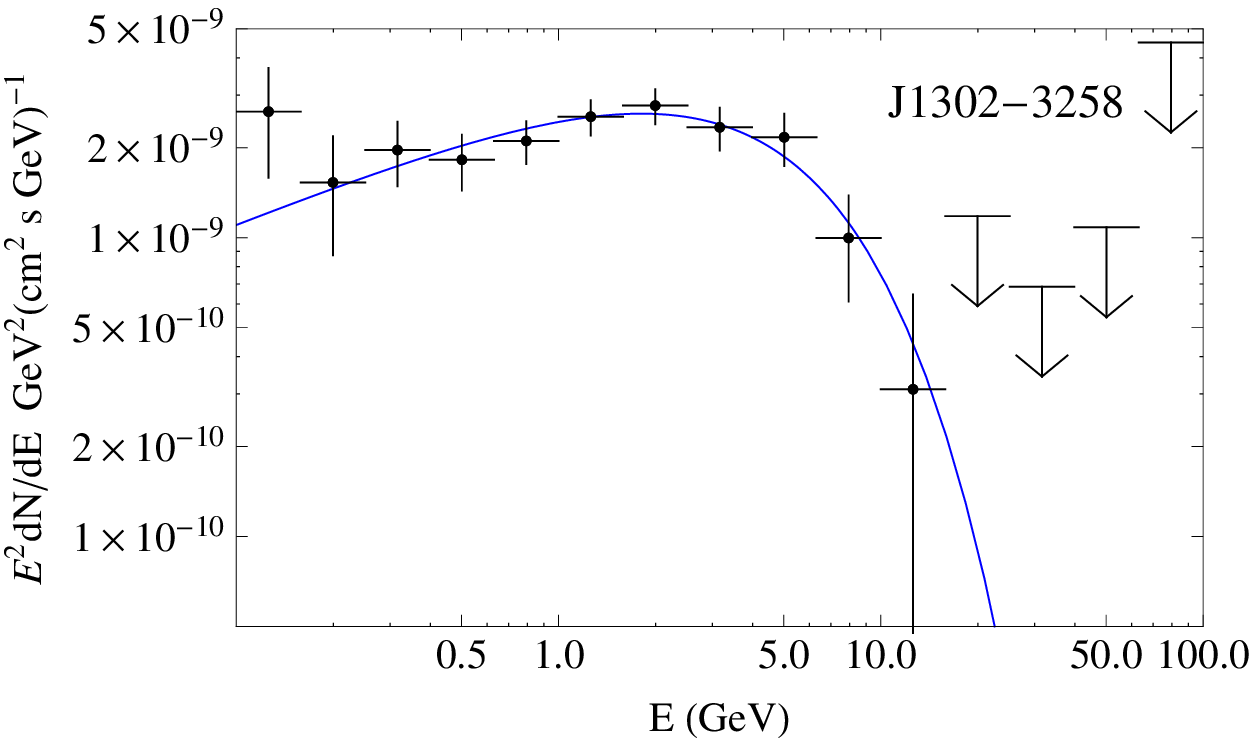} 
\caption{The gamma-ray spectra of MSPs (continued).}
\label{fig:MSPs3}
\end{figure*}

\begin{figure*}
\includegraphics[width=3.40in,angle=0]{plots/MSP/J1311m3430.eps} 
\includegraphics[width=3.40in,angle=0]{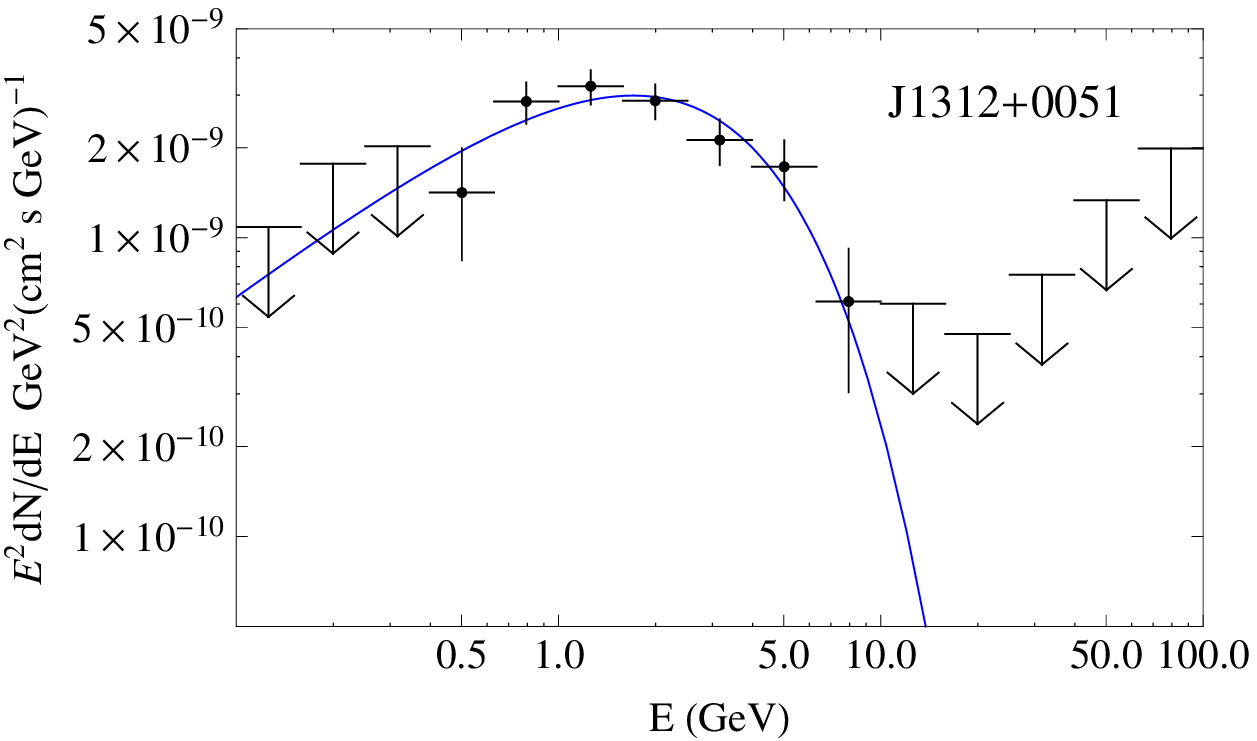} \\
\includegraphics[width=3.40in,angle=0]{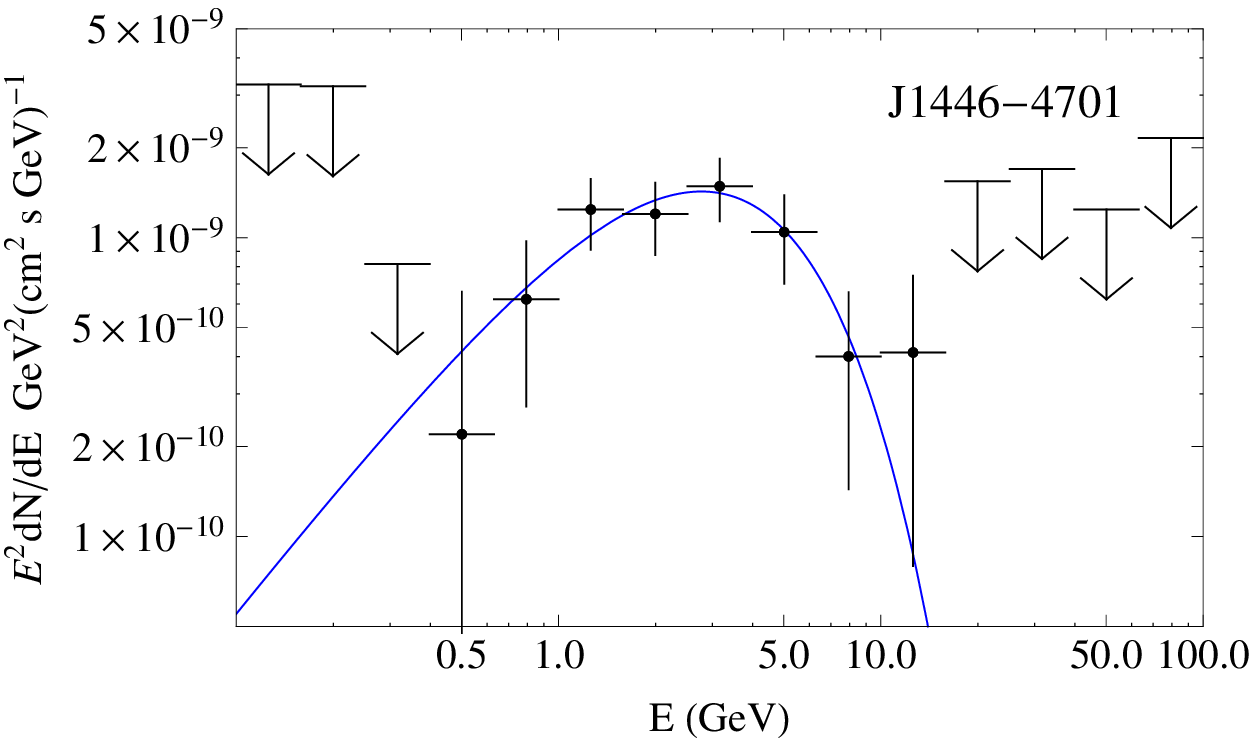} 
\includegraphics[width=3.40in,angle=0]{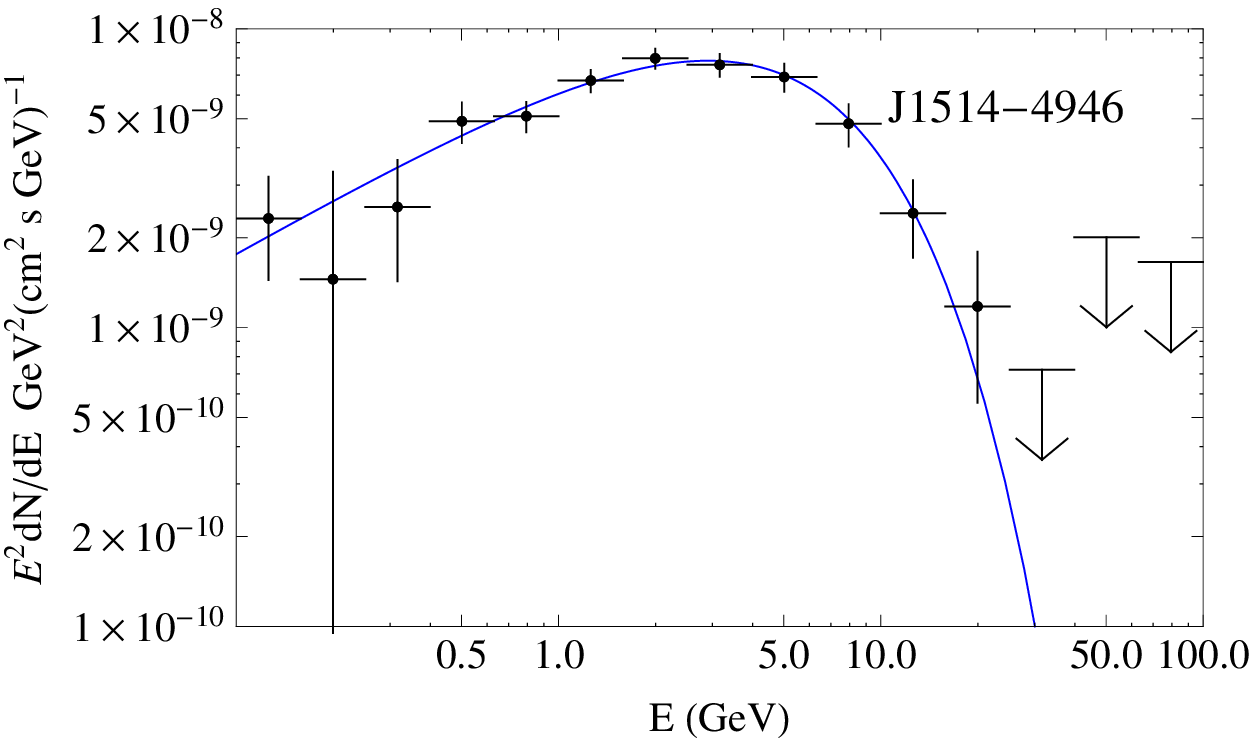} \\
\includegraphics[width=3.40in,angle=0]{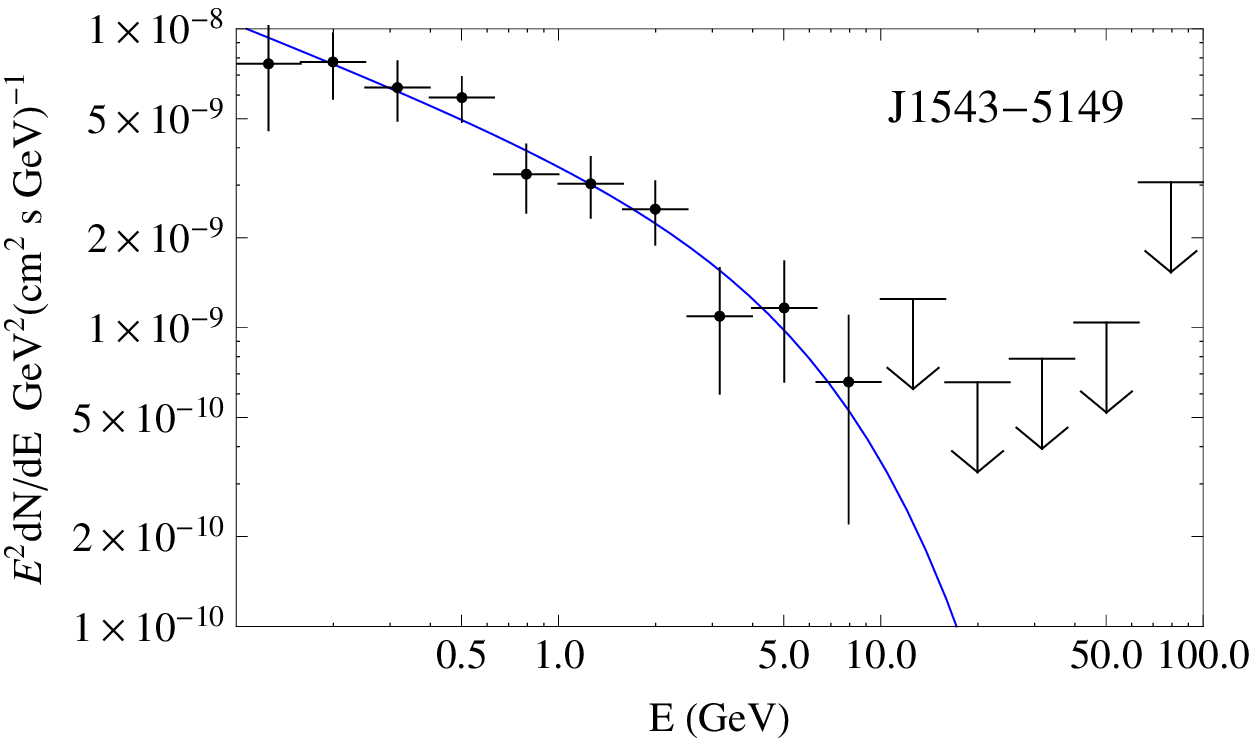} 
\includegraphics[width=3.40in,angle=0]{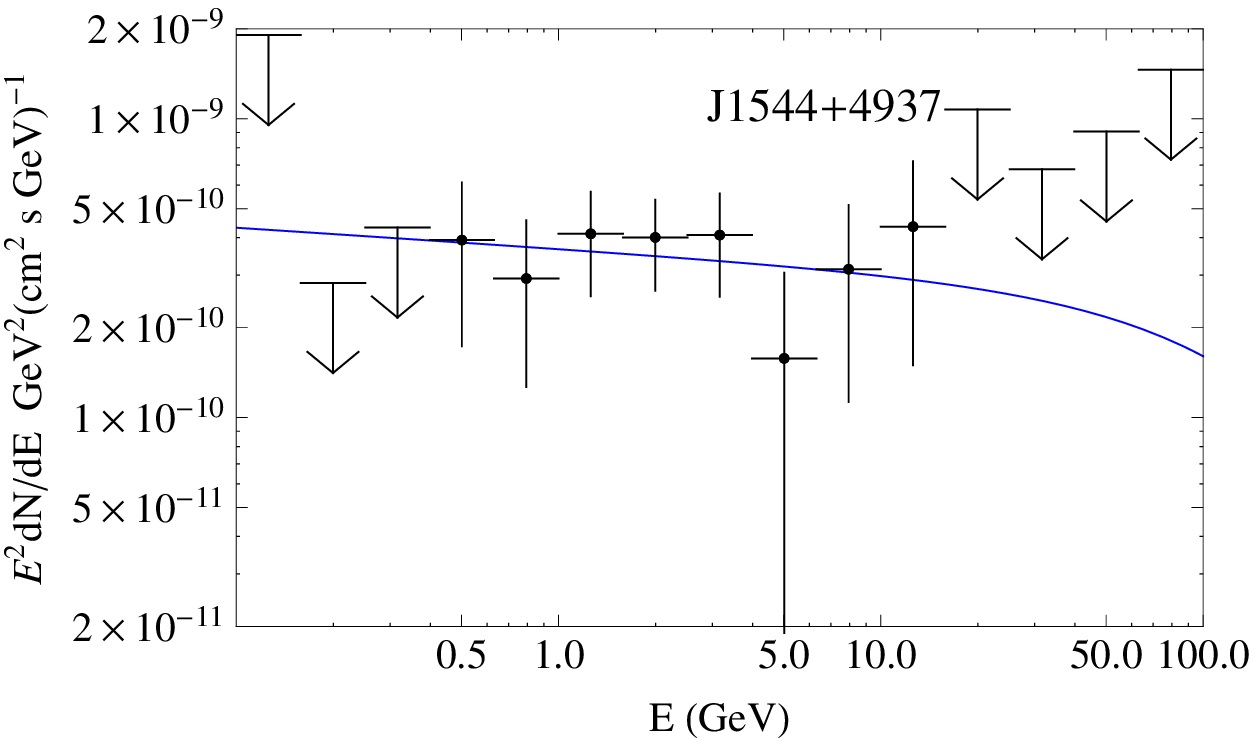} \\
\includegraphics[width=3.40in,angle=0]{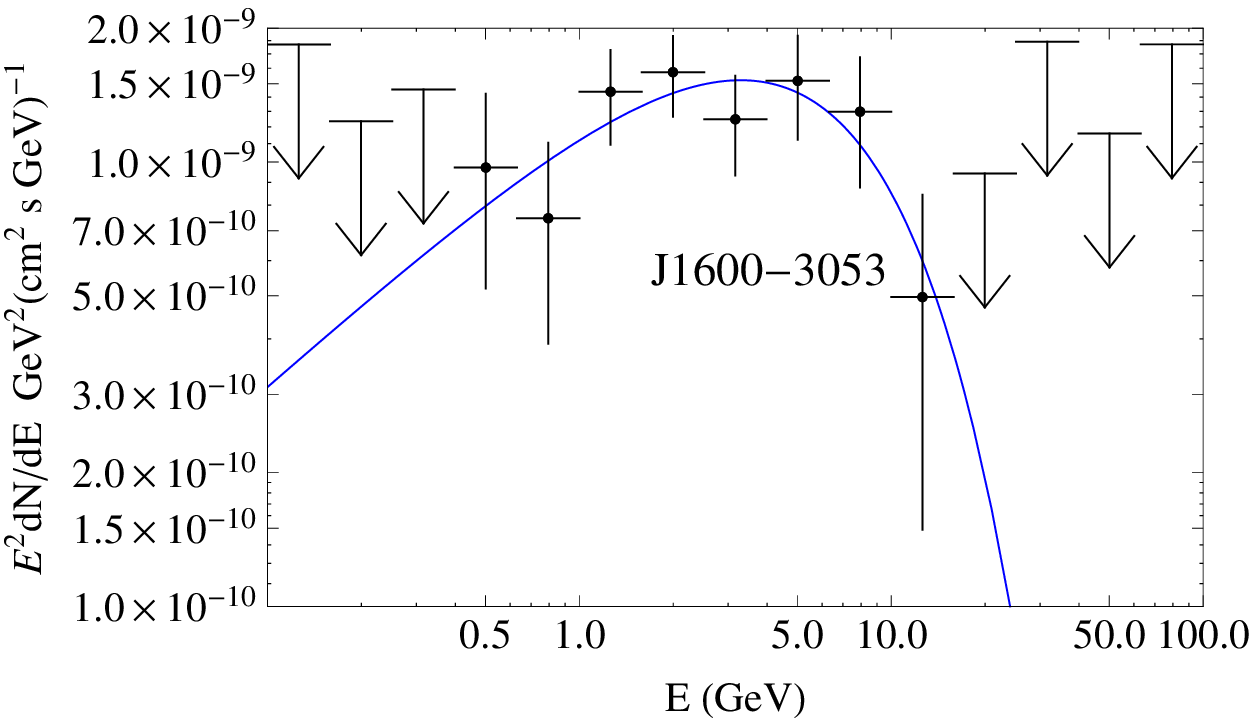} 
\includegraphics[width=3.40in,angle=0]{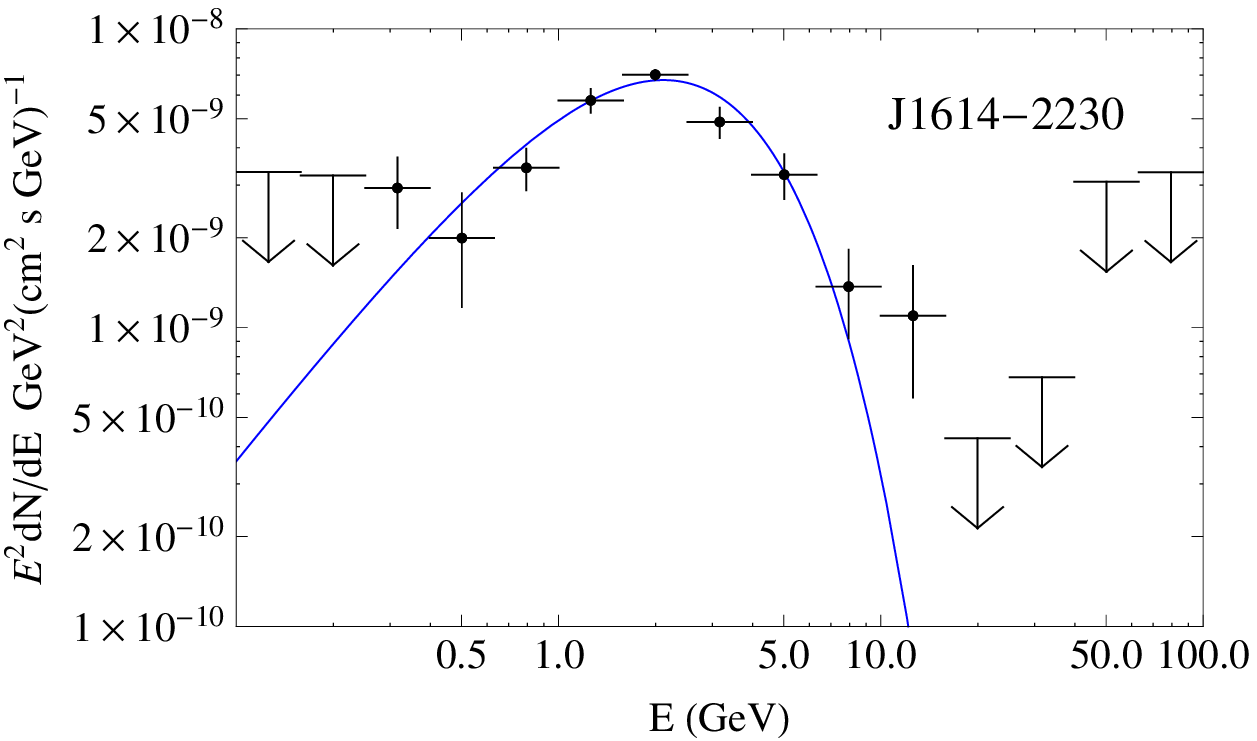} 
\caption{The gamma-ray spectra of MSPs (continued).}
\label{fig:MSPs4}
\end{figure*}

\begin{figure*}
\includegraphics[width=3.40in,angle=0]{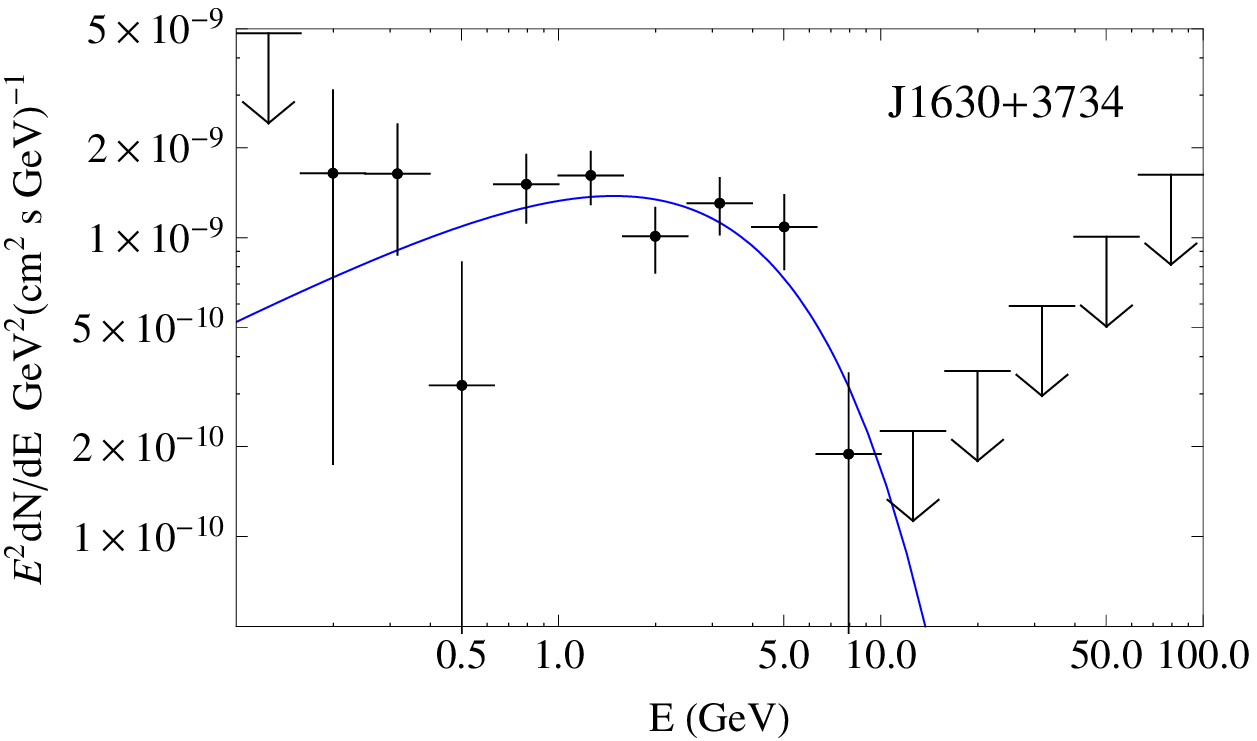} 
\includegraphics[width=3.40in,angle=0]{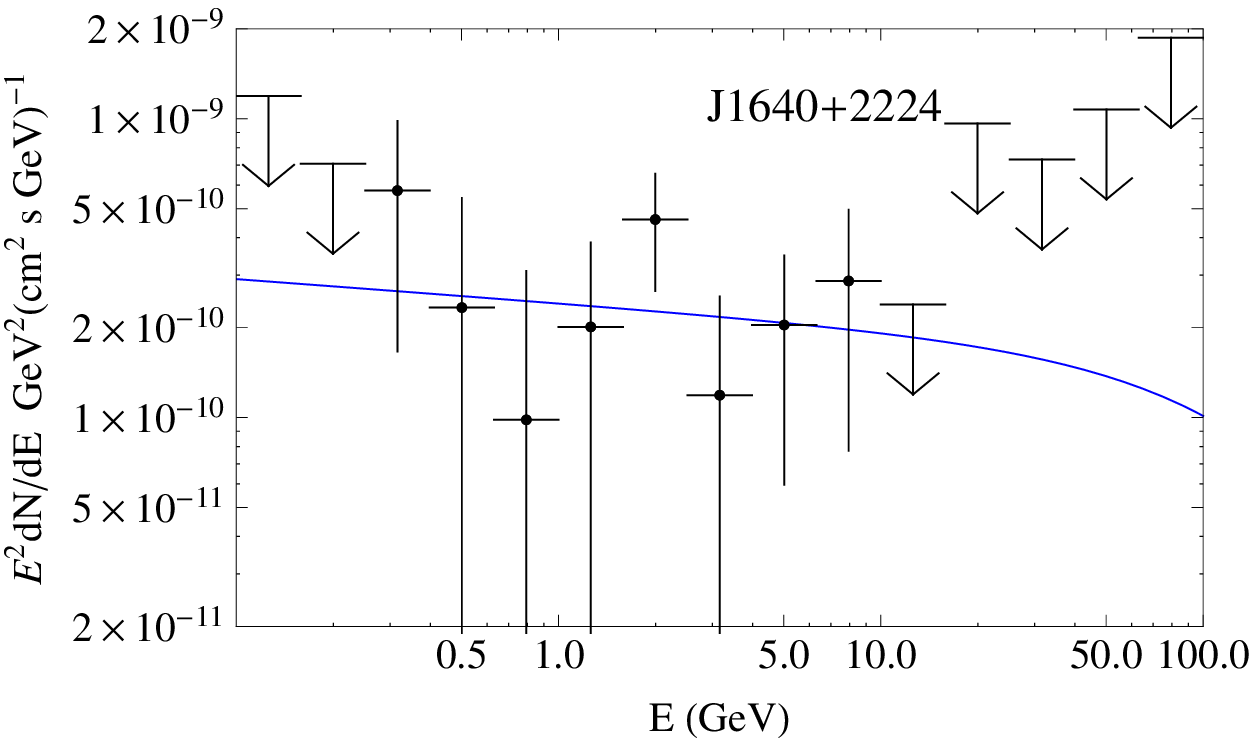} \\
\includegraphics[width=3.40in,angle=0]{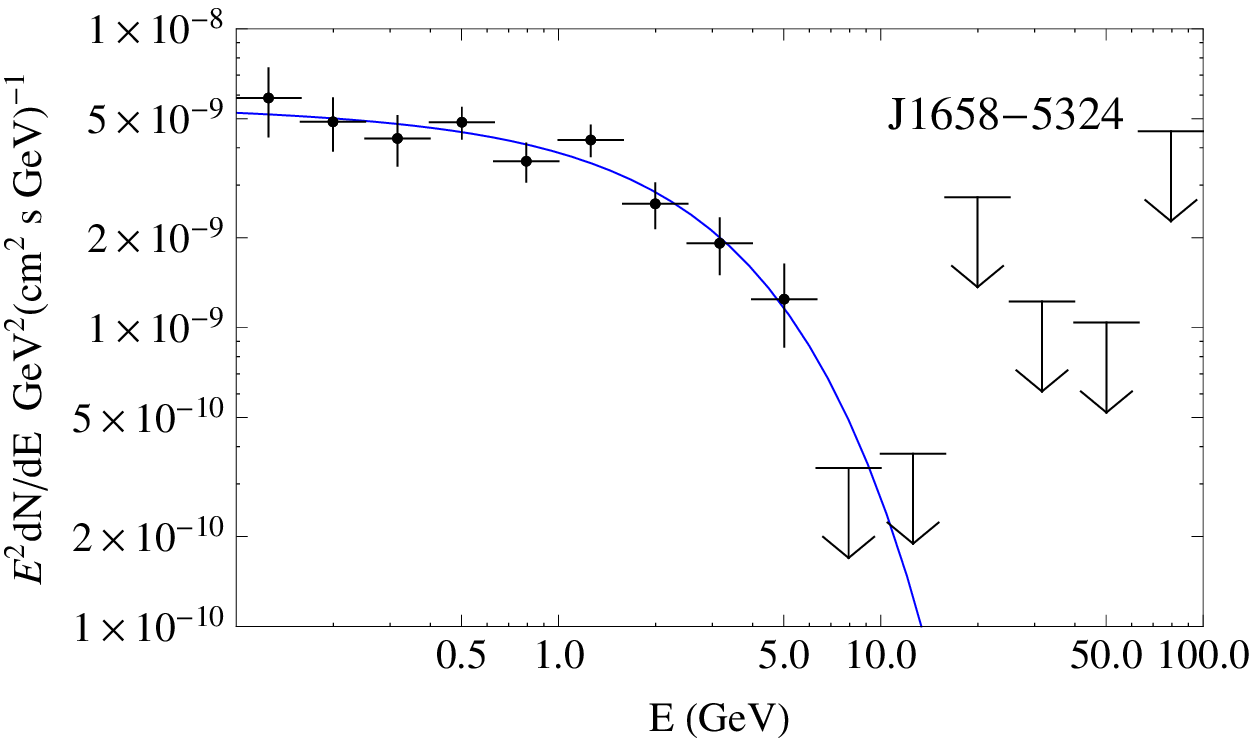} 
\includegraphics[width=3.40in,angle=0]{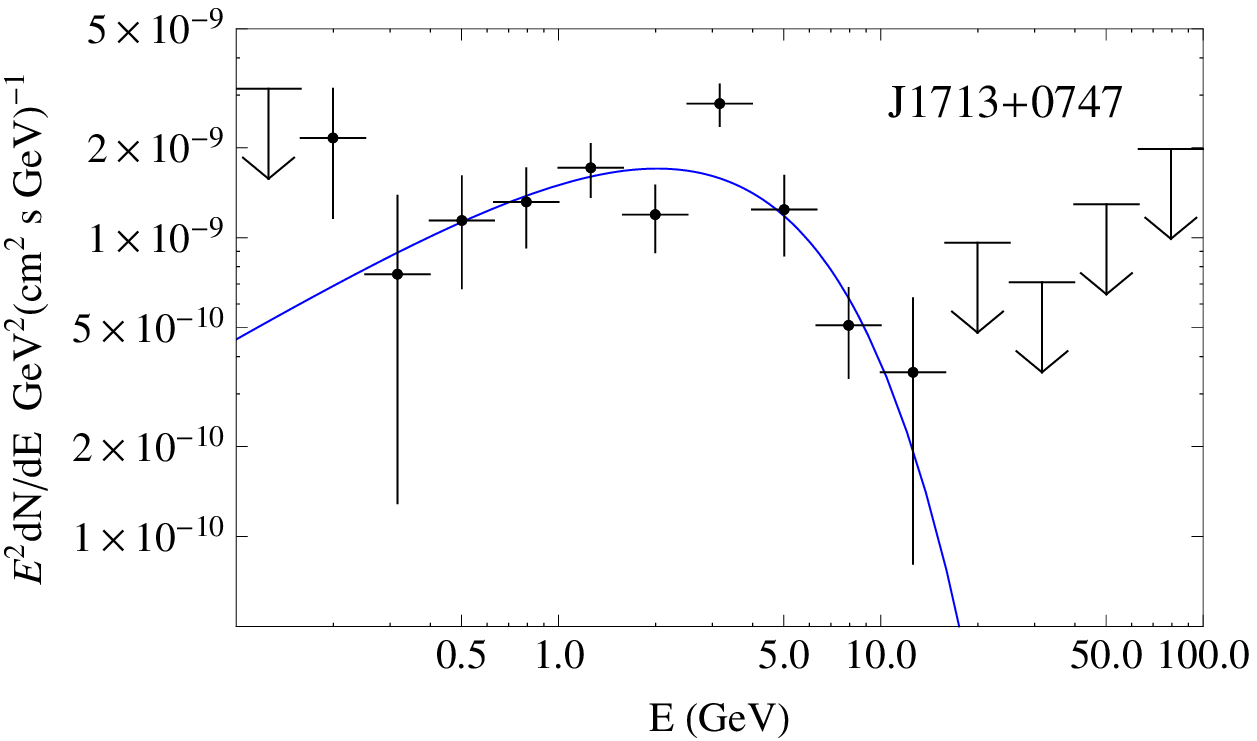} \\
\includegraphics[width=3.40in,angle=0]{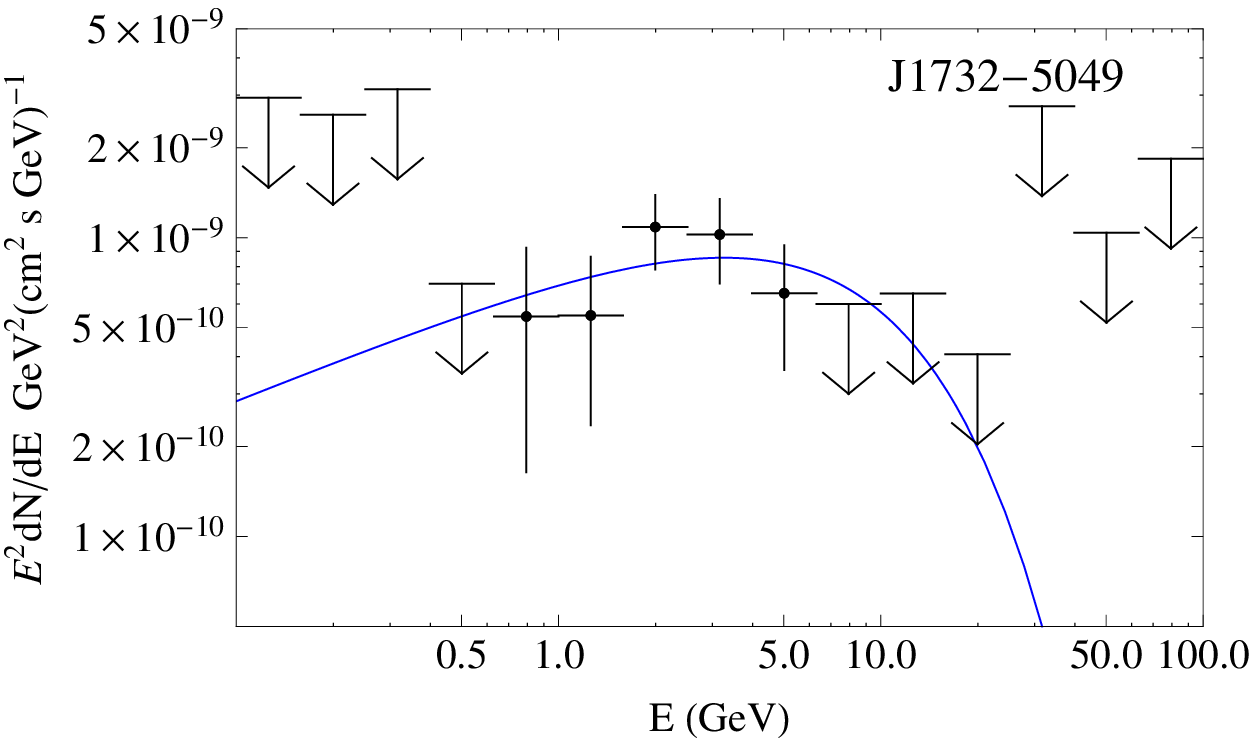} 
\includegraphics[width=3.40in,angle=0]{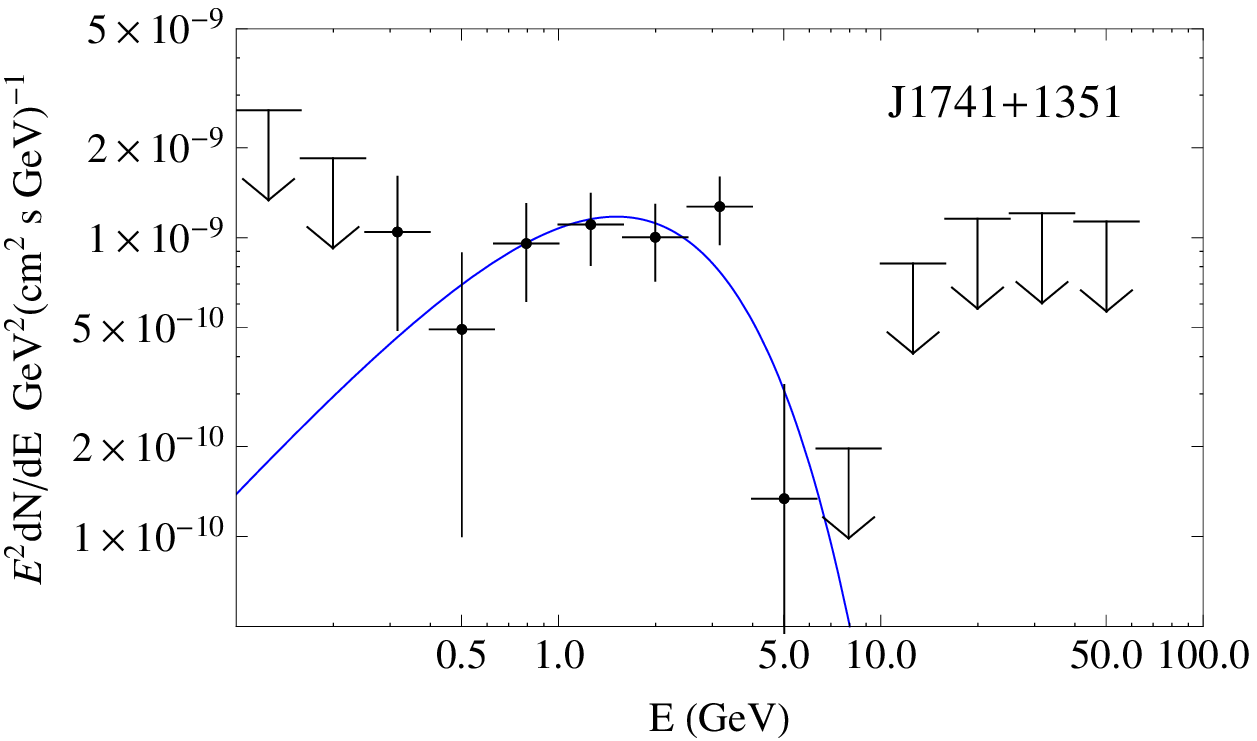} \\
\includegraphics[width=3.40in,angle=0]{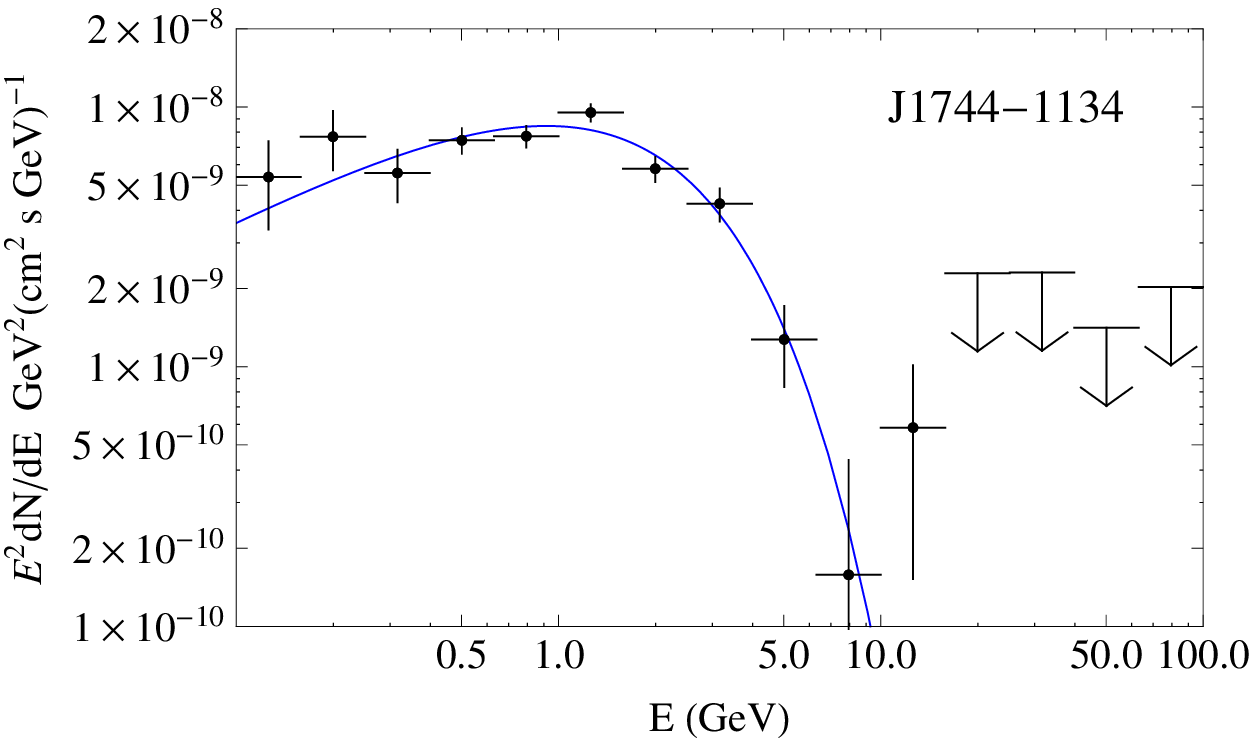} 
\includegraphics[width=3.40in,angle=0]{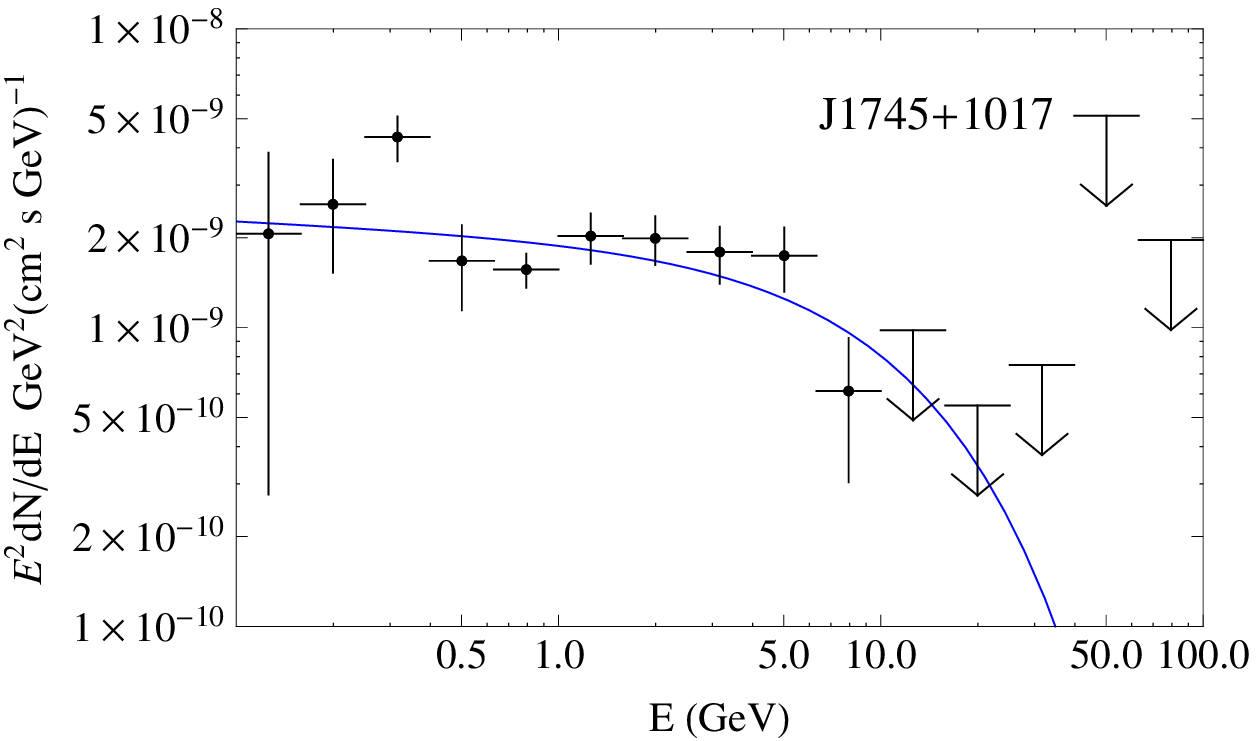} 
\caption{The gamma-ray spectra of MSPs (continued).}
\label{fig:MSPs5}
\end{figure*}

\begin{figure*}
\includegraphics[width=3.40in,angle=0]{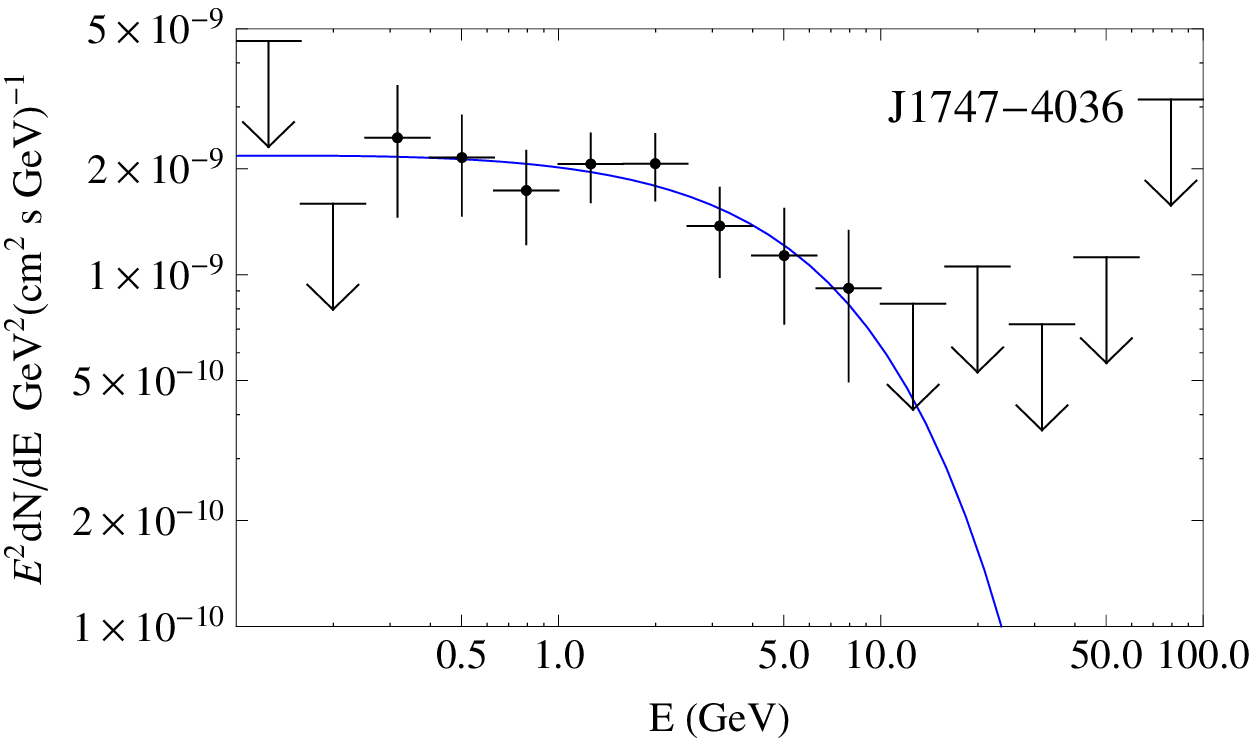} 
\includegraphics[width=3.40in,angle=0]{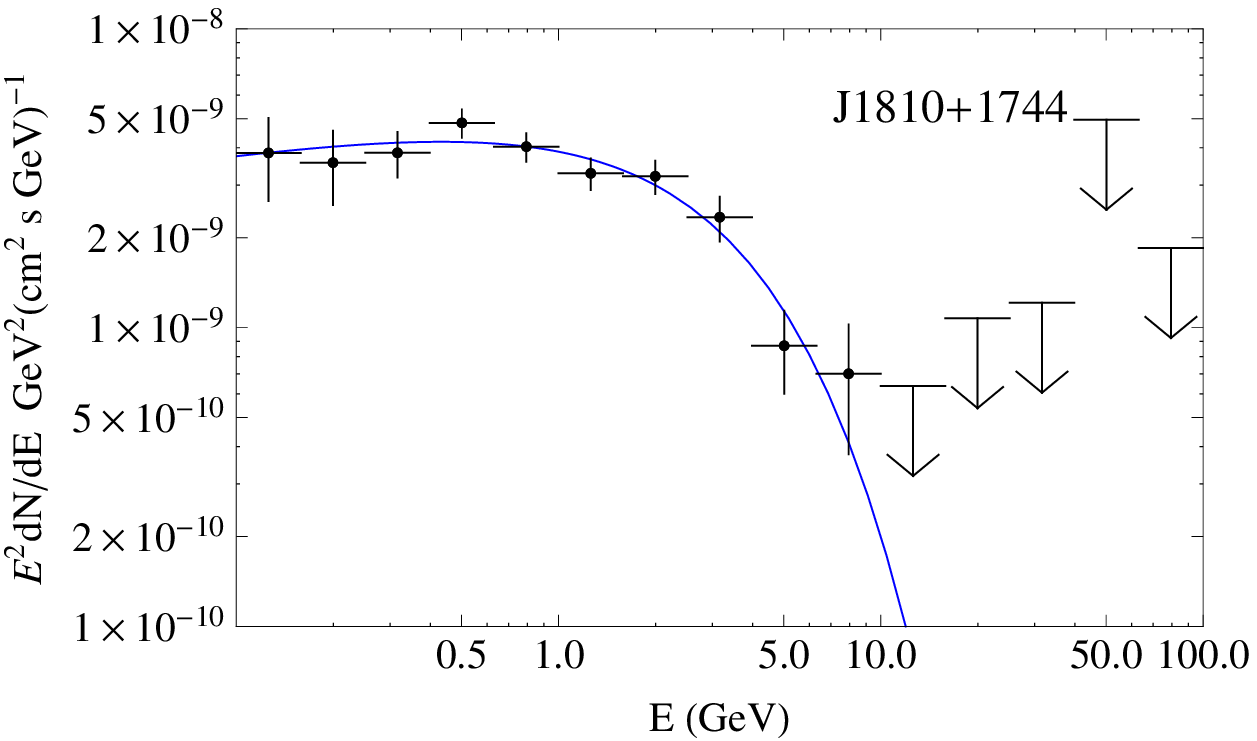} \\
\includegraphics[width=3.40in,angle=0]{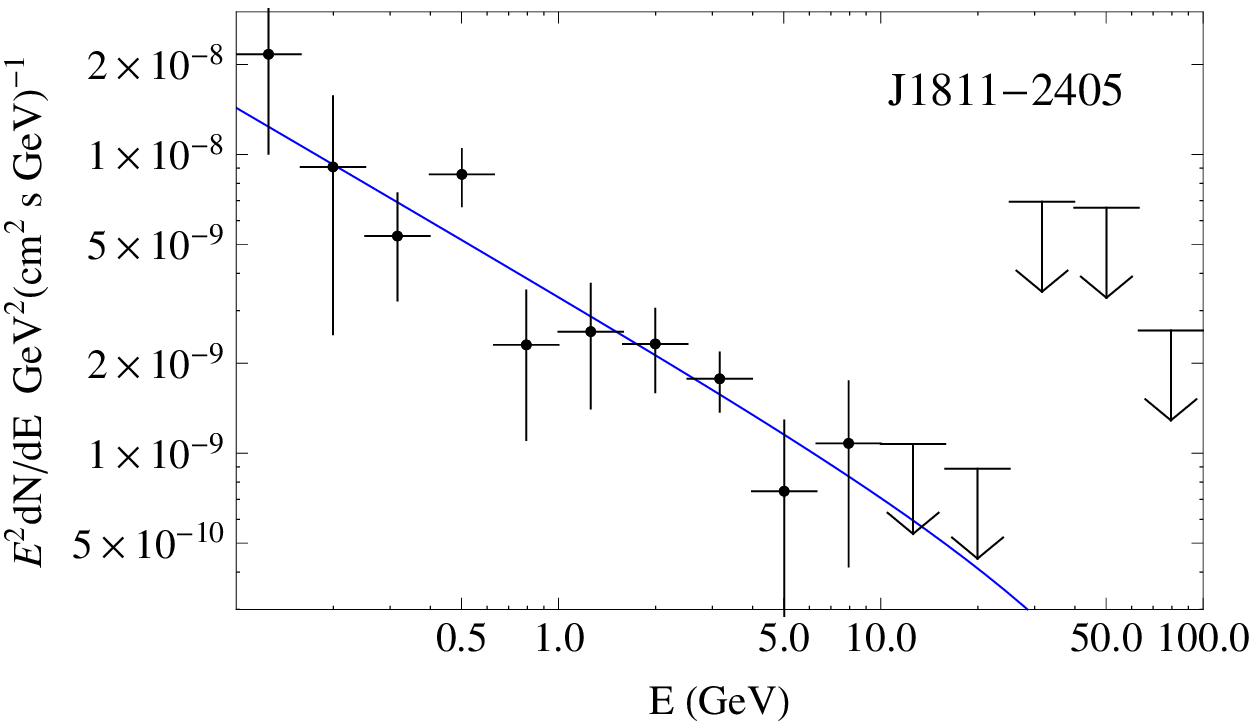} 
\includegraphics[width=3.40in,angle=0]{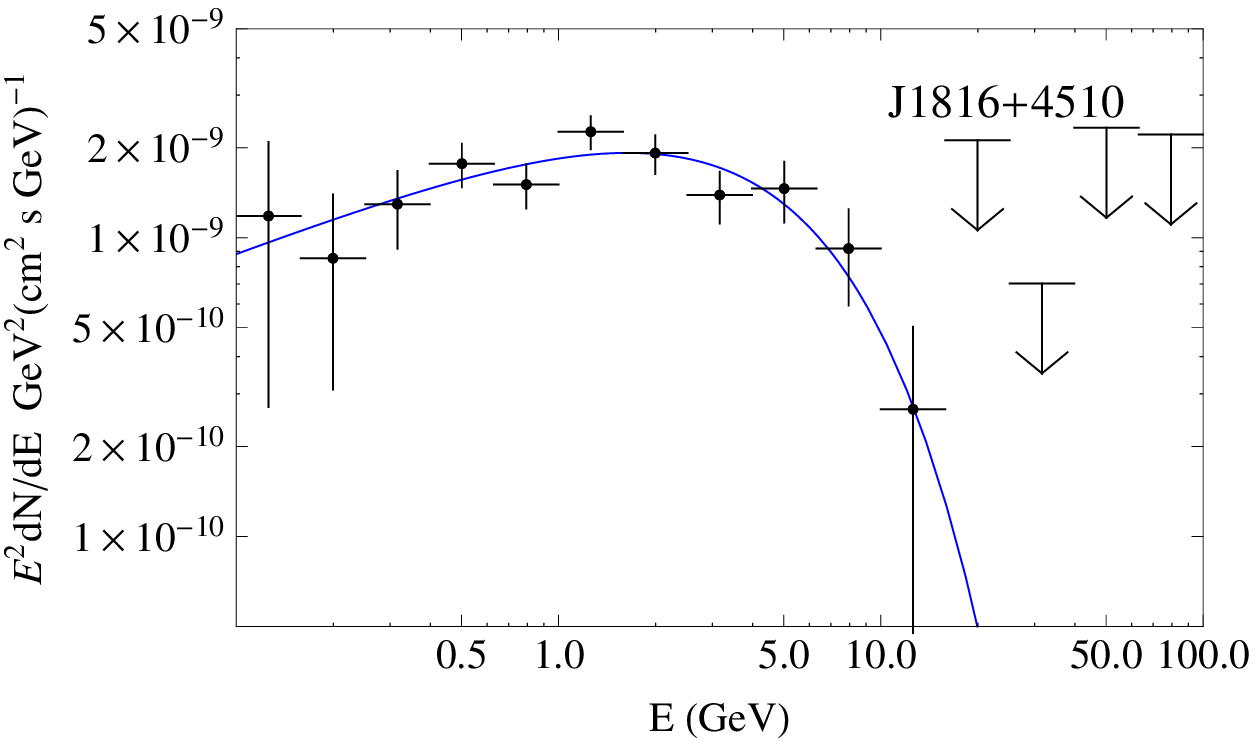} \\
\includegraphics[width=3.40in,angle=0]{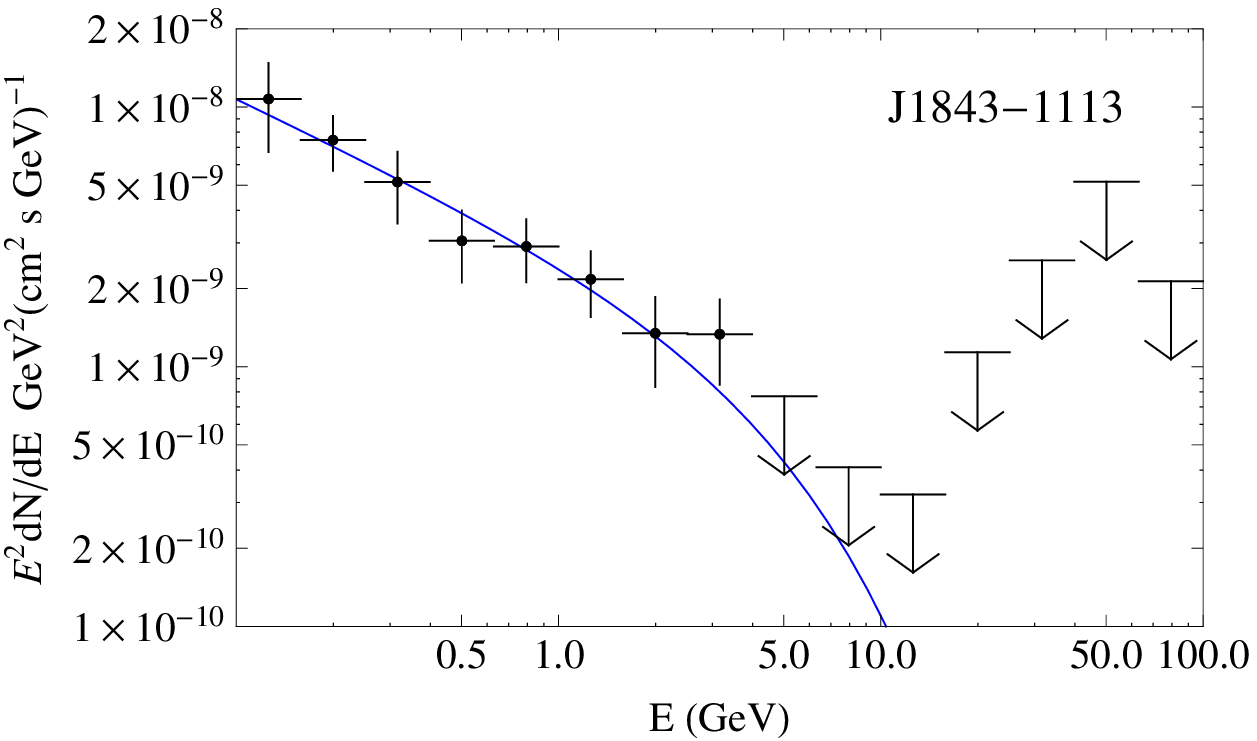} 
\includegraphics[width=3.40in,angle=0]{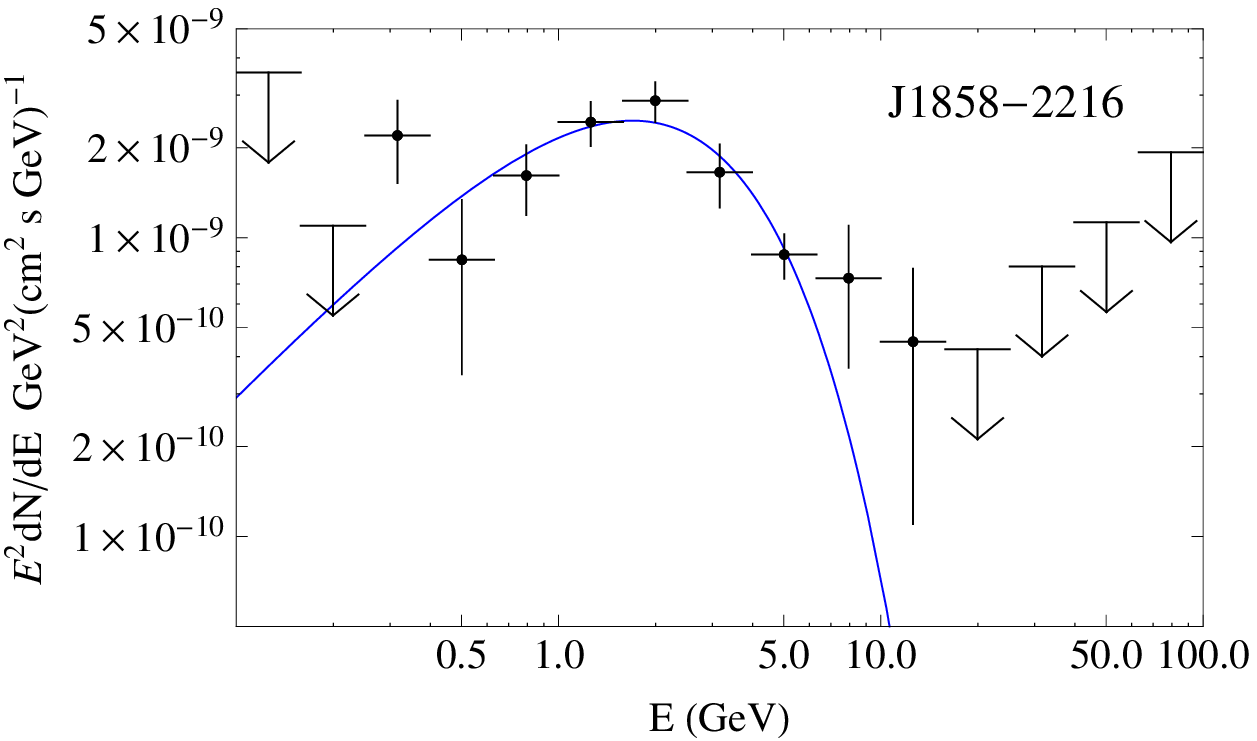} \\
\includegraphics[width=3.40in,angle=0]{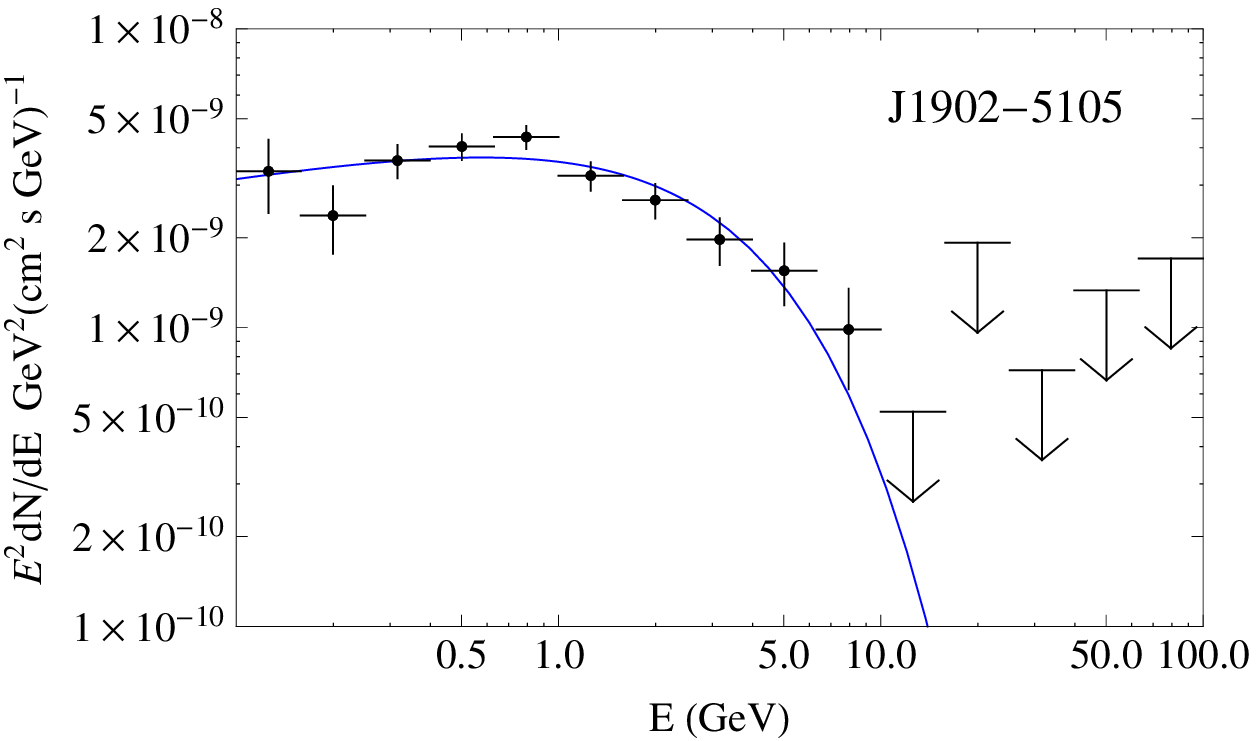} 
\includegraphics[width=3.40in,angle=0]{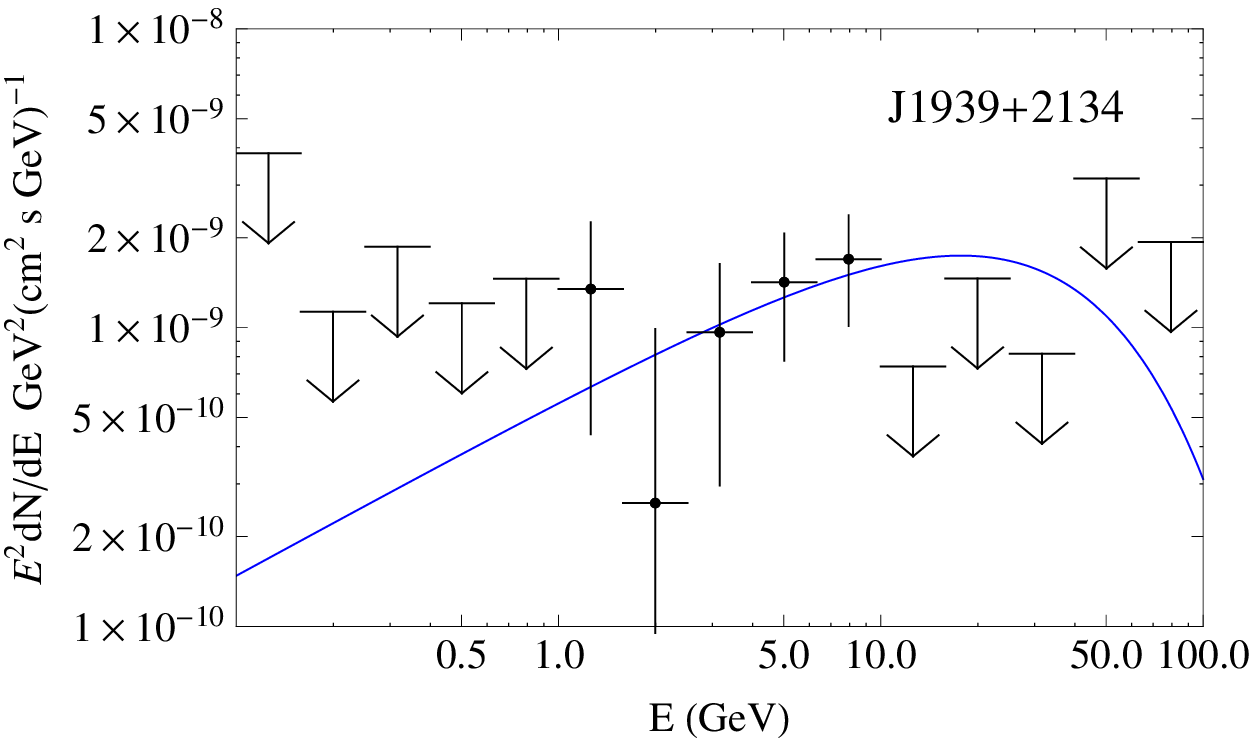} 
\caption{The gamma-ray spectra of MSPs (continued).}
\label{fig:MSPs6}
\end{figure*}

\begin{figure*}
\includegraphics[width=3.40in,angle=0]{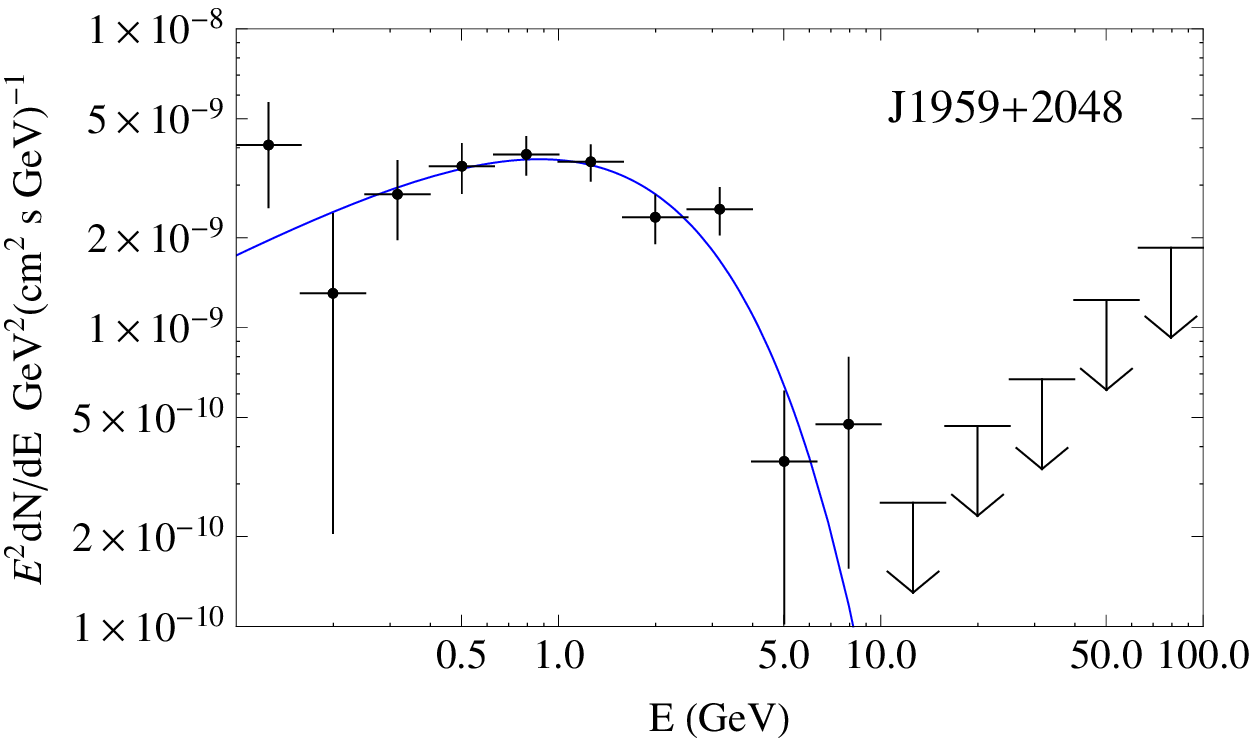} 
\includegraphics[width=3.40in,angle=0]{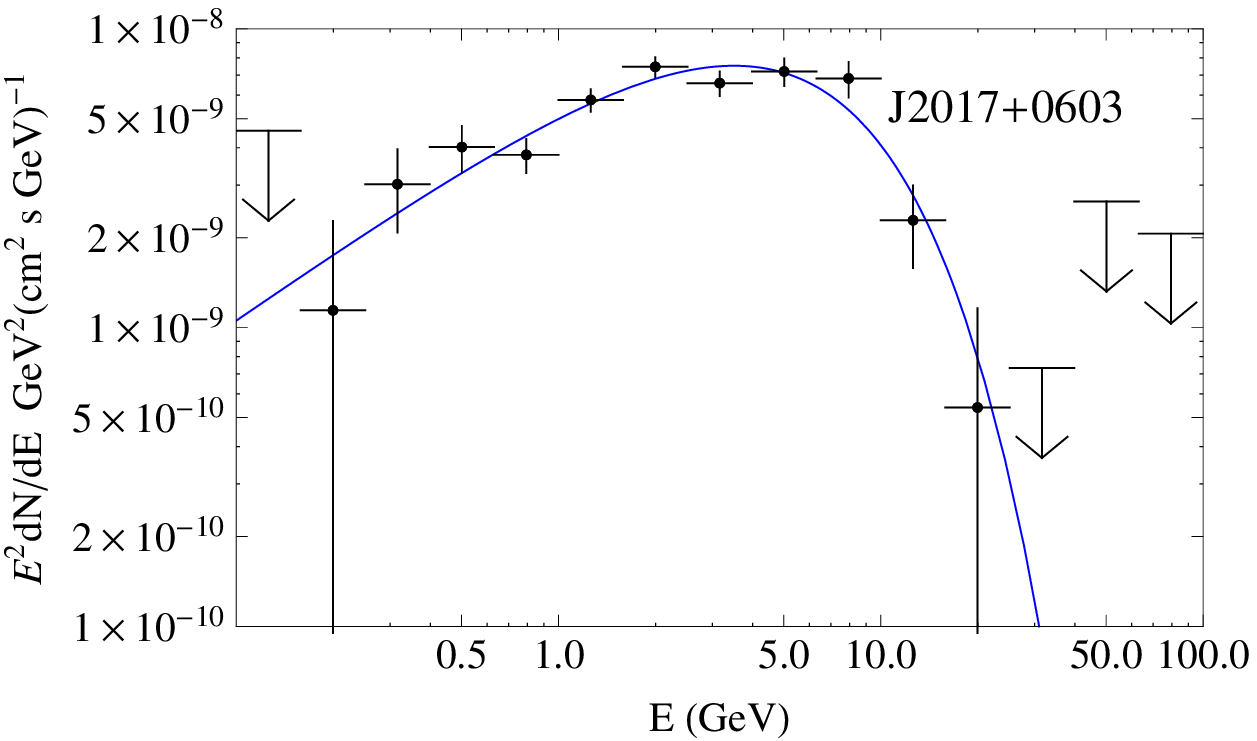} \\
\includegraphics[width=3.40in,angle=0]{plots/MSP/J2043p1711.eps} 
\includegraphics[width=3.40in,angle=0]{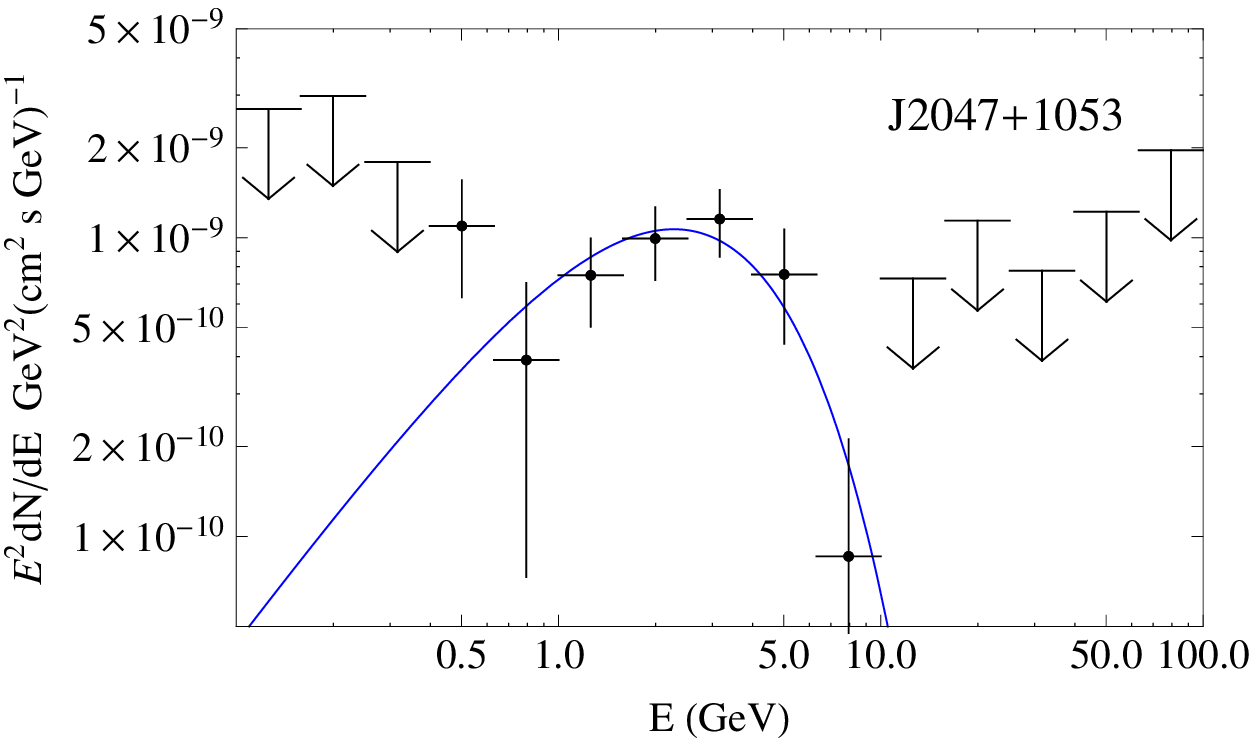} \\
\includegraphics[width=3.40in,angle=0]{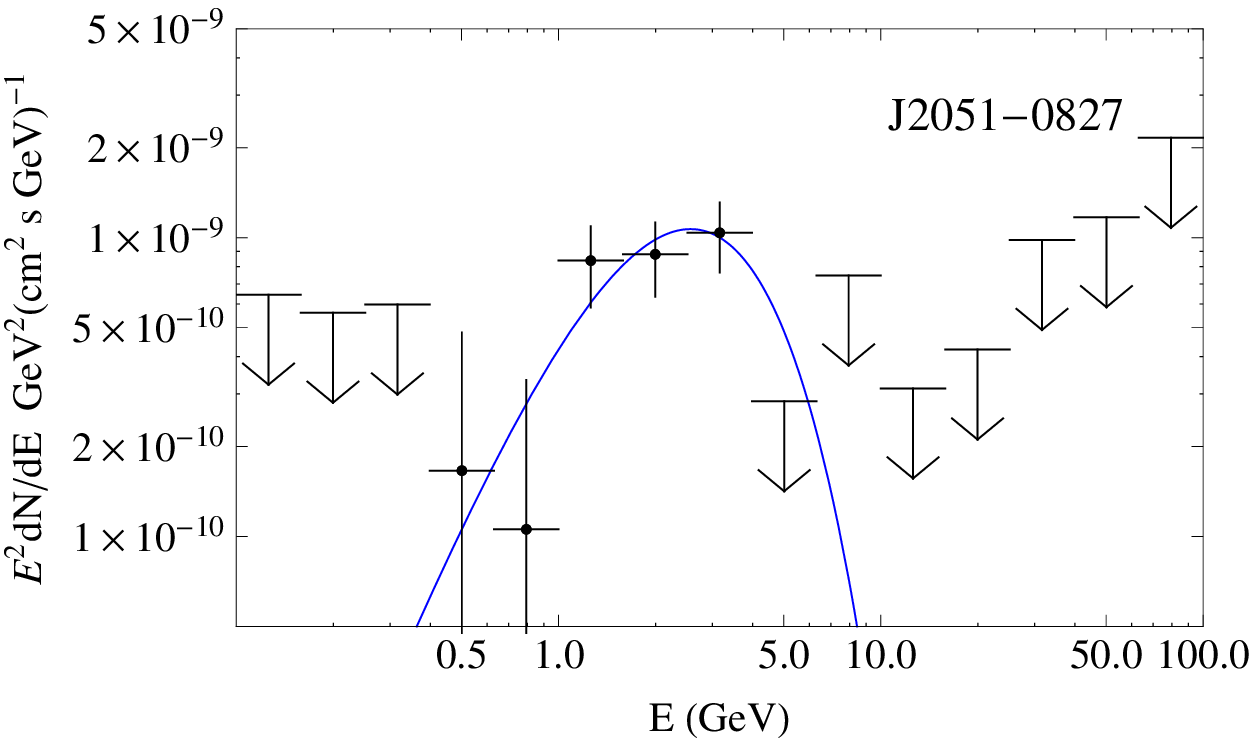} 
\includegraphics[width=3.40in,angle=0]{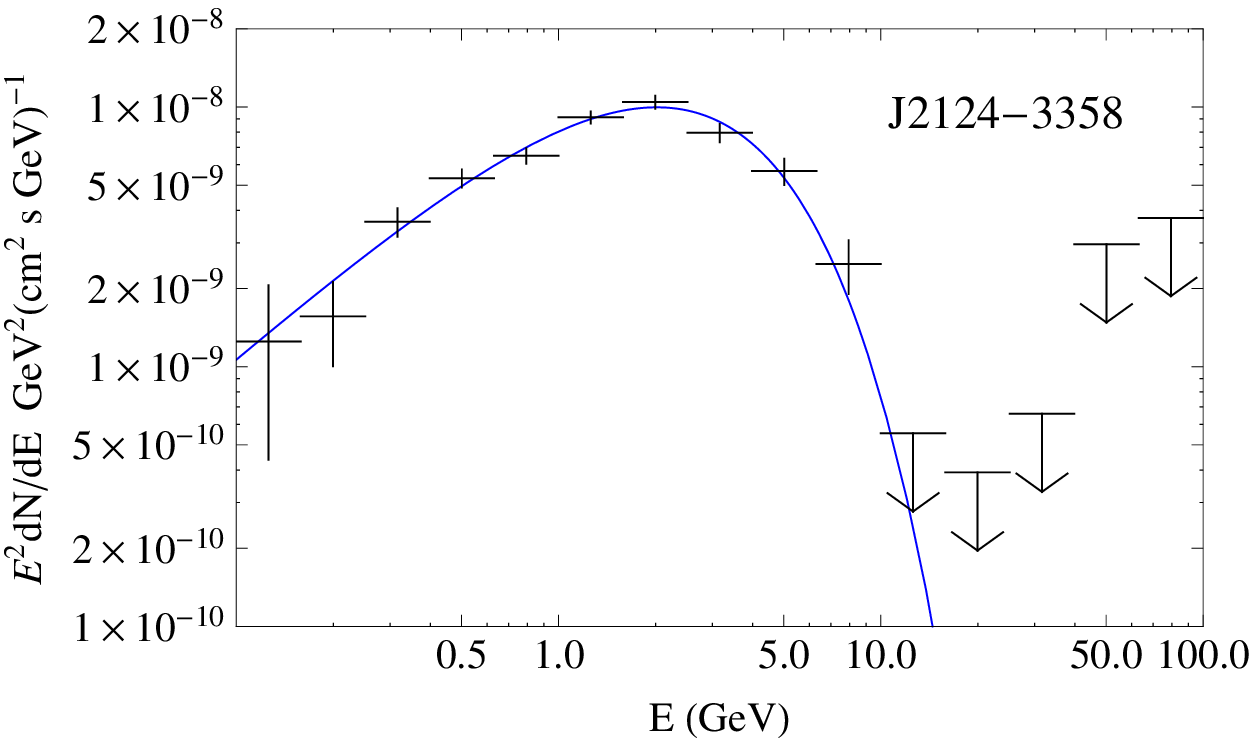} \\
\includegraphics[width=3.40in,angle=0]{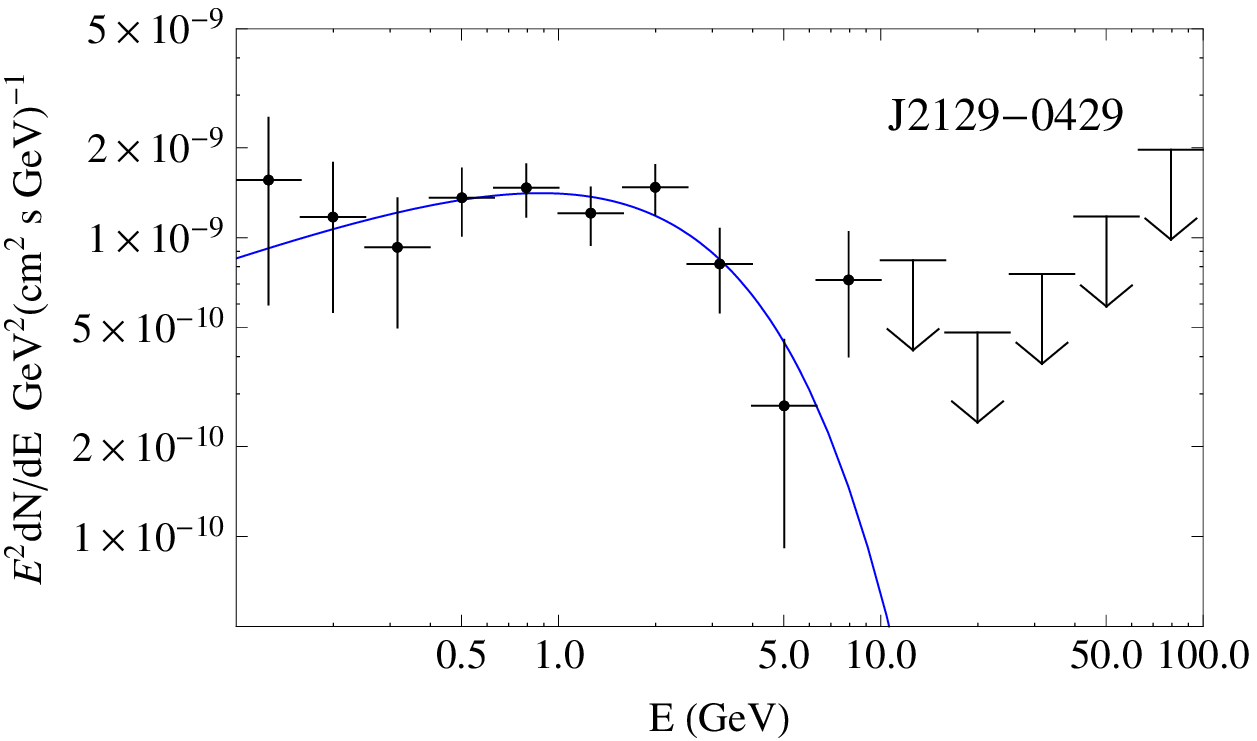} 
\includegraphics[width=3.40in,angle=0]{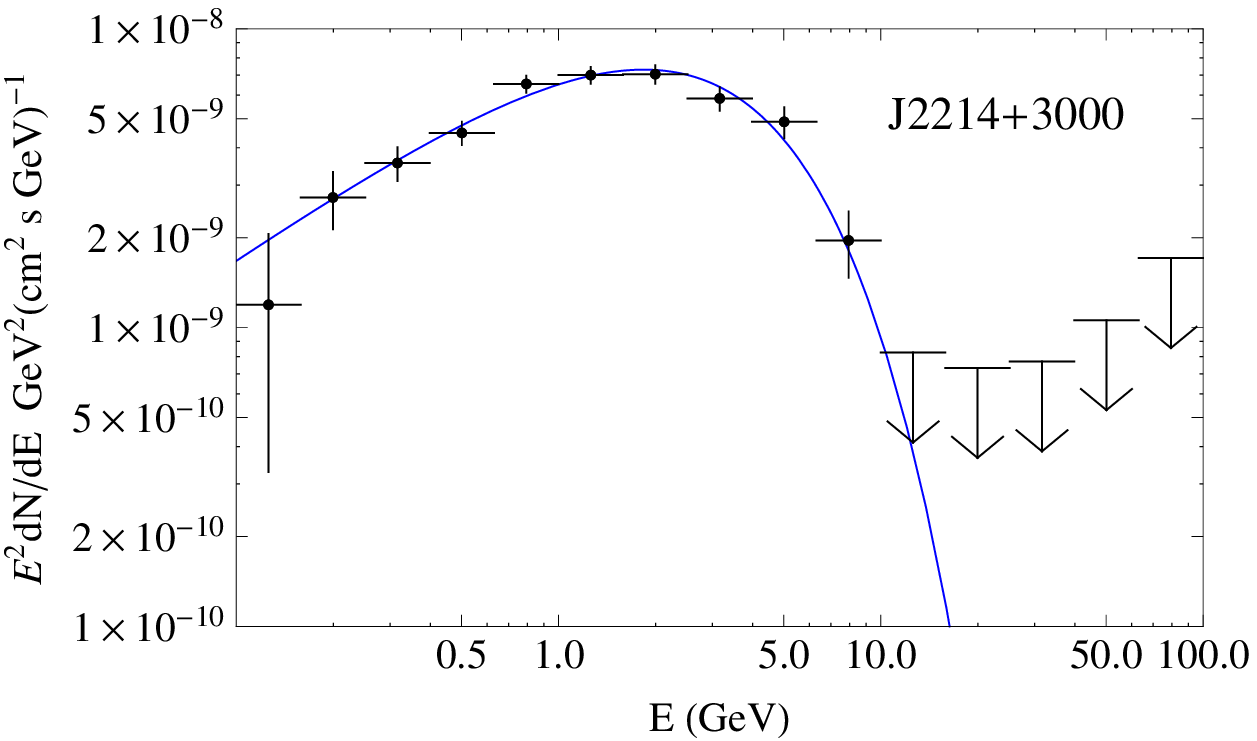} 
\caption{The gamma-ray spectra of MSPs (continued).}
\label{fig:MSPs7}
\end{figure*}

\begin{figure*}
\includegraphics[width=3.40in,angle=0]{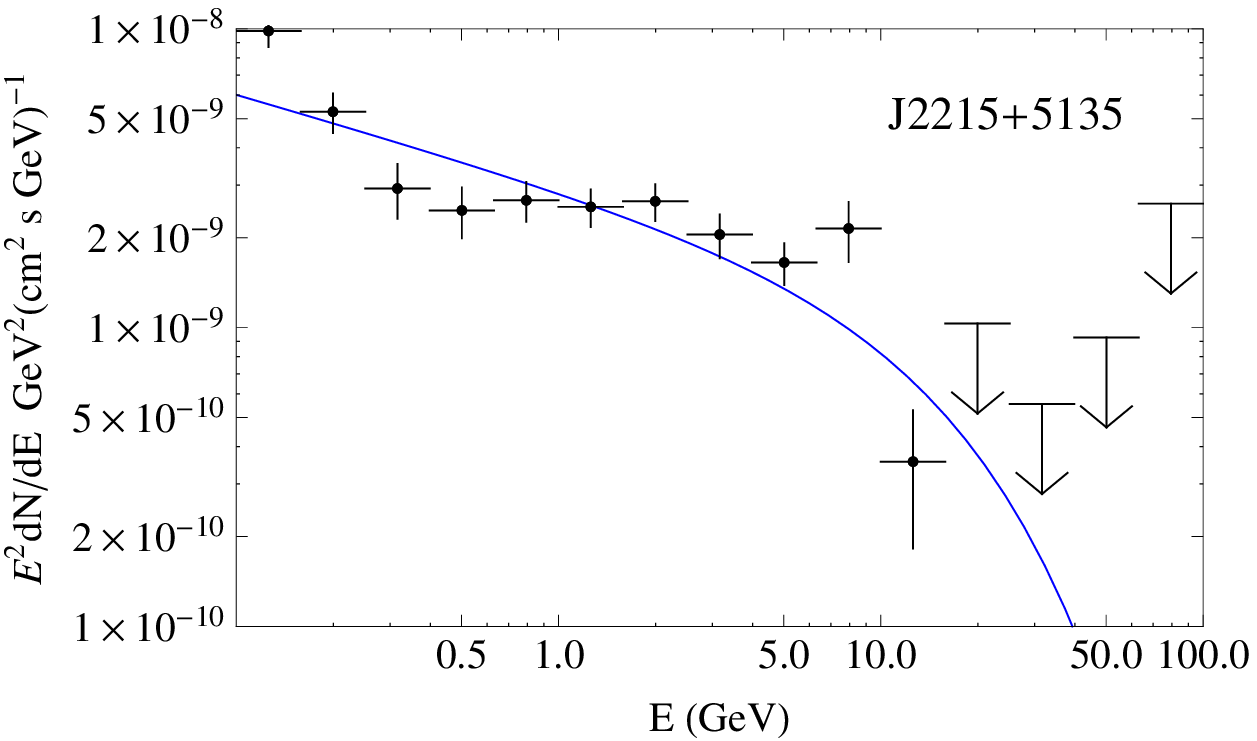} 
\includegraphics[width=3.40in,angle=0]{plots/MSP/J2241m5236.eps} \\
\includegraphics[width=3.40in,angle=0]{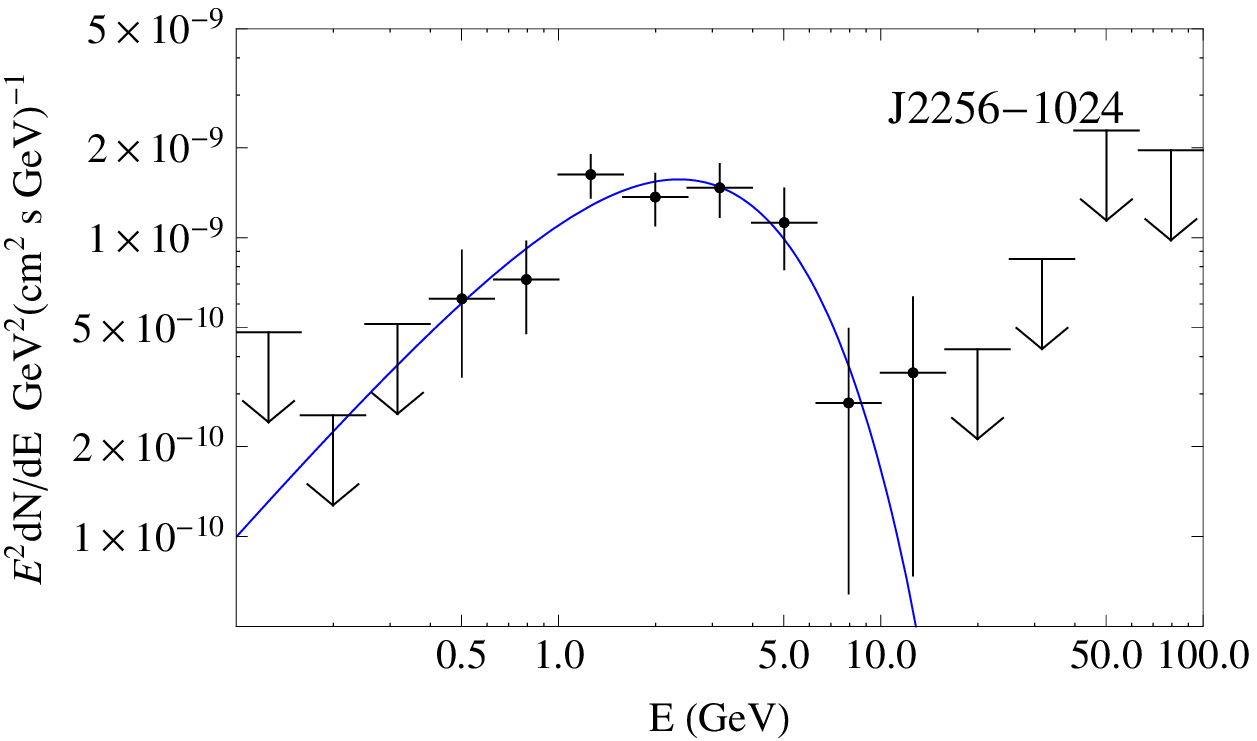} 
\includegraphics[width=3.40in,angle=0]{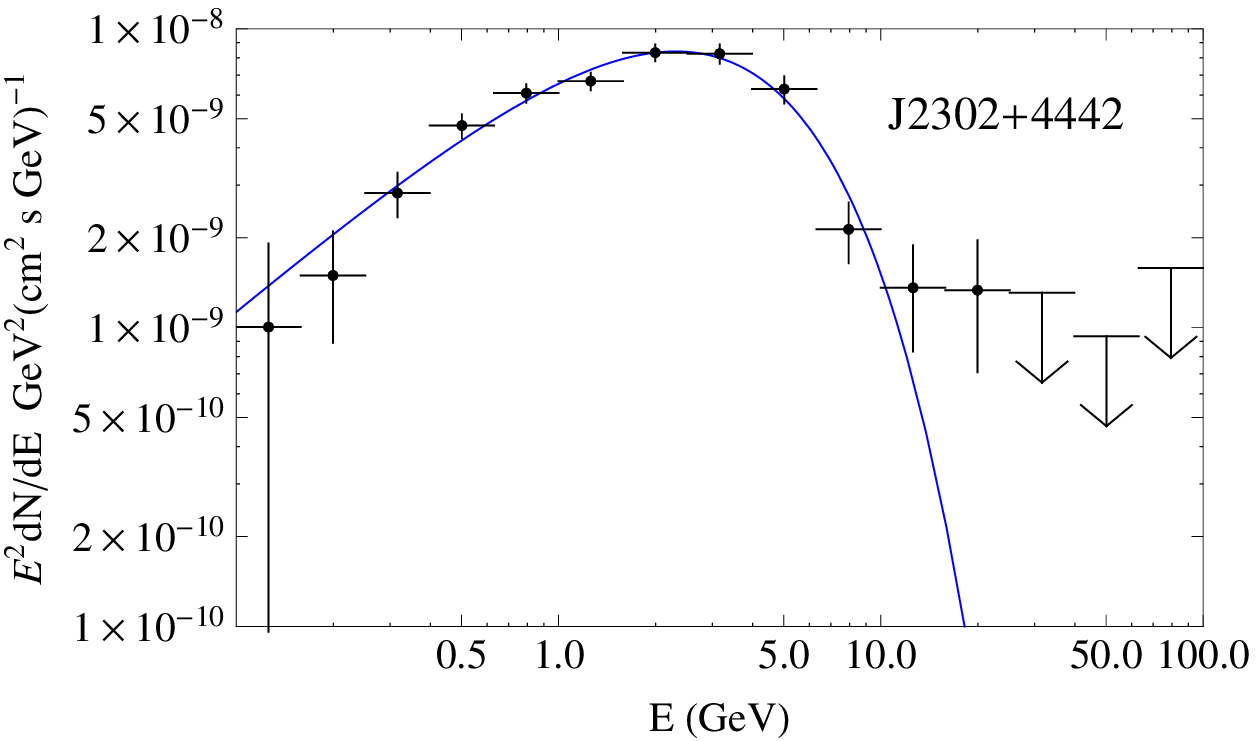} \\
\includegraphics[width=3.40in,angle=0]{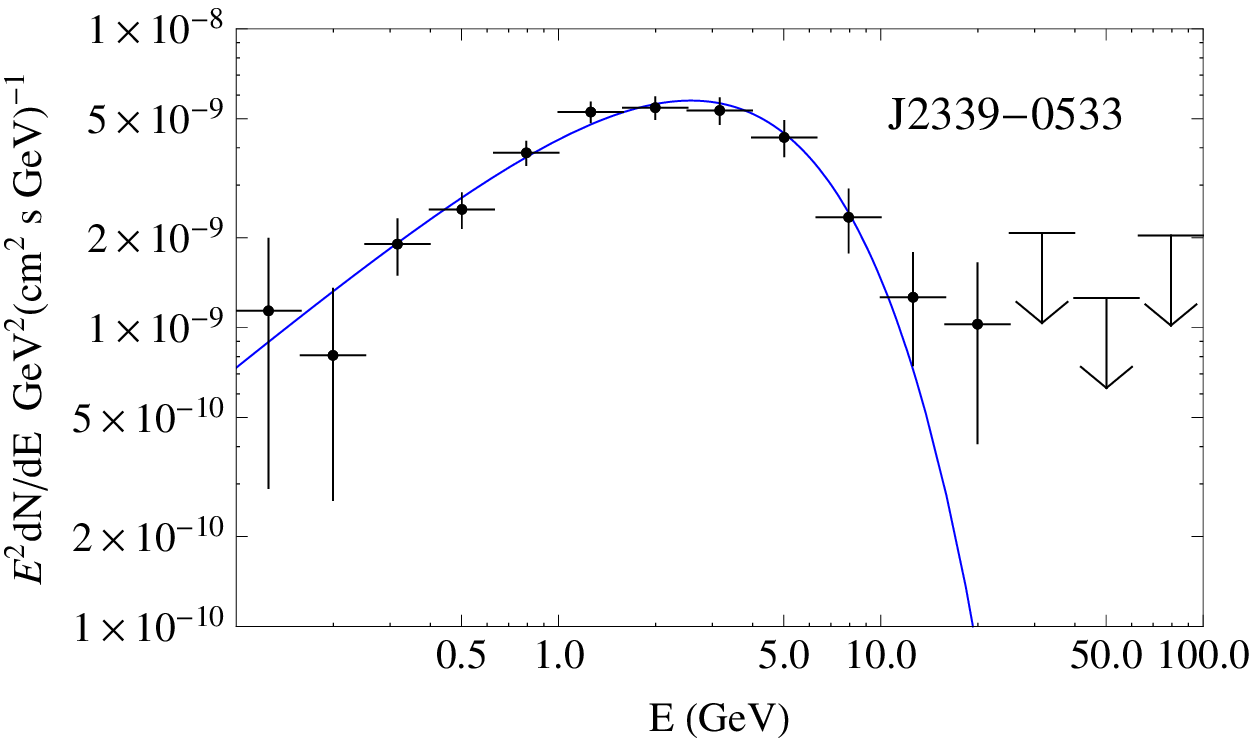} 
\caption{The gamma-ray spectra of MSPs (continued).}
\label{fig:MSPs8}
\end{figure*}

\begin{figure*}
\includegraphics[width=3.40in,angle=0]{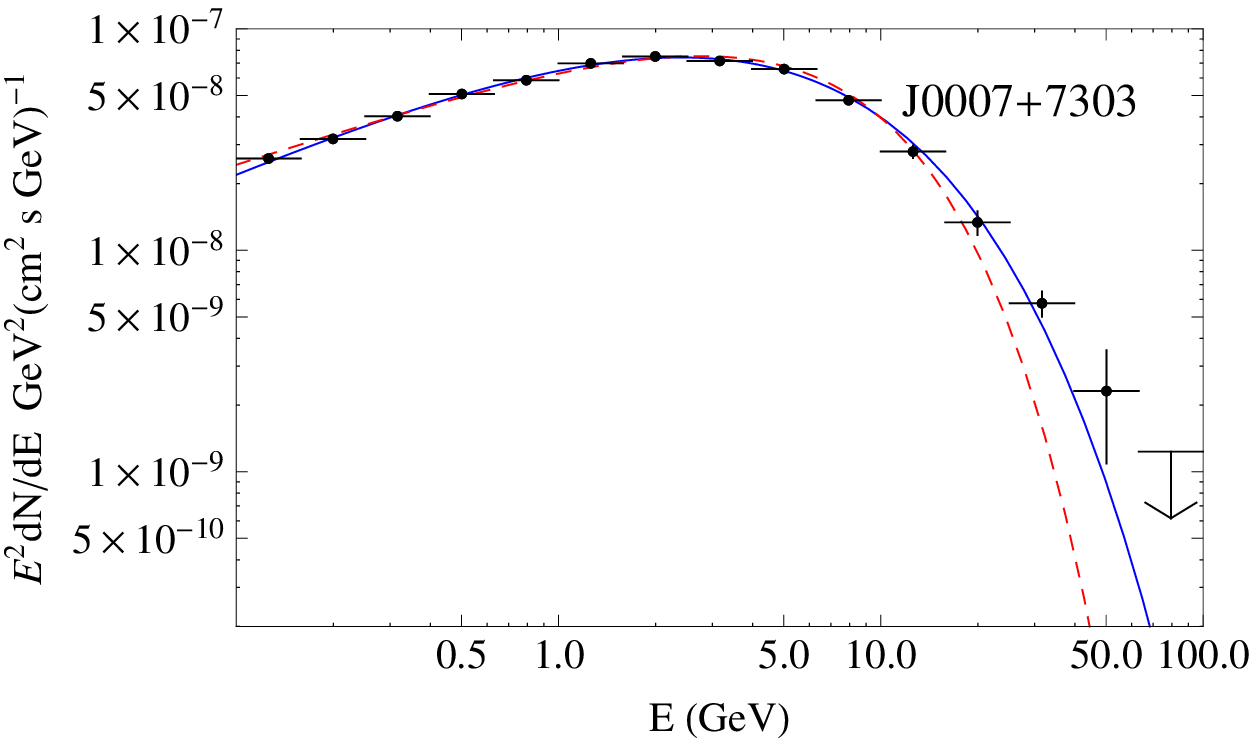}
\includegraphics[width=3.40in,angle=0]{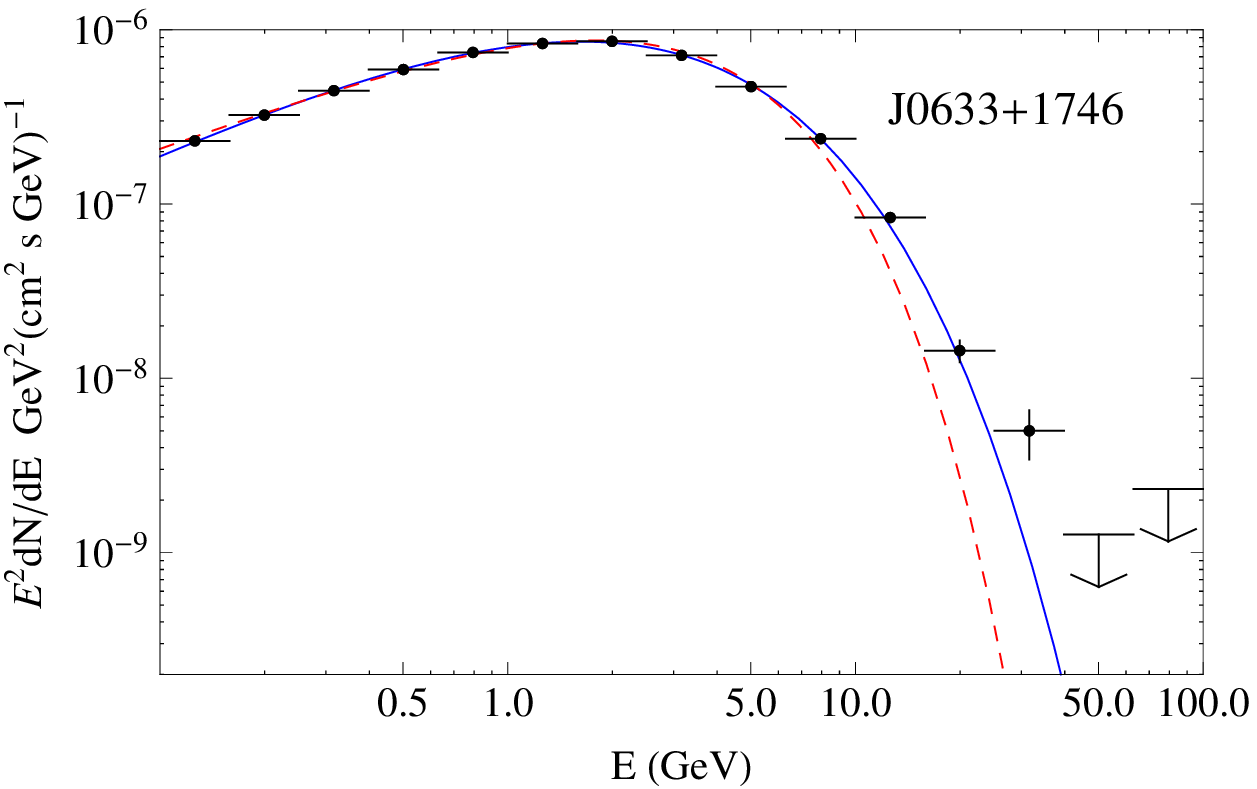} \\
\includegraphics[width=3.40in,angle=0]{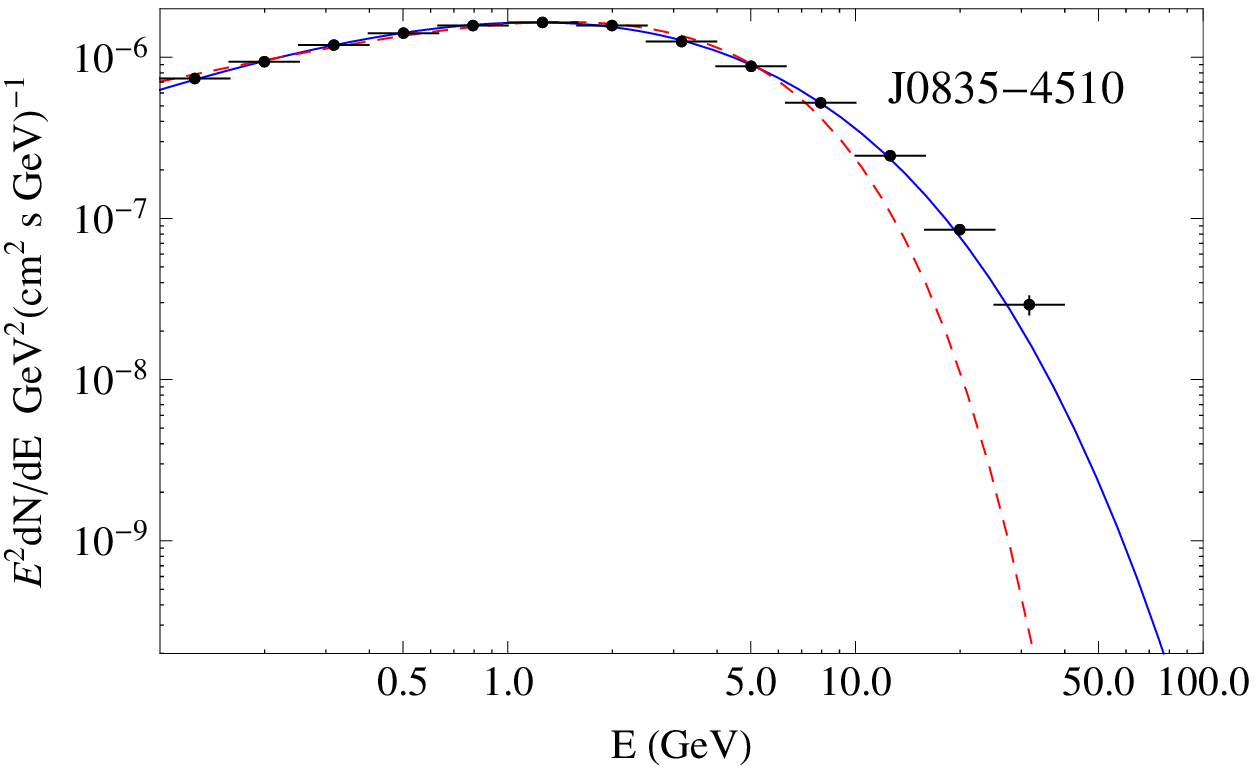} 
\includegraphics[width=3.40in,angle=0]{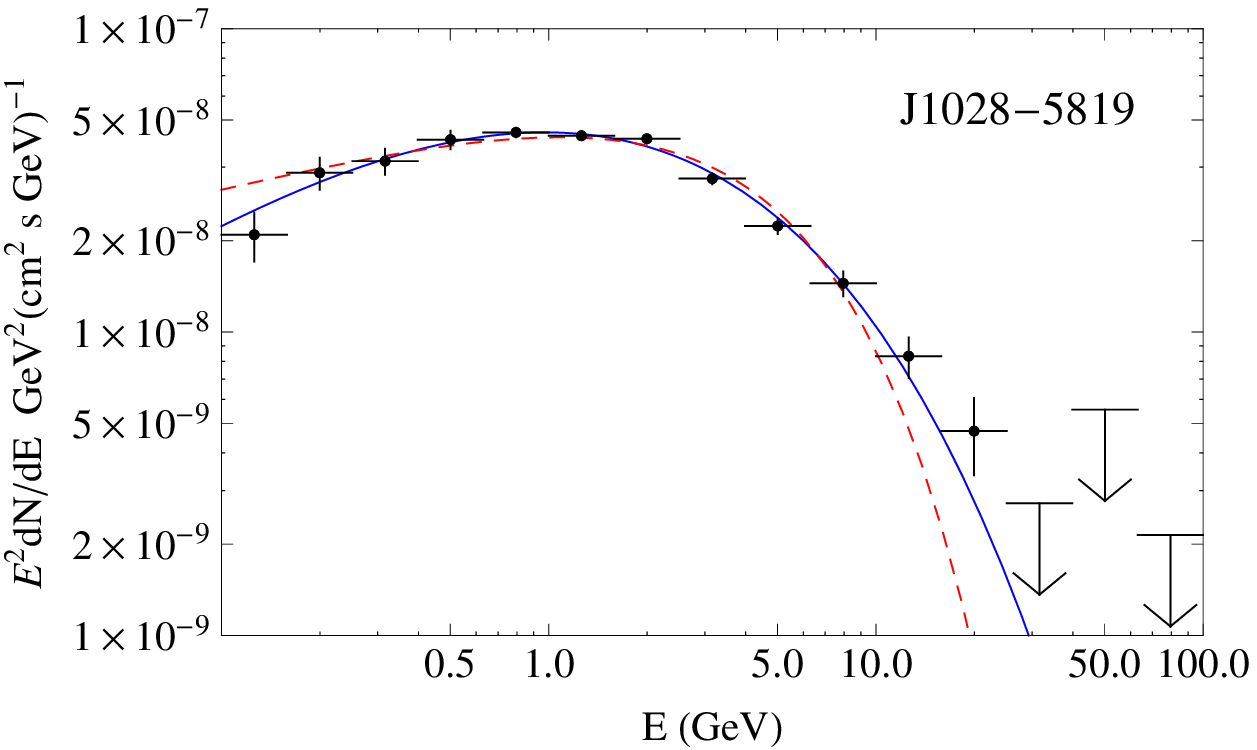} \\
\includegraphics[width=3.40in,angle=0]{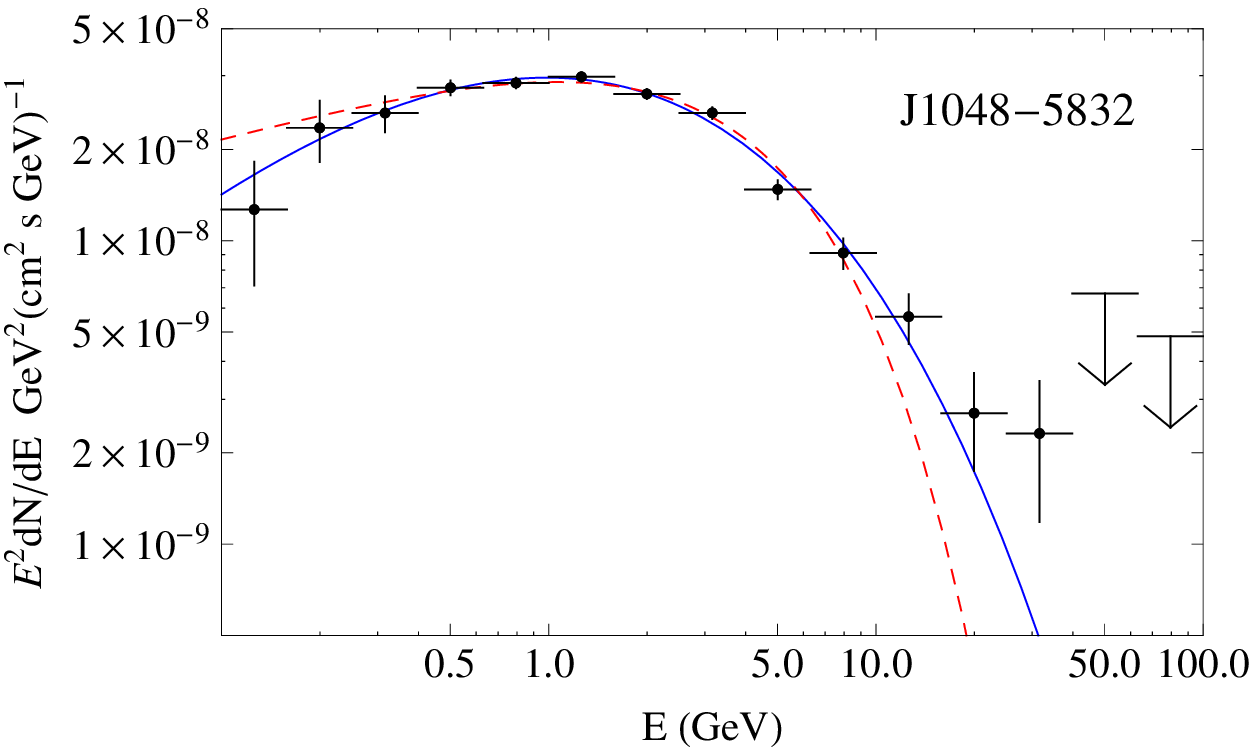} 
\includegraphics[width=3.40in,angle=0]{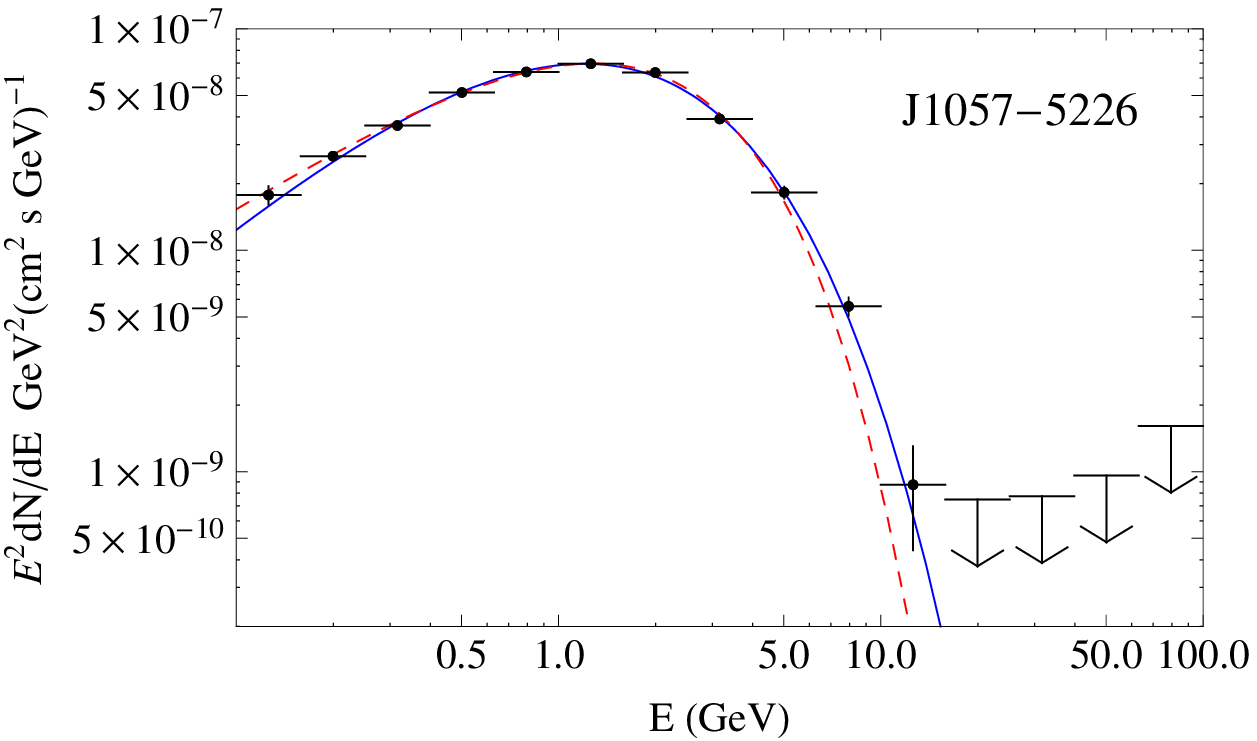} \\
\includegraphics[width=3.40in,angle=0]{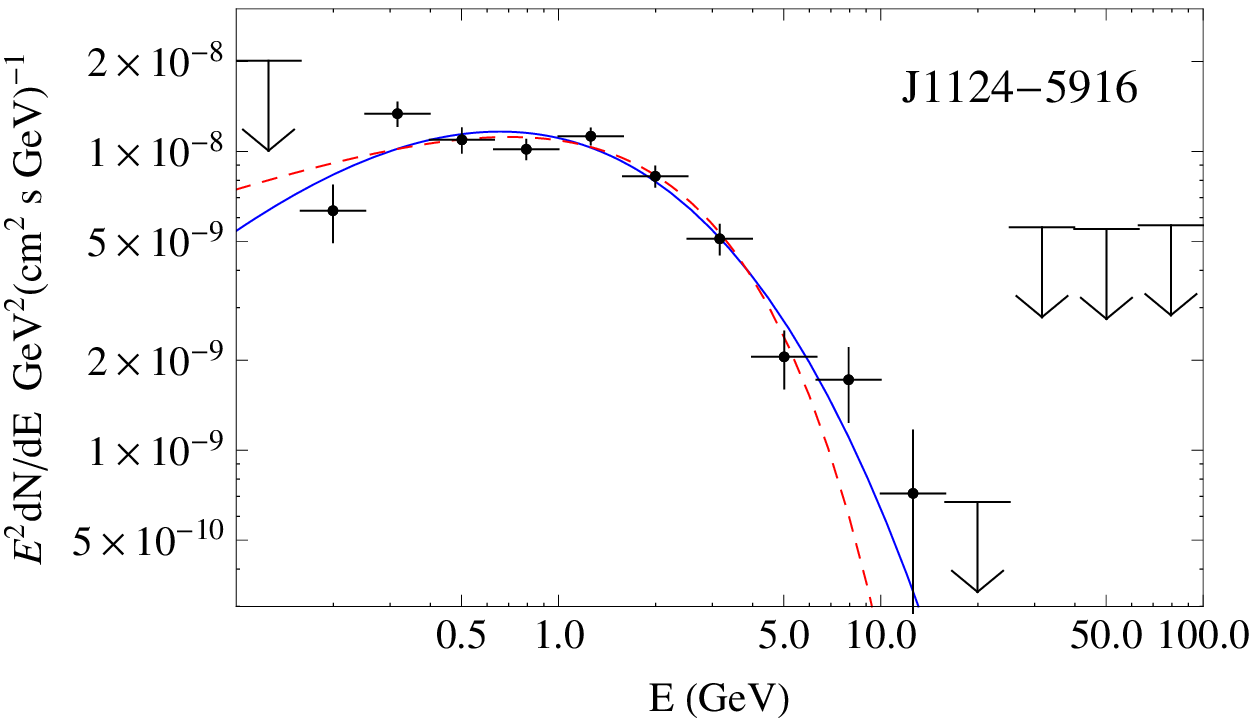} 
\includegraphics[width=3.40in,angle=0]{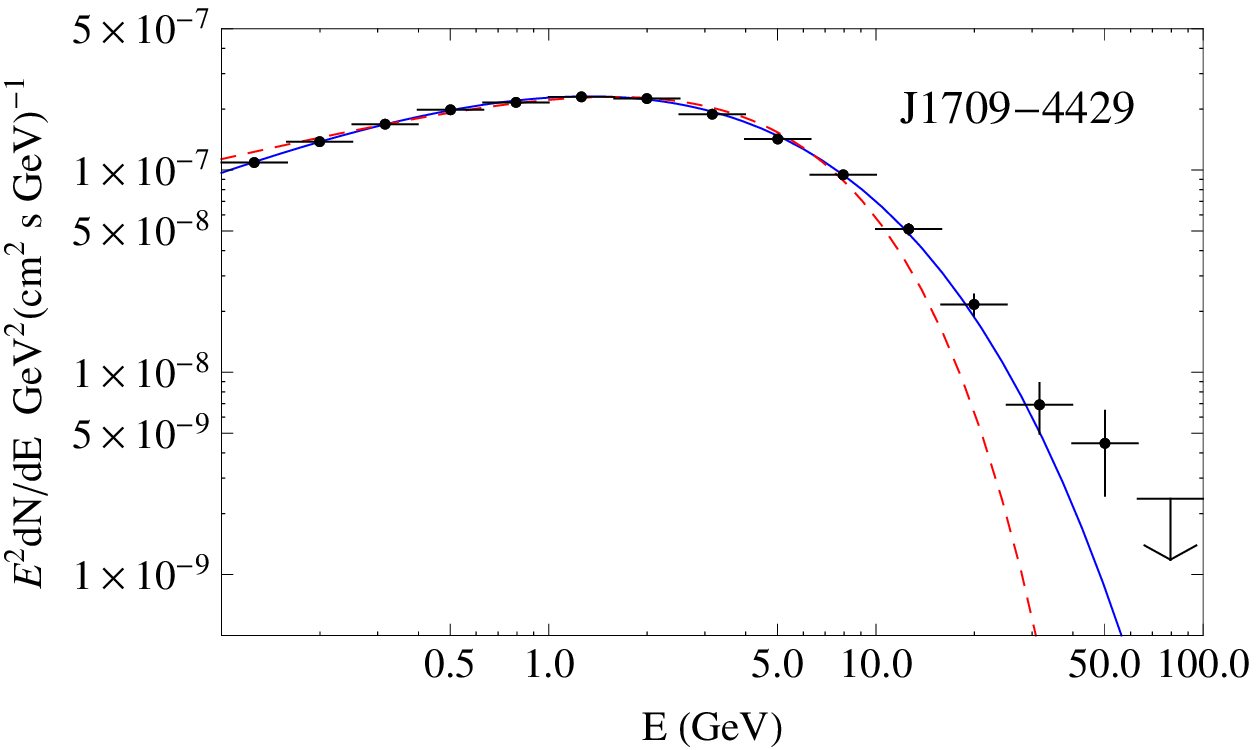} 
\caption{The spectra of several young pulsars. Dashed red lines denote the best fit using the exponentially cut-off power-law parameterization (Eq.~\ref{eq:MSP_Spect}), while the solid blue lines denote the best fit using the sub-exponential cut-off form (Eq.~\ref{eq:MSP_Spect2}).}
\label{youngspec1}
\end{figure*}

\begin{figure*}
\includegraphics[width=3.40in,angle=0]{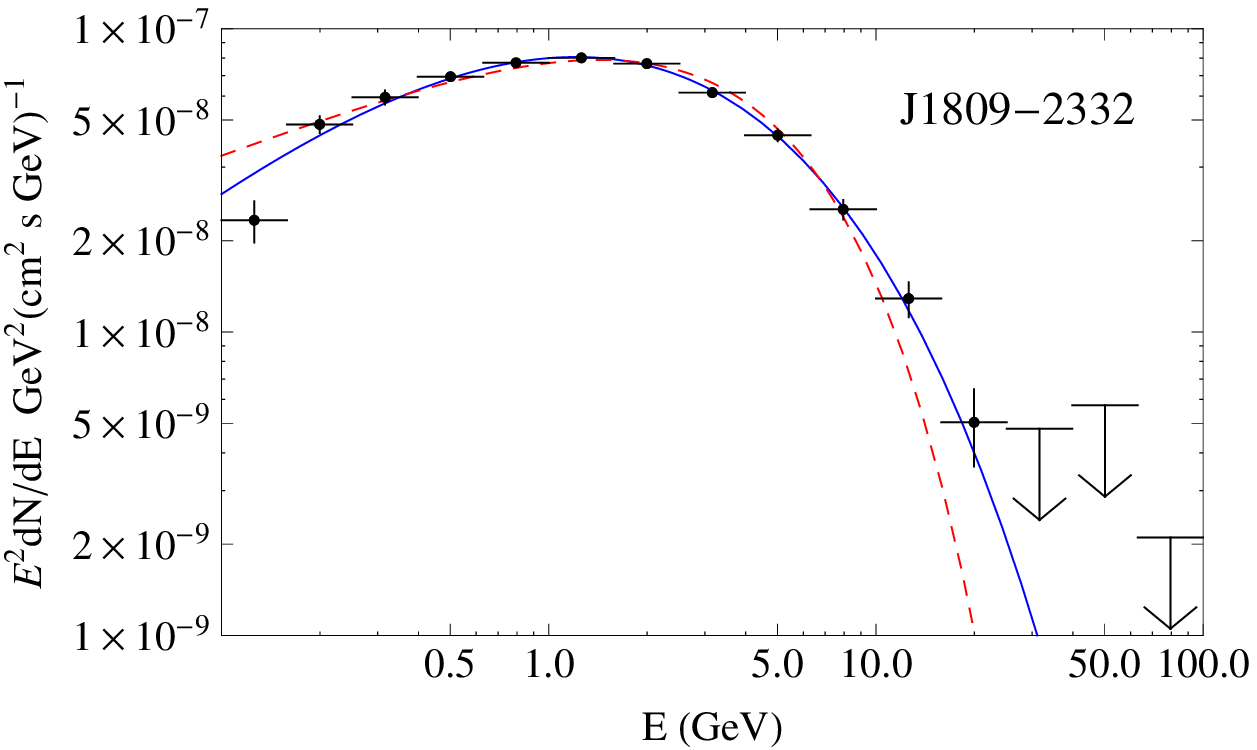} 
\includegraphics[width=3.40in,angle=0]{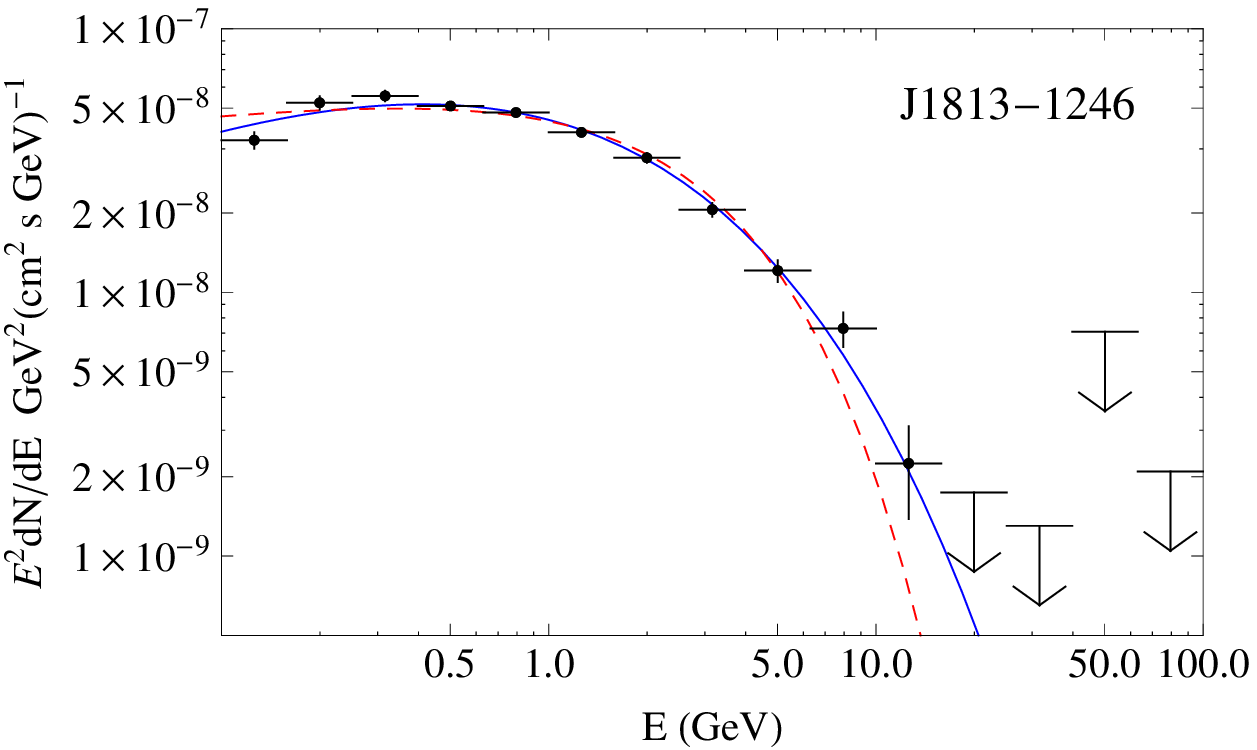} \\
\includegraphics[width=3.40in,angle=0]{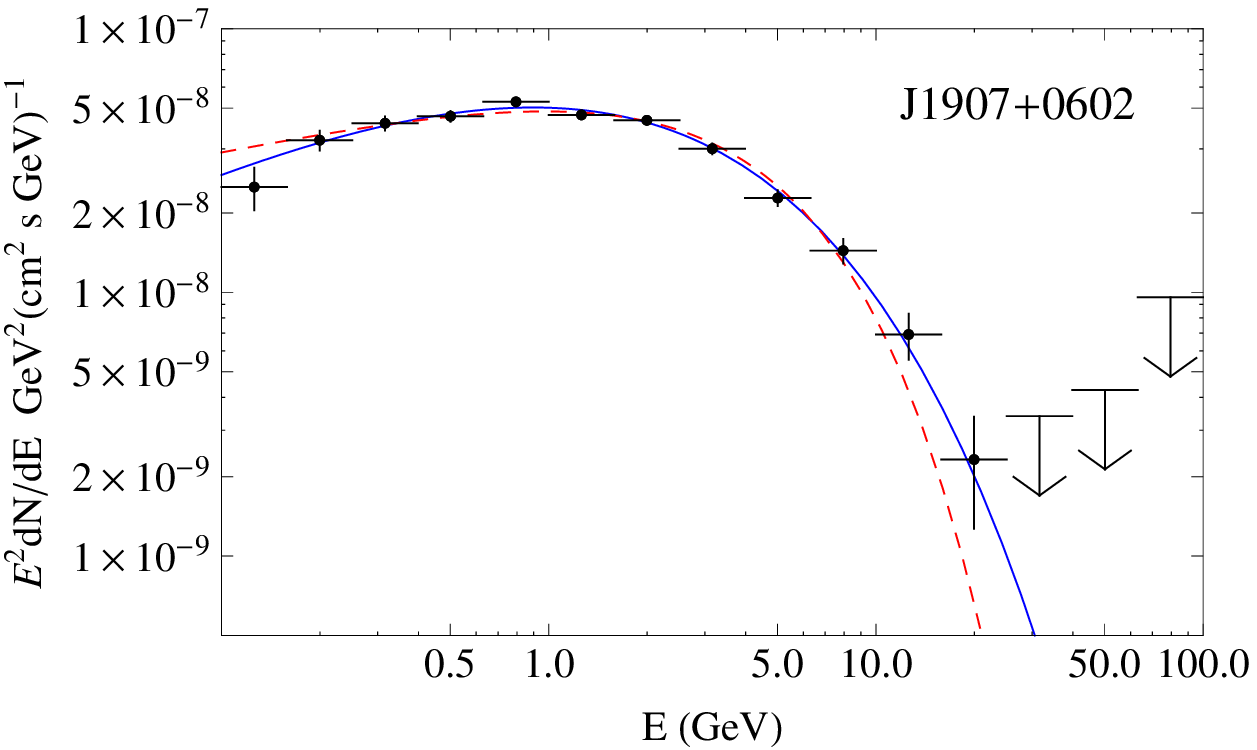} 
\includegraphics[width=3.40in,angle=0]{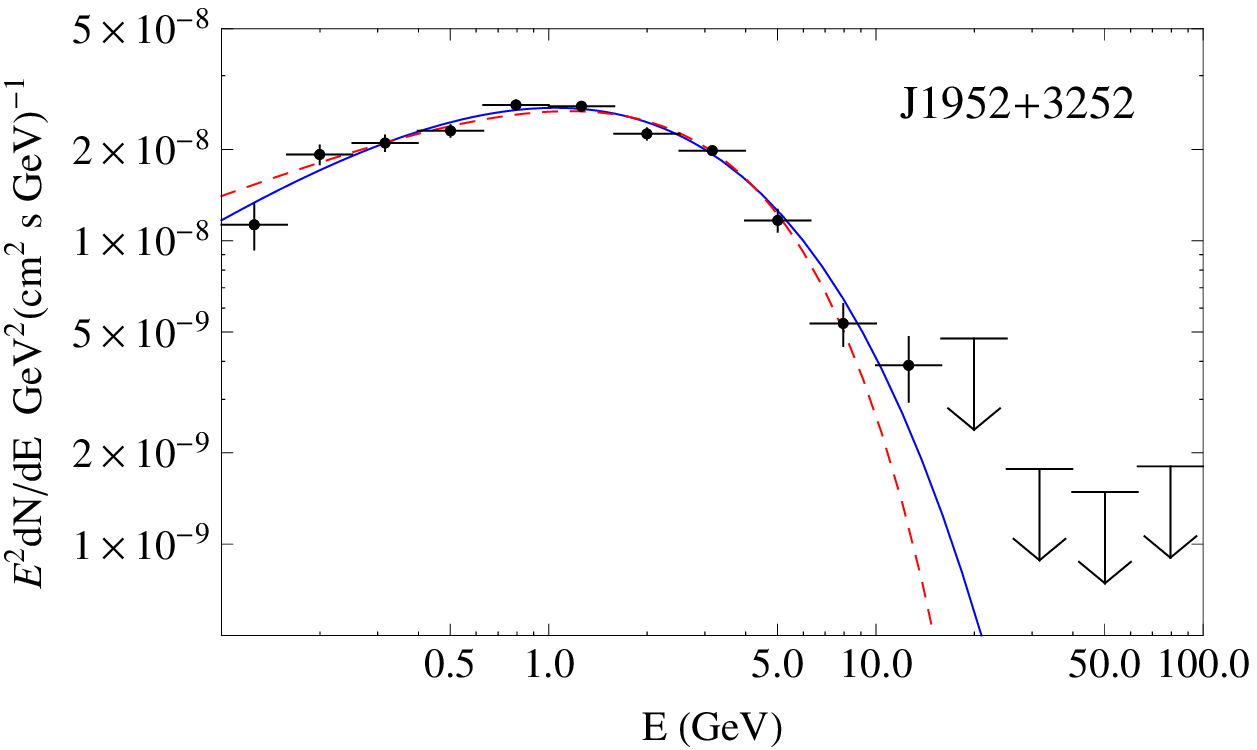} \\
\includegraphics[width=3.40in,angle=0]{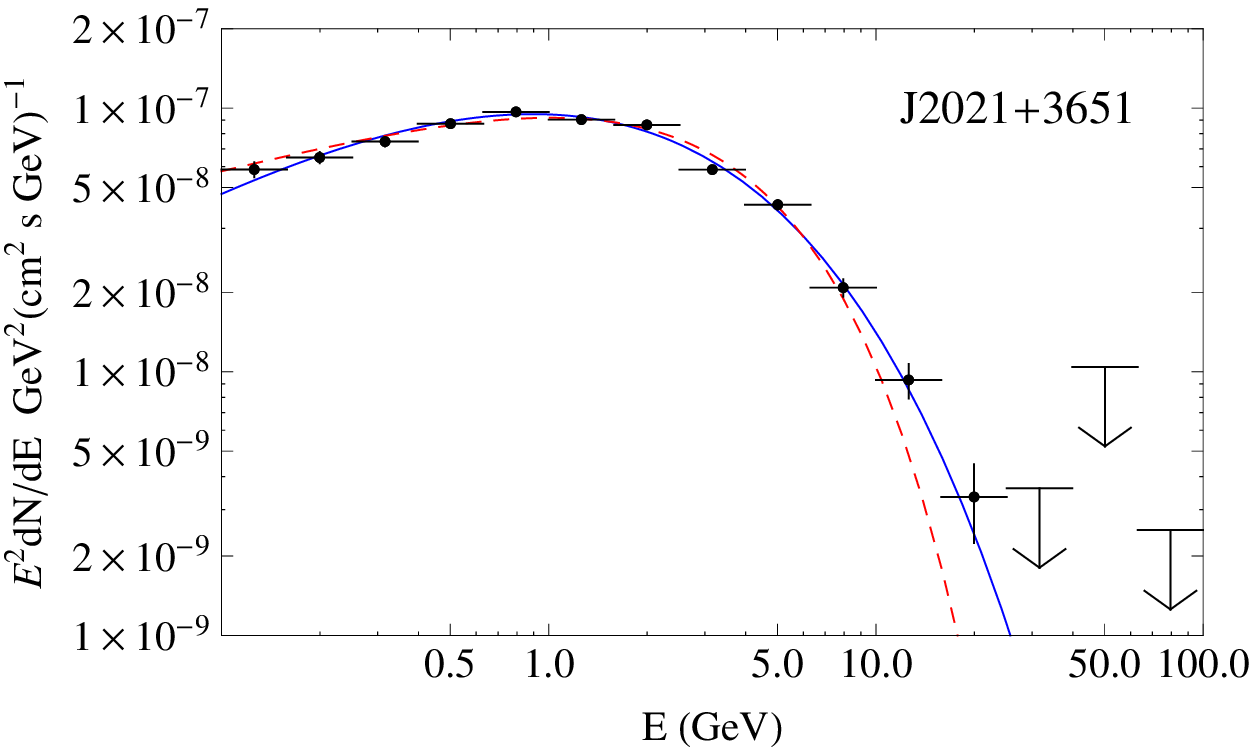} 
\includegraphics[width=3.40in,angle=0]{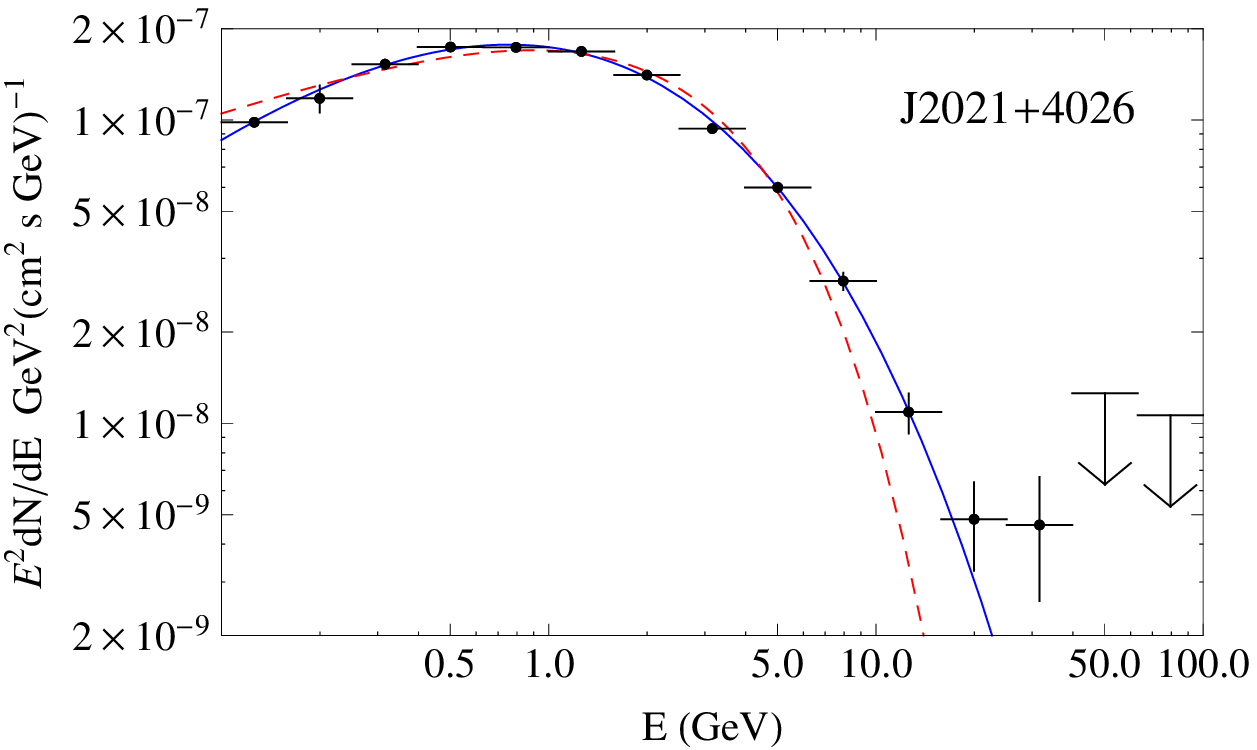} \\
\includegraphics[width=3.40in,angle=0]{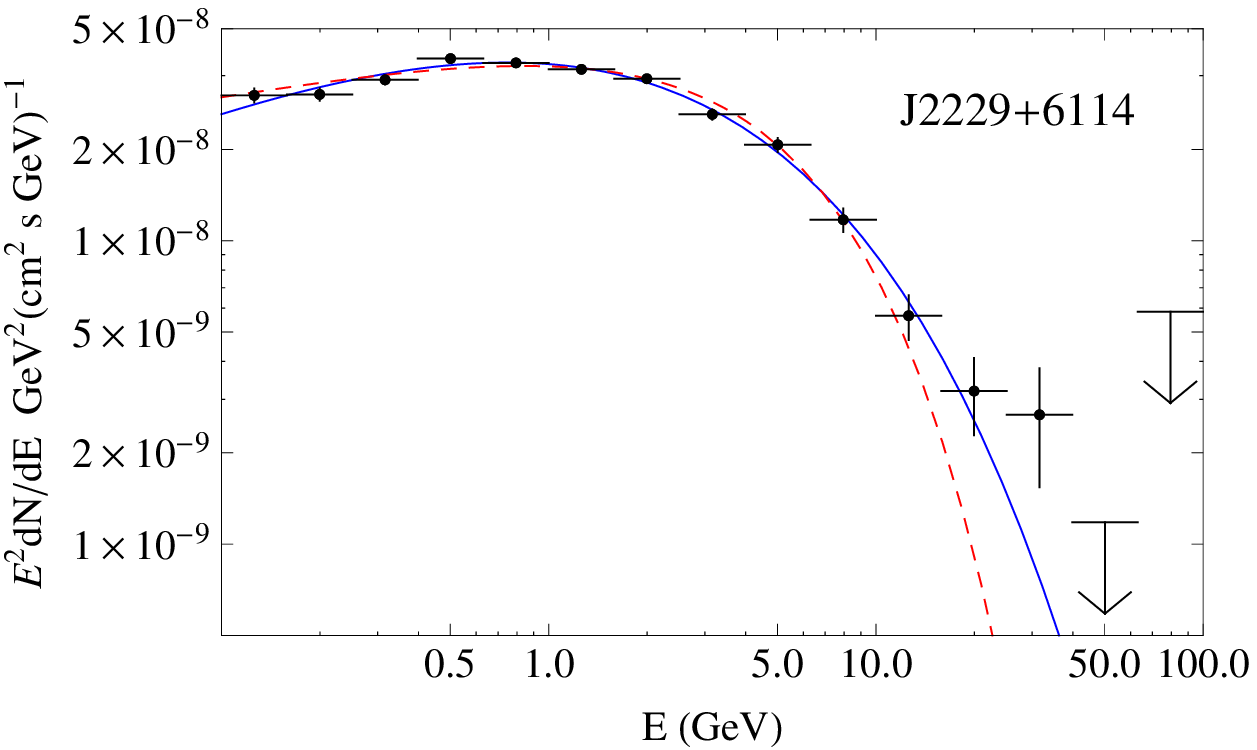} 
\includegraphics[width=3.40in,angle=0]{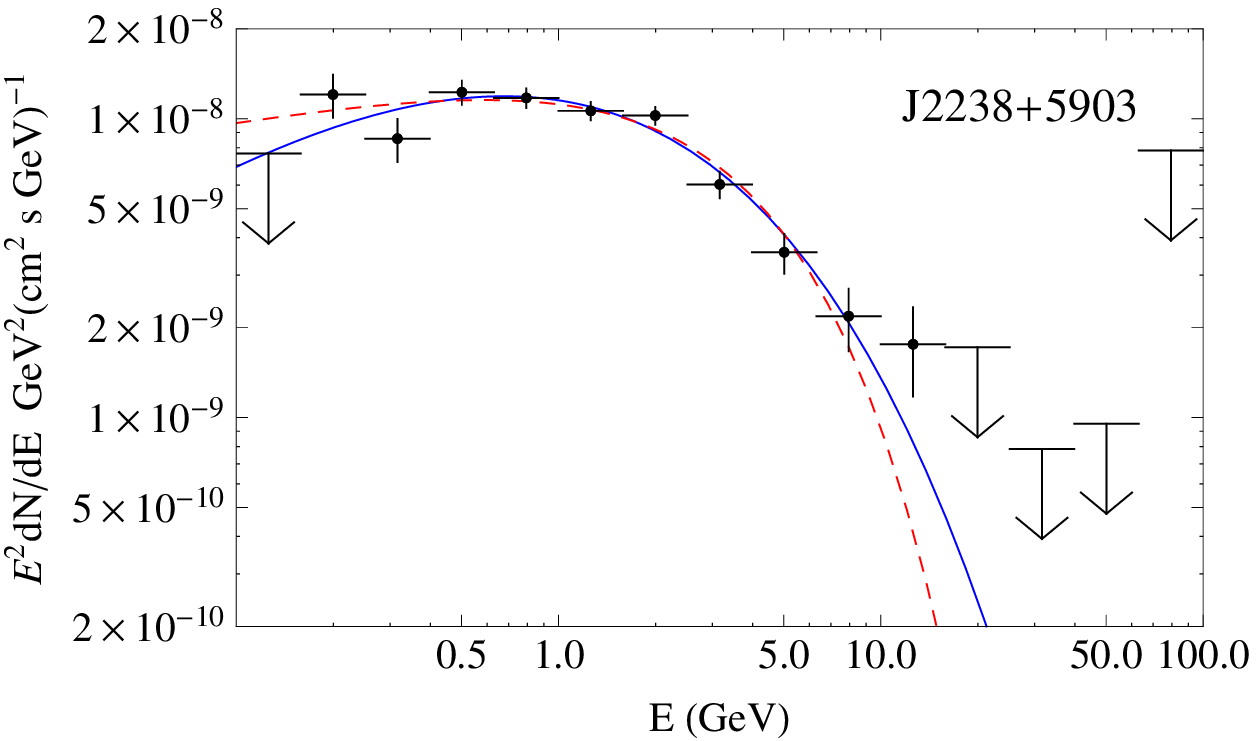}  
\caption{The spectra of several young pulsars (continued). Dashed red lines denote the best fit using the exponentially cut-off power-law parameterization (Eq.~\ref{eq:MSP_Spect}), while the solid blue lines denote the best fit using the sub-exponential cut-off form (Eq.~\ref{eq:MSP_Spect2}).}
\label{youngspec2}
\end{figure*}

\bibliography{MSP_properties7}
\bibliographystyle{apsrev}

\end{document}